\documentclass[a4paper,11pt]{article}

\usepackage{jheppub} 
\usepackage{amsmath}
\usepackage{amssymb}
\usepackage{booktabs} 
\usepackage{mathrsfs}
\usepackage{bbm}
\usepackage{bm}
\usepackage{graphicx}
\usepackage{epsfig}
\usepackage{dsfont}
\usepackage{amsfonts}
\usepackage{color}
\usepackage{array,longtable}
\usepackage{multirow}
\usepackage{slashed}
\usepackage[utf8]{inputenc}

\usepackage{lineno}
\nolinenumbers

\usepackage{orcidlink}
\usepackage{adjustbox}
\usepackage{lmodern}

\allowdisplaybreaks[4]

\title{\bf \boldmath Study of $\eta^\prime \to \eta \pi\pi $ Decays in Large-$N_C$ Chiral Perturbation Theory}

\author[a]{Feng-Zhi Chen\,\orcidlink{0000-0002-5586-0070},}
\author[b,1]{Xin-Qiang Li,\note{Corresponding author.}\orcidlink{0000-0002-3962-3577}}
\author[b,c]{Ya-Dong Yang\,\orcidlink{0000-0002-9327-0557}}
\author[b]{and Yuan-He Zou\,\orcidlink{0009-0009-3999-0244}}

\affiliation[a]{Department of Physics, College of Physics and Optoelectronic Engineering, Jinan University,\\ Guangzhou 510632, China}
\affiliation[b]{Institute of Particle Physics and Key Laboratory of Quark and Lepton Physics~(MOE), Central China Normal University,\\ Wuhan, Hubei 430079, China}
\affiliation[c]{Institute of Particle and Nuclear Physics, Henan Normal University,\\ Xinxiang, Henan 453007, China}

\emailAdd{fzchen@jnu.edu.cn}
\emailAdd{xqli@mail.ccnu.edu.cn}
\emailAdd{yangyd@mail.ccnu.edu.cn}
\emailAdd{yuanhe99zou@mails.ccnu.edu.cn}

\abstract{We investigate the $\eta^\prime \to \eta \pi\pi$ decays within the framework of large-$N_{C}$ chiral perturbation theory, by calculating the decay amplitudes up to next-to-next-to-leading order in a simultaneous expansion in powers of external momenta, quark masses, and $1/N_C$. Projecting the amplitudes onto partial waves allows us to implement a unitarization procedure to account for the $S$- and $D$-wave $\pi\pi$ final-state interactions. The relevant low-energy constants are determined by fitting our theoretical results to the precise experimental data from the A2 collaboration. A comparison of fits with and without $\pi\pi$ final-state interactions demonstrates that including these effects significantly improves the agreement of our theoretical predictions with the experimental measurements. Consequently, the Dalitz-plot parameters are extracted as $a=-0.085(18)_{\mathrm{stat}}(4)_{\mathrm{syst}}$, $b=-0.081(10)_{\mathrm{stat}}(6)_{\mathrm{syst}}$, and $d=-0.045(6)_{\mathrm{stat}}(8)_{\mathrm{syst}}$. Our results provide therefore a refined theoretical description of the $\eta^\prime \to \eta \pi\pi$ decay dynamics.}


\begin{document} 
\maketitle
\flushbottom

\section{Introduction}
\label{sec:intro}

Chiral perturbation theory (ChPT)~\cite{Weinberg:1978kz,Gasser:1982ap,Gasser:1984gg,Gasser:1983yg} has been established as the rigorous effective field theory of quantum chromodynamics (QCD) in the low-energy regime; for comprehensive reviews, we refer the reader to refs.~\cite{Pich:1995bw,Ecker:1994gg,Scherer:2002tk,Bernard:2006gx}. Based on the spontaneous breaking of the $SU(3)_L \otimes SU(3)_R$ chiral symmetry, ChPT systematically describes the dynamics of the Goldstone boson octet ($\pi, K, \eta$) through an expansion in powers of momenta and light quark masses. However, including the $\eta^\prime$ meson in this framework presents a fundamental challenge. Due to the $U(1)_A$ axial anomaly, $\eta^\prime$ is not a Goldstone boson in the chiral limit, and its mass remains substantial even when the masses of light quarks are neglected. Consequently, the standard $SU(3)$ ChPT does not treat $\eta^\prime$ as an explicit degree of freedom. This difficulty is resolved in the large-$N_C$ limit of QCD ($N_C \to \infty$), where the anomaly $U(1)_A$ is suppressed by $1/N_C$. In this limit, the symmetry group extends to $U(3)_L \otimes U(3)_R$, and $\eta^\prime$ becomes the ninth Goldstone boson. This leads to the formulation of large-$N_C$ ChPT~\cite{Kaiser:2000gs}, which employs a simultaneous triple expansion in powers of the external momenta $p^2$, the quark masses $m_q$, and the $1/N_C$, allowing for a consistent description of the generic nonet pseudo-scalar mesons.

The hadronic decays $\eta^\prime \to \eta \pi \pi$ offer a unique laboratory to test these theoretical frameworks and investigate low-energy QCD dynamics~\cite{Escribano:2010wt,Gonzalez-Solis:2018xnw,Beisert:2002ad,Borasoy:2005du,Borasoy:2006uv,Kubis:2009sb,Isken:2017dkw,Akdag:2021efj}. A distinctive feature of these decays is that the vector resonances are forbidden by G-parity conservation, implying that the decays are dominated only by the scalar resonances, specifically $a_0(980)$ and $f_0(980)/f_0(500)$~\cite{Escribano:2010wt,Gonzalez-Solis:2018xnw,Fariborz:1999gr,Singh:1975aq}. From a theoretical perspective, the leading-order (LO) contribution in large-$N_C$ ChPT is chirally suppressed, resulting in decay rates and spectral shapes that disagree with the available experimental data. Previous studies have already shown that the next-to-leading-order (NLO) contributions are dominant~\cite{Escribano:2010wt,Gonzalez-Solis:2018xnw,Herrera-Siklody:1999sbe}, and the $\pi\pi$ final-state interactions (FSI) play a critical role~\cite{Escribano:2010wt,Gonzalez-Solis:2018xnw}. Although the NLO calculations combined with resonance models have improved the theoretical description, a systematic analysis incorporating both higher-order effects and a rigorous unitarization procedure is desirable to match the increasing precision of the experimental measurements. The presence of $\pi^0$, $\eta$ and $\eta^\prime$ in these processes is also ideal for studying their mixing properties~\cite{Guo:2015xva,Bickert:2016fgy,Feldmann:1998vh} and extracting the light quark masses~\cite{Borasoy:2006uv,Guevara:2025weo}.

On the experimental side, the decays $\eta^\prime \to \eta \pi \pi$ have been measured with increasing precision. The charged channel $\eta^\prime \to \eta \pi^+ \pi^-$ has been investigated by the VES~\cite{Dorofeev:2006fb} and BESIII~\cite{BESIII:2010niv,BESIII:2017djm} collaborations. Meanwhile, the neutral channel $\eta^\prime \to \eta \pi^0 \pi^0$ has been studied by the GAMS-$4\pi$~\cite{Blik:2009zz}, A2~\cite{Adlarson:2017wlz} and BESIII~\cite{BESIII:2017djm,BESIII:2022tas} collaborations. In particular, a pronounced cusp structure around the $\pi^+\pi^-$ mass threshold has been observed in the $\pi^0\pi^0$ invariant mass spectrum of $\eta^\prime \to \eta \pi^0 \pi^0$ decay~\cite{Adlarson:2017wlz,BESIII:2017djm}. These high-statistics datasets, particularly from the A2 and BESIII collaborations, provide strong constraints on the Dalitz-plot parameters and necessitate theoretical predictions beyond the standard approximations. Future high-precision measurements expected from the CLAS and GlueX collaborations at JLab~\cite{CLAS:2003umf,GlueX:2020idb,Dudek:2012vr,Gan:2020aco}, the upcoming High Intensity Heavy-ion Accelerator Facility (HIAF)~\cite{Chen:2024wad}, as well as the planned Super
Tau-Charm Factories (STCF)~\cite{Achasov:2023gey,Charm-TauFactory:2013cnj,Fang:2021wes}, further motivate the need for refined theoretical studies.

Motivated by the above observations, we will present in this work a comprehensive study of the $\eta^\prime \to \eta \pi \pi$ decays within the framework of large-$N_C$ ChPT up to the next-to-next-to-leading order (NNLO) in a simultaneous expansion in powers of external momenta $p^2$, quark masses $m_q$, and $1/N_C$. To our knowledge, this is the first calculation of these processes up to this order in such a combined expansion. We will systematically choose the relevant effective operators and compute their contributions to the decay amplitudes. To restore unitarity and account for the significant $\pi\pi$ rescattering effects, we will project the decay amplitudes onto partial waves and unitarize the $S$ and $D$ waves using the $N/D$ method~\cite{Oller:1998zr,Pallante:1999qf,Oset:1999cq,Nieves:1999bx,Oller:2000fj,Jamin:2000wn,Oller:2000ma,Guo:2011pa,Chew:1960iv}. The involved low-energy constants (LECs) are determined by fitting our theoretical results to the precise data from the A2 collaboration~\cite{Adlarson:2017wlz}. The Dalitz-plot parameters are then extracted, with the numerical results given by $a=-0.085(18)_{\mathrm{stat}}(4)_{\mathrm{syst}}$, $b=-0.081(10)_{\mathrm{stat}}(6)_{\mathrm{syst}}$, and $d=-0.045(6)_{\mathrm{stat}}(8)_{\mathrm{syst}}$. With these efforts, a refined theoretical description of $\eta^\prime \to \eta \pi\pi$ decays is provided, which could be confronted with future high-precision measurements at the JLab~\cite{Gan:2020aco}, HIAF~\cite{Chen:2024wad} and STCF~\cite{Achasov:2023gey,Charm-TauFactory:2013cnj} facilities.

The paper is organized as follows. In section~\ref{sec:theory}, we outline the large-$N_C$ ChPT formalism and present the analytical expressions for the decay amplitudes up to NNLO. Section~\ref{sec:kinematics} details the kinematics, partial-wave decomposition, and the unitarization procedure. In section~\ref{sec:fitdata}, we present our numerical results, including the determination of the LECs and the extraction of the Dalitz-plot parameters. A summary is finally given in section~\ref{sec:conclusion}. For convenience, explicit expressions for both the $\eta^\prime\to\eta\pi\pi$ decay and the $\pi\pi$ rescattering amplitudes are collected in appendices~\ref{app:decayamp} and \ref{app:pionscattering}, respectively. 

\section{\texorpdfstring{Large-$\boldsymbol{N_C}$}{Large-Nc} chiral perturbation theory}
\label{sec:theory}

\subsection{Lagrangian and power counting}

To incorporate the $\eta'$ meson into the low-energy effective theory, we consider the framework of large-$N_C$ ChPT, where the anomaly $U(1)_A$ is naturally suppressed in the large-$N_C$ limit. The QCD Lagrangian for light quarks coupling to external fields is expressed as
\begin{align}\label{eq:QCD}
    \mathcal{L} = \mathcal{L}_{\text{QCD}}^{0} + \bar{q}\gamma^\mu (v_\mu + \gamma_5 a_\mu)q + \bar{q}(i\gamma_5 p - s)q -\theta \omega\,,
\end{align}
where $\mathcal{L}_{\text{QCD}}^{0}$ is the Lagrangian for light quarks $q = (u, d, s)^T$ in the exact chiral limit and when the vacuum angle $\theta$ is set to zero. The external fields $v_\mu, a_\mu, s$, and $p$ denote vector, axial-vector, scalar, and pseudo-scalar sources, respectively. They are all hermitean $3\times3$ matrices in flavor space, with the mass matrix of the three light quarks contained in the scalar external field $s$. In the combined chiral and large-$N_C$ limits, the theory exhibits a $U(3)_L \otimes U(3)_R$ symmetry, which is spontaneously broken to $U(3)_V$. The resulting pseudo-scalar nonet, comprising the standard octet and the singlet $\eta_0$, acts as the dynamical degrees of freedom. Furthermore, the term $\theta$ in eq.~\eqref{eq:QCD} reflects the non-trivial features connected with the axial $U(1)_A$ anomaly, with $\omega$ being the topological charge density. In such an effective theory, the anomaly $U(1)_A$ manifests itself as a mass term for the singlet $\eta_0$ via the Witten-Veneziano mechanism~\cite{Witten:1979vv,Veneziano:1979ec}, explicitly breaking the symmetry $U(3)_A$. The nine resulting Goldstone bosons are embedded in the $3\times3$ unitary matrix
\begin{align}\label{eq:U}
U(x) = \exp\left(\frac{i\phi(x)}{F_0}\right)\,,
\end{align}
with
\begin{align}\label{eq:phi}
	\phi = \sum_{a=0}^8 \phi_a \lambda_a =
	\begin{pmatrix}
        \pi^0 + \frac{1}{\sqrt{3}}\eta_8 + \sqrt{\frac{2}{3}}\eta_0 \quad & \sqrt{2}\pi^+ \quad & \sqrt{2}K^+ \\[0.15cm]
		\sqrt{2}\pi^- \quad & -\pi^0 + \frac{1}{\sqrt{3}}\eta_8 + \sqrt{\frac{2}{3}}\eta_0 \quad & \sqrt{2}K^0 \\[0.15cm]
		\sqrt{2}K^- \quad & \sqrt{2}\bar{K}^0 \quad & -\frac{2}{\sqrt{3}}\eta_8 + \sqrt{\frac{2}{3}}\eta_0
	\end{pmatrix}\,,
\end{align}
where $F_0$ is the decay constant of the pion in the chiral limit and is counted as of $\mathcal{O}(\sqrt{N_C})$. $\lambda_0 = \sqrt{2/3}\,\mathds{1}$ and $\lambda_a$ ($a=1,\dots,8$) are the Gell-Mann matrices.

The effective Lagrangian $\mathcal{L}_{\text{eff}}$ is constructed through a simultaneous expansion in the powers of the meson momenta $p^2$, the quark masses $m_q$, and the $1/N_C$~\cite{Kaiser:2000gs,Escribano:2010wt,Gonzalez-Solis:2018xnw}. By introducing the expansion parameter $\delta$, with $\mathcal{O}(\delta) \sim \mathcal{O}(p^2/\Lambda_{\chi}^2) \sim \mathcal{O}(m_q/\Lambda_{\chi}) \sim \mathcal{O}(1/N_C)$, where $\Lambda_{\chi} \sim 1~\mathrm{GeV}$ is the scale where ChPT breaks down, the Lagrangian is organized as
\begin{equation}\label{eq:Leff}
    \mathcal{L}_{\text{eff}} = \mathcal{L}^{(\delta=0)} + \mathcal{L}^{(\delta=1)} + \mathcal{L}^{(\delta=2)} + \cdots\,.
\end{equation}
Following the same counting scheme as in ref.~\cite{Kaiser:2000gs}, we can determine the combined order of a given operator as
\begin{equation}
    D_\delta = \frac{D_p}{2} + D_{N_C^{-1}}\,,
\end{equation}
where $D_p$ represents the standard chiral dimension (including both derivatives and quark masses). The $1/N_C$ order is characterized by
\begin{equation}
    D_{N_C^{-1}} = -2 + N_{\text{ft}} + N_\theta\,,
\end{equation}
where $N_{\text{ft}}$ denotes the number of flavor traces, and $N_\theta$ counts the total insertions of the singlet field $\psi \equiv \sqrt{6}\eta_0/F_0$, the external source $\theta$ associated with the topological charge density $\omega$ and its derivative $D_\mu \theta$. In this way, we can identify the large-$N_C$ scaling behavior of the LECs associated with the corresponding operators.

The LO ($\delta=0$) Lagrangian reads~\cite{Leutwyler:1996sa,Kaiser:2000gs}
\begin{align}\label{eq:LO}
    \mathcal{L}^{(\delta=0)} = \frac{F_0^2}{4}\langle D_\mu U^\dagger D^\mu U\rangle + \frac{F_0^2}{4}\langle U^\dagger\chi + \chi^\dagger U \rangle - \frac{1}{2}\tau (\psi+\theta)^2\,,
\end{align}
where $\langle\cdots\rangle$ indicates the trace over the flavor indices, $\chi = 2B_0(s+ip)$ with $B_0$ being related to the scalar singlet quark condensate, and $\tau$ denotes the topological susceptibility of purely gluonic theory. The two LECs $B_0$ and $\tau$ are both of $\mathcal{O}(N_C^0)$. At NLO ($\delta=1$), the Lagrangian incorporates both $\mathcal{O}(1/N_C)$ and the chiral corrections of $\mathcal{O}(p^2)$ and $\mathcal{O}(N_C\,p^4)$. Specific to the $\eta^\prime\to\eta\pi\pi$ decays we are considering, its explicit expression is given by~\cite{Kaiser:2000gs}
\begin{align}\label{eq:NLO}
    \mathcal{L}^{(\delta=1)} = & \frac{F_0^2}{12}\Lambda_1 D_\mu\psi D^\mu\psi -i\frac{F_0^2}{12}\Lambda_2 (\psi+\theta) \langle U^{\dagger} \chi-\chi^{\dagger} U\rangle + L_2\langle D_\mu U^\dagger D_\nu U D^\mu U^\dagger D^\nu U\rangle \nonumber\\[0.15cm]
    & + (2L_2+L_3) \langle D_\mu U(D^\mu U)^\dagger D_\nu U(D^\nu U)^\dagger \rangle + L_5\langle D_\mu U(D^\mu U)^\dagger(\chi U^\dagger+U\chi^\dagger) \rangle \nonumber\\[0.15cm]
    & + L_8\langle U\chi^\dagger U\chi^\dagger+\chi U^\dagger \chi U^\dagger \rangle\,,
\end{align}
where the covariant derivatives are defined, respectively, by 
\begin{align}
    D_\mu U &= \partial_\mu U-i\left(v_\mu+a_\mu\right) U + i U\left(v_\mu-a_\mu\right)\,, \nonumber \\[0.15cm]
    D_\mu\psi &= \partial_\mu \psi - 2\langle a_\mu \rangle\,, \nonumber \\[0.15cm]
    D_\mu\theta &= \partial_\mu \theta +2\langle a_\mu \rangle\,.
\end{align}
The counting rules imply that the LECs scale as $\Lambda_{1,2}\sim\mathcal{O}(N_C^{-1})$ and $L_{2,3,5,8}\sim\mathcal{O}(N_C)$, respectively. The NNLO Lagrangian, $\mathcal{L}^{(\delta=2)}$, includes the $\mathcal{O}(p^6)$ single-trace operators~\cite{Bijnens:2001bb,Ebertshauser:2001nj,Bijnens:1999sh} as well as the $\mathcal{O}(p^4)$ and $\mathcal{O}(p^2)$ contributions involving multiple traces and insertions of the gauge invariant sum $\psi+\theta$~\cite{Herrera-Siklody:1996tqr}. Specifically, the terms relevant to the decays $\eta^\prime\to\eta\pi\pi$ are given, respectively, by
\begin{align}
	\mathcal{L}^{(\delta=2, N_C p^6)} =& C_1\langle u\cdot u h_{\mu\nu} h^{\mu\nu} \rangle + C_5\langle (u\cdot u)^2\chi_+ \rangle + C_8\langle u\cdot u u_\mu\chi_+ u^\mu \rangle + C_{10}\langle \chi_+ u_\mu u_\nu u^\mu u^\nu \rangle \nonumber\\[0.15cm]
	& + C_{12}\langle \chi_+ h_{\mu\nu}h^{\mu\nu}\rangle + C_{14}\langle u\cdot u \chi_+^2\rangle + C_{17}\langle \chi_+ u_\mu \chi_+ u^\mu \rangle + C_{19}\langle\chi_+^3\rangle \nonumber\\[0.15cm]
	& + iC_{22}\langle \chi_-\{h_{\mu\nu},u^\mu u^\nu\}\rangle + iC_{25}\langle h_{\mu\nu} u^\mu\chi_- u^\nu\rangle + C_{26}\langle u\cdot u \chi_-^2\rangle\nonumber\\[0.15cm]
	& + C_{29}\langle u_\mu\chi_- u^\mu \chi_-\rangle + C_{31}\langle\chi_-^2\chi_+ \rangle \,,\label{eq:NNLOp6} \\[0.2cm]
	\mathcal{L}^{(\delta=2, p^4)} =& L_1\langle D_\mu U^\dagger D^\mu U \rangle^2 + L_2\langle D_\mu U^\dagger D_\nu U\rangle\langle D^\mu U^\dagger D^\nu U \rangle \nonumber\\[0.15cm]
	& + L_4\langle D_\mu U^\dagger D^\mu U \rangle\langle U^\dagger\chi+\chi^\dagger U \rangle + L_6\langle U^\dagger\chi+\chi^\dagger U \rangle^2 \nonumber\\[0.15cm]
	& + L_7\langle U^\dagger\chi-\chi^\dagger U  \rangle^2 + L_{13}\langle U^\dagger D_\mu U \rangle\langle U^\dagger D^\mu U D_\nu U^\dagger D^\nu U \rangle \nonumber\\[0.15cm]
	& + L_{18}\langle U^\dagger D_\mu U \rangle\langle D^\mu U^\dagger\chi-D^\mu U\chi^\dagger \rangle + iL_{31}D_\mu \theta \langle U^\dagger D^\mu U U^\dagger D_\nu UU^\dagger D^\nu U \rangle \nonumber\\[0.15cm]
	& + iL_{46} D_\mu \theta\langle  U^\dagger D^\nu U (U^\dagger\chi+\chi^\dagger U)\rangle + iL_{53} D^\mu D_\mu \theta\langle U^\dagger\chi-\chi^\dagger U \rangle \nonumber\\[0.15cm]
	& + iL_{21}(\psi+\theta)\langle D_\mu U^\dagger D^\mu U(U^\dagger \chi -\chi^\dagger U) \rangle \nonumber\\[0.15cm]
	& + iL_{25}(\psi+\theta)\langle U^\dagger\chi U^\dagger\chi-\chi^\dagger U\chi^\dagger U \rangle\,,\label{eq:NNLOp4}\\[0.2cm]
	\mathcal{L}^{(\delta=2, N_C^{-1}p^2)} &= -\frac{F_0^2}{4} v_1^{(2)}(\psi+\theta)^2\langle D_\mu U^\dagger D^\mu U \rangle -\frac{F_0^2}{4} v_2^{(2)}(\psi+\theta)^2\langle U^\dagger\chi+\chi^\dagger U \rangle\,,\label{eq:NNLOp2}
\end{align}
where $u=\sqrt{U}$, and $C_i$, $L_i$, $v_{1,2}^{(2)}$ denote the relevant LECs. Specifically, $C_i$ stem from the standard $SU(3)$ ChPT Lagrangian of $\mathcal{O}(p^6)$ with single-trace terms, $L_i$ are obtained from the standard $SU(3)$ ChPT Lagrangian of $\mathcal{O}(p^4)$ with two traces or involve the interactions with the singlet field, while $v_{1,2}^{(2)}$ are of $\mathcal{O}(N_C^{-2})$ and arise from expanding the potentials in refs.~\cite{Leutwyler:1996sa,Kaiser:2000gs} up to terms of order $(\psi + \theta)^2$. The chiral building blocks present in eqs.~\eqref{eq:NNLOp6}--\eqref{eq:NNLOp2} are defined, respectively, as
\begin{align}
	\chi_{\pm} &= u^{\dagger} \chi u^{\dagger} \pm u \chi^{\dagger} u\,, \nonumber\\[0.15cm]
	u_\mu &= i\left[u^{\dagger}\left(\partial_\mu-i r_\mu\right) u-u\left(\partial_\mu-i l_\mu\right) u^{\dagger}\right] = i u^{\dagger} D_\mu U u^{\dagger}\,, \nonumber\\[0.15cm]
	h_{\mu \nu} &= \nabla_\mu u_\nu+\nabla_\nu u_\mu\,, \nonumber\\[0.15cm]
	\nabla_\mu X &= \partial_\mu X+\left[\Gamma_\mu, X\right]\,, \nonumber\\[0.15cm]
	\Gamma_\mu &= \frac{1}{2}\left[u^{\dagger}\left(\partial_\mu-i r_\mu\right) u+u\left(\partial_\mu-i l_\mu\right) u^{\dagger}\right]\,,
\end{align}
where $l_\mu=(v_\mu-a_\mu)/2$ and $r_\mu=(v_\mu+a_\mu)/2$ represent the external left- and right-handed fields, respectively.

\subsection{\texorpdfstring{$\eta-\eta^\prime$}{eta-eta'} mixing and relevant parameters}

A detailed treatment of $\eta-\eta'$ mixing is required to describe the decays $\eta' \to \eta \pi \pi$. Defining the flavor basis as $\eta_A \equiv (\eta_8, \eta_0)^T$, we can write the most general Lagrangian quadratic in $\eta_A$ and up to NNLO as~\cite{Bickert:2016fgy}
\begin{align}\label{eq:eff}
    \mathcal{L}_{2} = \frac{1}{2}\partial_\mu \eta_A^T \mathcal{K}_A \partial^\mu \eta_A - \frac{1}{2}\eta_A^T M_A^2 \eta_A + \frac{1}{2}\square \eta_A^T \mathcal{C}_A \square\eta_A\,, 
\end{align}
where $\square\equiv \partial_\mu\partial^\mu$ is the d'Alembert operator, and $\mathcal{K}_A$, $M_A^2$, and $\mathcal{C}_A$ denote the kinetic, mass-squared, and higher-derivative matrices, respectively. Without loss of generality, the three symmetric $2\times2$ matrices can be parametrized, respectively, as
\begin{align}
		\mathcal{K}_A &=\begin{pmatrix}
		1+k_8 & k_{80}\\[0.10cm]
		k_{80} & 1+k_0
		\end{pmatrix}\,,\\[0.15cm]
		M_A^2 &=\begin{pmatrix}
		m_8^2 & m_{80}^2\\[0.10cm]
		m_{80}^2 & m_0^2
		\end{pmatrix}\,,\label{eq:massmatrix} \\[0.15cm]
		\mathcal{C}_A &=\begin{pmatrix}
		c_8 & c_{80}\\[0.10cm]
		c_{80} & c_0
		\end{pmatrix}\,.
\end{align}
At the LO approximation, these matrices are simplified as
\begin{align}\label{eq:lomass}
	\mathcal{K}_A\to \mathds{1}\,, \qquad M_A^2\to M_A^{(0)\,2}=\begin{pmatrix}
		m_8^{(0)\,2} & m_{80}^{(0)\,2}\\[0.2cm]
		m_{80}^{(0)\,2} & m_0^{(0)\,2}
	\end{pmatrix}\,, \qquad
	\mathcal{C}_A\to 0\,, 
\end{align}
where the components of the lowest-order mass-squared matrix are given, respectively, by
\begin{align}
	m_8^{(0)\,2} &= \frac{1}{3}(4m_K^2-m_\pi^2)\,,\notag\\[0.15cm]
	m_0^{(0)\,2} &= M_0^2+\frac{1}{3}(2m_K^2+m_\pi^2)\,, \notag\\[0.15cm]
	m_{80}^{(0)\,2} &= \frac{2\sqrt{2}}{3}(m_\pi^2-m_K^2)\,.
\end{align}
Here, since the higher-order mass corrections are sub-leading at this order, we can take $m_\pi=\sqrt{2B_0\hat{m}}$ and $m_K=\sqrt{B_0(\hat{m}+m_s)}$, with $\hat{m}=m_u=m_d$, as the physical masses of pion and kaon, respectively. The parameter $M_0=\sqrt{6\tau}/F_0$ represents the mass of $\eta_0$ in the chiral and large-$N_C$ limits, which is determined by the potential term in eq.~\eqref{eq:LO}. It can be seen that the mixing already appears at the LO approximation, due to the non-diagonal nature of the mass-squared matrix $M_A^{(0)\,2}$ at this order.

To canonically normalize the kinetic terms in eq.~\eqref{eq:eff}, the higher-derivative term (involving $\square$) must first be eliminated via a field redefinition
\begin{align}
	\eta_A=\left(\mathds{1}+\frac{1}{2}\mathcal{C}_A\square\right)\eta_B\,.
\end{align}
Substituting this into eq.~\eqref{eq:eff}, applying the relation $\eta_B^T \square \eta_B=\partial_\mu(\eta_B^T \partial^\mu \eta_B)-\partial_\mu \eta_B^T \partial^\mu \eta_B$ and neglecting the total-derivative terms, we can see that the $\mathcal{C}_A$ term is canceled by a term originating from the first term in eq.~\eqref{eq:eff}. Keeping only the terms generated by the field transformation up to NNLO, we can write the resulting effective kinetic matrix for the field $\eta_B$ as
\begin{align} \label{eq:kB_new}
	\mathcal{K}_B &= \mathcal{K}_A+\frac{1}{2}(\mathcal{C}_A  M_A^2+M_A^2 \mathcal{C}_A) = \mathcal{K}_A+\frac{1}{2}\begin{pmatrix}
		C_{11}&C_{12}\\[0.15cm]
		C_{21}&	C_{22}
	\end{pmatrix}\nonumber\\[0.2cm]
	&\equiv \begin{pmatrix}
		1+\delta_8^{(1)}+\delta_8^{(2)}&\delta_{80}^{(1)}+\delta_{80}^{(2)}\\[0.15cm]
		\delta_{80}^{(1)}+\delta_{80}^{(2)}&	1+\delta_0^{(1)}+\delta_0^{(2)}
	\end{pmatrix}\,,
\end{align}
where the intermediate coefficients are given, respectively, by
\begin{align}
	C_{11} &= 2c_8m_8^2+2c_{80}m_{80}^2\,,\nonumber\\[0.15cm]
	C_{12} &= C_{21}=(c_0+c_8)m_{80}^2+c_{80}(m_0^2+m_8^2)\,,\nonumber\\[0.15cm]
	C_{22} &= 2c_0 m_0^2+2c_{80}m_{80}^2\,.
\end{align}
Here $\delta_i^{(j)}$ represents the $j$-th order correction to the kinetic matrix elements, with the explicit results provided in refs.~\cite{Guo:2015xva,Bickert:2016fgy}. Analogously, the elements of the mass-sqared matrices $M_B^2=M_A^2$, as defined by eq.~\eqref{eq:massmatrix}, can be expanded perturbatively as
\begin{align}
	m_8^2 &= m_{8}^{(0)\,2}+m_{8}^{(1)\,2}+m_{8}^{(2)\,2}\,,\nonumber\\[0.15cm]
	m_0^2 &= m_{0}^{(0)\,2}{}+m_{0}^{(1)\,2}+m_{0}^{(2)\,2}\,,\nonumber\\[0.15cm]
	m_{80}^2 &= m_{80}^{(0)\,2}+m_{80}^{(1)\,2}+m_{80}^{(2)\,2}\,,
\end{align}
where the superscripts denote the orders of the mass corrections. 

Next, to achieve canonical normalization of the kinetic matrix $\mathcal{K}_B$, we introduce a symmetric matrix $\sqrt{Z}$ and redefine the field as $\eta_B=\sqrt{Z}\eta_C$. To this end, the transformation matrix must be required to satisfy the condition
\begin{align} \label{eq:tiaojian1}
	\sqrt{Z}^T \mathcal{K}_B \sqrt{Z} &= \mathds{1}\,. 
\end{align}
We parametrize $\sqrt{Z}$ perturbatively as
\begin{align} \label{eq:zmatrix}
	\sqrt{Z}= \begin{pmatrix}
		1+d_8^{(1)}+d_8^{(2)}&d_{80}^{(1)}+d_{80}^{(2)}\\[0.15cm]
		d_{80}^{(1)}+d_{80}^{(2)}&	1+d_0^{(1)}+d_0^{(2)}
	\end{pmatrix}\,.
\end{align}
Substituting eqs.~\eqref{eq:kB_new} and \eqref{eq:zmatrix} into eq.~\eqref{eq:tiaojian1} and solving order by order, we can obtain the NLO coefficients as
\begin{align}
	d_8^{(1)}=-\frac{1}{2}\delta_8^{(1)}\,, \qquad
	d_{80}^{(1)}=-\frac{1}{2}\delta_{80}^{(1)}\,, \qquad
	d_{0}^{(1)}=-\frac{1}{2}\delta_{0}^{(1)}\,.
\end{align}
At NNLO, the coefficients are found to be given, respectively, as
\begin{align}
	d_8^{(2)} &= \frac{1}{8}\left[3\delta_8^{(1)\,2}+3\delta_{80}^{(1)\,2}-4\delta_8^{(2)}\right]\,, \nonumber\\[0.15cm]
	d_{80}^{(2)} &= \frac{1}{8}\left[3\delta_{0}^{(1)}\delta_{80}^{(1)}+3\delta_8^{(1)}\delta_{80}^{(1)}-4\delta_{80}^{(2)}\right]\,, \nonumber\\[0.15cm]
	d_{0}^{(2)} &= \frac{1}{8}\left[3\delta_{0}^{(1)\,2}+3\delta_{80}^{(1)\,2}-4\delta_0^{(2)}\right]\,.
\end{align}
With these transformations, the kinetic terms are now canonically normalized, and the Lagrangian expressed in terms of the field $\eta_C$ takes the form 
\begin{align}
	\mathcal{L}_{2}=\frac{1}{2}\partial_\mu \eta_C^T \partial^\mu \eta_C-\frac{1}{2}\eta_C^T M_C^2 \eta_C\,,
\end{align}
where the mass-squared matrix $M_C^2$ is defined by
\begin{align}
	M_C^2=\sqrt{Z}^T M_A^2 \sqrt{Z}
        \equiv\begin{pmatrix}
		\hat{m}_8^2&\hat{m}_{80}^2\\[0.15cm]
		\hat{m}_{80}^2&\hat{m}_0^2
	\end{pmatrix}\,, \label{eq:massmatrix1}
\end{align}
with the elements given explicitly by
\begin{align}
    \hat{m}_8^2 = & m_8^{(0)\,2} \left(1 + 2d_8^{(1)} + d_8^{(1)\,2} + 2d_8^{(2)}\right) 
    + 2m_{80}^{(0)\,2} \left(2d_{80}^{(1)} + 2d_8^{(1)}d_{80}^{(1)} + 2d_{80}^{(2)}\right) \nonumber\\[0.15cm]
    & + m_{8}^{(1)\,2} \left(1 + 2d_{8}^{(1)}\right) + m_0^{(0)\,2} d_{80}^{(1)\,2} 
    + m_8^{(1)\,2} \left(1 + 2d_8^{(1)}\right) + 2m_{80}^{(1)\,2} d_{80}^{(1)}\,, \nonumber\\[0.2cm]
    \hat{m}_{80}^2 = & m_8^{(0)\,2} \left(d_{80}^{(1)} + d_8^{(1)} d_{80}^{(1)} + d_{80}^{(2)}\right) \nonumber\\[0.15cm]
    & + m_{80}^{(0)\,2} \left(1 + d_8^{(1)} + d_8^{(2)} + d_0^{(1)} + d_8^{(1)} d_0^{(1)} + d_0^{(2)} + d_{80}^{(1)\,2}\right) \nonumber\\[0.15cm]
    & + m_{8}^{(1)\,2} d_{80}^{(1)} + m_{0}^{(0)\,2} \left(d_{80}^{(1)} + d_{80}^{(2)} + d_{80}^{(1)} d_0^{(1)}\right) \nonumber\\[0.15cm]
    & + m_{80}^{(1)\,2} \left(1 + d_8^{(1)} + d_0^{(1)}\right) + m_{0}^{(1)\,2} d_{80}^{(1)} + m_{80}^{(2)\,2} \,, \nonumber\\[0.2cm]
    \hat{m}_0^2 = & m_8^{(0)\,2} d_{80}^{(1)\,2} + m_{80}^{(0)\,2} \left(2 d_{80}^{(1)} + 2 d_{80}^{(2)} + 2 d_{80}^{(1)} d_0^{(1)}\right) + m_0^{(2)\,2} + 2m_{80}^{(1)\,2} d_{80}^{(1)} \nonumber\\[0.15cm]
    & + m_{0}^{(0)\,2} \left(1 + 2 d_0^{(1)} + 2 d_0^{(2)} + d_0^{(1)}\right) + m_{0}^{(1)\,2} \left(1 + 2 d_0^{(1)}\right) \,,
\end{align}
up to the second order in the corrections $d_i^{(j)}$ and $m_i^{(j)}$. 

Although the kinetic terms are now normalized, the mass-squared matrix $M_C^2$ in eq.~\eqref{eq:massmatrix1} is still non-diagonal. To diagonalize it without disturbing the kinetic terms, we introduce an orthogonal rotation matrix $\mathcal{R}$ and define the physical field $\eta_D$ through $\eta_C=\mathcal{R}\,\eta_D$, where
\begin{align}
	\mathcal{R}\equiv \begin{pmatrix}
	    \cos\theta & \sin\theta \\[0.15cm]
		  -\sin\theta & \cos\theta 
	\end{pmatrix}\,.  \label{eq:transmatrix}
\end{align}
Identifying $\eta_D$ with the physical mass eigenstates, $\eta_D=(\eta,\eta^\prime)^T$, we obtain the resulting diagonal mass-squared matrix as
\begin{align}
	M_P^2=\mathcal{R}^T\sqrt{Z}^TM_A^2\,\sqrt{Z}\,\mathcal{R}\equiv\begin{pmatrix}
		m_\eta^2&0\\[0.15cm]
		0&m_{\eta^\prime}^2
	\end{pmatrix}\,. \label{eq:mass}
\end{align}
The final Lagrangian in terms of the physical fields is given by
\begin{align}
	\mathcal{L}&=\frac{1}{2}\partial_\mu\eta_D^T\partial^\mu \eta_D-\frac{1}{2}\eta_D^T M_P^2 \eta_D\nonumber\\[0.15cm]
	&=\frac{1}{2}\partial_\mu \eta \partial^\mu \eta-\frac{1}{2}m_\eta^2 \eta^2+\frac{1}{2}\partial_\mu\eta^\prime\partial^\mu \eta^\prime-\frac{1}{2}m_{\eta^\prime}^2\eta^{\prime 2}\,.
\end{align}
From the diagonalization condition specified by eq.~\eqref{eq:mass}, we derive the relation
\begin{align}
	\sin2\theta^{(2)}=\frac{2 \hat{m}_{80}^{(2)\,2}}{m_{\eta^\prime}^2-m_\eta^2}\,, \label{eq:mixingangle}
\end{align}
where the superscript ``$(2)$'' denotes corrections up to the NNLO in $\delta$. The complete relation between the flavor-basis field $\eta_A$ and the physical field $\eta_D$ is given by
\begin{align}
	\eta_A=T\eta_D=\left(1+\frac{1}{2}\mathcal{C}_A \square\right)\sqrt{Z}\mathcal{R}\eta_D \,,
\end{align}
where the transformation matrix $T$ is parametrized as
\begin{align}
	T=\begin{pmatrix}
		-A \sin \theta+B_8 \cos \theta & A \cos \theta+B_8 \sin \theta \\[0.15cm]
		A \cos \theta-B_0 \sin \theta & A \sin \theta+B_0 \cos \theta
	\end{pmatrix}\,,\label{eq:tranmatrix}
\end{align}
with the coefficients
\begin{align}
		A & =-\delta_{80}^{(1)}\left(\frac{1}{2}-\frac{3}{8} \delta_1^{(1)}-\frac{3}{8} \delta_8^{(1)}\right)-\frac{1}{2} \delta_{80}^{(2)}+\frac{c_{81}}{2} \square\,,\nonumber\\[0.15cm]
		B_i & =1-\frac{1}{2} \delta_i^{(1)}+\frac{3}{8} \delta_i^{(1)\,2}+\frac{3}{8} \delta_{80}^{(1)\,2}-\frac{1}{2} \delta_i^{(2)}+\frac{c_i}{2} \square\,.
\end{align}

Equipped with eq.~\eqref{eq:mixingangle}, we can now calculate the $\eta-\eta^\prime$ mixing angle $\theta$ order by order in the combined expansion parameter $\delta$ within the framework of large-$N_C$ ChPT. To this end, we use the numerical values of the relevant input parameters~\cite{ParticleDataGroup:2024cfk},
\begin{align}
    m_{\pi^0} = 0.135~\mathrm{GeV}\,, \quad m_{\eta} = 0.548~\mathrm{GeV}\,, \quad m_{\eta^\prime} = 0.958~\mathrm{GeV}\,, \quad m_K = 0.495~\mathrm{GeV}\,,
\end{align}
as well as the LECs~\cite{Guo:2015xva},
\begin{align}
   \Lambda_1 = -0.17\,, \quad \Lambda_2 = 0.06\,, \quad L_5 = 1.2\times10^{-3}\,, \quad L_8 = 0.5\times10^{-3}\,.
\end{align}  
These parameters will also be used as input for the subsequent calculation of the $\eta^\prime \to \eta \pi \pi$ decay amplitudes. At LO and NLO, the $\eta-\eta^\prime$ mixing angle is determined as $\theta^{(0)} = -19.5^\circ$ and $\theta^{(1)} = -16.69^\circ$, with the corresponding mass parameter of $\eta_0$ given by $M_0^{(0)} = 0.820~\text{GeV}$ and $M_0^{(1)} = 0.859~\text{GeV}$, respectively. However, a direct numerical determination of the NNLO mixing angle $\theta^{(2)}$ through eq.~\eqref{eq:mixingangle} is infeasible due to the large number of unknown LECs. As an alternative, we adopt the experimental value $\theta^{(2)}=\theta_{\mathrm{exp}} = -13.3(5)^\circ$~\cite{KLOE:2006guu} as an external input. Furthermore, we fix $M_0^{(2)} = 0.82~\text{GeV}$ and identify the NNLO masses of $\eta^\prime$ and $\eta$ with their physical values. These constraints allow us to determine the unknown LECs as $v_2^{(2)} = 0.24$, $L_{18} = 1.98\times10^{-3}$, and $L_{25} = 0.93\times10^{-3}$. It must be stressed that these are the effective parameters specific to our current truncation scheme rather than the universal fundamental determinations. As noted in ref.~\cite{Guo:2015xva}, $M_0$ and $v_2^{(2)}$ enter the mass Lagrangian exclusively through the linear combination $M_{0,\text{eff}}^2 = M_0^2 + 6v_2^{(2)}(2m_K^2 + m_\pi^2)$. Although $v_2^{(2)}$ could be theoretically isolated via processes such as $\eta_0\eta_0 \to \pi\pi$ scattering, their strong anti-correlation cannot be resolved by using current mass data on $\eta$ and $\eta^\prime$ alone. Therefore, our explicit choice of $M_0^{(2)} = 0.82~\text{GeV}$ serves as a necessity to break such a theoretical degeneracy in the $M_0^2-v_2^{(2)}$ plane. The primary phenomenological motivation for this procedure is to construct an effective scheme that strictly reproduces the precise physical kinematics of the $\eta-\eta^\prime$ system. This is in turn crucial to ensure accurate phase-space boundaries and vertex couplings for our subsequent calculations of the $\eta^\prime \to \eta \pi \pi$ decay amplitudes.   

\section{\texorpdfstring{$\boldsymbol{\eta^\prime\to\eta\pi\pi}$}{eta' to eta pi pi} decay amplitudes}
\label{sec:kinematics}

\subsection{Kinematics and Dalitz-plot parametrisation}

The decay process $\eta^\prime(p_{\eta^\prime}) \to \eta(p_\eta) \pi(p_1)\pi(p_2)$ is fully described in terms of three Lorentz-invariant Mandelstam variables $s=(p_1+p_2)^2 \equiv m_{\pi\pi}^2$, $t=(p_{\eta^\prime}-p_1)^2 \equiv m_{\eta\pi}^2$, and $u=(p_{\eta^\prime}-p_2)^2$, which fulfill the relation $s+t+u = m_{\eta^\prime}^2 + m_\eta^2 + 2m_\pi^2$. The corresponding differential decay width is expressed as
\begin{align} \label{eq:threebody}
	d\Gamma = \frac{1}{2m_{\eta^\prime} \mathcal{N}} |\mathcal{M}|^2 d\Phi_3(p_{\eta^\prime}; p_\eta, p_1, p_2)\,, 
\end{align}
where $\mathcal{M}$ denotes the decay amplitude, $d\Phi_3$ is the three-body phase-space element, and $\mathcal{N}$ accounts for the number of identical particles in the final state, with $\mathcal{N}=1$ for the charged and $\mathcal{N}=2$ for the neutral decay mode, respectively. By expressing the kinematics in eq.~\eqref{eq:threebody} in terms of the Mandelstam variables $s$, $t$, and $u$, one obtains~\cite{Gonzalez-Solis:2018xnw}
\begin{align}
	\Gamma(\eta^\prime\to\eta\pi\pi) = \frac{1}{256\pi^3 m_{\eta^\prime}^3 \mathcal{N}} \int ds\, dt\, |\mathcal{M}(s,t,u)|^2\,,
\end{align}
where the integration limits for the variable $t$ are given by
\begin{align}
	t_{\text{max}/\text{min}}(s) = \frac{1}{2}\left[m_{\eta^\prime}^2 + m_{\eta}^2 + 2m_\pi^2 - s \pm \frac{\lambda^{1/2}(s,m_{\eta^\prime}^2,m_{\eta}^2)\lambda^{1/2}(s,m_{\pi}^2,m_{\pi}^2)}{s}\right]\,, \label{eq:intlimt1}
\end{align}
while the variable $s$ lies within the range $4m_\pi^2 \le s \le (m_{\eta^\prime}-m_{\eta})^2$, with $\lambda(x,y,z) = x^2+y^2+z^2-2xy-2yz-2zx$ being the K\"{a}ll\'{e}n function.

Experimentally, the decay width of the three-body process $\eta^\prime \to \eta \pi \pi$ is more conveniently expressed in terms of the Dalitz variables $X$ and $Y$ as~\cite{Gonzalez-Solis:2018xnw}
\begin{align}
	\Gamma\left(\eta^{\prime} \rightarrow \eta \pi \pi\right) = \frac{m_\pi Q^2}{128 \sqrt{3} \pi^3 m_{\eta^{\prime}}\left(2 m_\pi+m_\eta\right) \mathcal{N}} \int dX\, dY\, |\mathcal{M}(X, Y)|^2\,, \label{eq:decaywidth}
\end{align}
with 
\begin{equation}
X=\frac{\sqrt{3}}{Q}\left(T_{\pi_{1}}-T_{\pi_{2}}\right)\,, \qquad Y=\frac{m_\eta+2 m_\pi}{m_\pi} \frac{T_\eta}{Q}-1\,,
\end{equation}
where $Q = T_\eta + T_{\pi_1} + T_{\pi_2} = m_{\eta^\prime} - m_\eta - 2 m_\pi$ is the total kinetic energy available in the decay, and $T_\eta$, $T_{\pi_{1,2}}$ denote the kinetic energies of the mesons in the $\eta^\prime$ rest frame. Explicitly, they are given, respectively, by
\begin{equation}
    T_\eta=\frac{\left(m_{\eta^{\prime}}-m_\eta\right)^2-s}{2 m_{\eta^{\prime}}}\,, \quad 
    T_{\pi_{1}}=\frac{\left(m_{\eta^{\prime}}-m_{\pi}\right)^2-t}{2 m_{\eta^{\prime}}}\,, \quad 
    T_{\pi_{2}}=\frac{\left(m_{\eta^{\prime}}-m_{\pi}\right)^2-u}{2 m_{\eta^{\prime}}}\,.
\end{equation}
The integration limits for the two Dalitz variables are given, respectively, by
\begin{align}
    X_{\min/\max}(Y) &= \mp \frac{\sqrt{3}}{2 m_{\eta^{\prime}} Q} \, h\bigg(\big(m_{\eta^{\prime}} - m_\eta\big)^2 - \frac{2 m_{\eta^{\prime}} m_\pi Q}{m_\eta + 2 m_\pi} (Y + 1)\bigg)\,, \\[0.2cm]
    Y_{\text{min}} &= -1\,, \qquad 
	Y_{\text{max}} = \frac{1}{2 m_{\eta^\prime} m_\pi} \left(m_\eta m_{\eta^\prime} - m_\eta^2 + 4 m_\pi^2\right)\,,
\end{align}
with the auxiliary function $h(s)$ defined by
\begin{align}
	h(s) = \frac{\sqrt{\lambda(s, m_{\eta^\prime}^2, m_\eta^2)\, \lambda(s, m_\pi^2, m_\pi^2)}}{s}\,.
\end{align}
These expressions allow us to perform phase-space integration over the Dalitz variables $X$ and $Y$, accounting for both kinematic constraints and the symmetrization requirement for identical pions in the final state. The above formulation also provides a convenient framework for both theoretical calculations of the decay amplitude $\mathcal{M}(X,Y)$ and comparisons with the experimental Dalitz distributions~\cite{Dorofeev:2006fb,BESIII:2010niv,BESIII:2017djm,Adlarson:2017wlz}. 

The Dalitz-plot parameters are obtained by expanding the amplitude squared of the decay in powers of the Dalitz variables $X$ and $Y$ around the center of the Dalitz plot:
\begin{align} \label{eq:amplitude_expanded}
	|\mathcal{M}(X,Y)|^2 = |\mathscr{N}|^2 \left(1 + aY + bY^2 + cX + dX^2 + \cdots \right)\,,
\end{align}
where $\mathscr{N}$ denotes an overall normalization factor, and $a$, $b$, $c$, $d$ represent the Dalitz-plot parameters with real values. For the final state consisting of two neutral pions, the symmetry of the wave function implies that terms containing odd powers of $X$ must vanish. In the case of the $\pi^+\pi^-$ final state, charge-conjugation symmetry obeyed by strong interaction, combined with crossing symmetry between the $t$ and $u$ channels, also ensures that terms with odd powers of $X$ are zero within the Standard Model. Based on these symmetry considerations, the Dalitz-plot parameters provide a convenient and quantitative characterization of the dynamics underlying these decays. We summarize in table~\ref{tab:DalitzExTheory} the experimental and theoretical determinations of the Dalitz-plot parameters for both the charged and neutral decay channels based on the currently available data. One can see that clear deviations are observed among these different determinations, which are intricately linked to the experimental data used to constrain the corresponding theories or models.

\begin{table}[t]
	\centering
	\small 
	\renewcommand{\arraystretch}{1.5}
	\setlength{\tabcolsep}{4pt} 
	\resizebox{0.99\textwidth}{!}{%
		\begin{tabular}{lccccc}
			\toprule
            $\eta^{\prime}\to\eta\pi^{+}\pi^{-}$ & $a$ & $b$ & $c$ & $d$ & \# events \\
			\midrule
			VES~\cite{Dorofeev:2006fb}          & $-0.127(16)(8)$ & $-0.106(28)(14)$ & $0.015(11)(14)$          & $-0.082(17)(8)$ & --- \\
			BESIII~\cite{BESIII:2010niv}        & $-0.047(11)(3)$ & $-0.069(19)(9)$  & $0.019(11)(3)$           & $-0.073(12)(3)$ & $43826$ \\
			BESIII~\cite{BESIII:2017djm}        & $-0.056(4)(3)$  & $-0.049(6)(6)$   & $2.7(2.4)(1.8)\times 10^{-3}$ & $-0.063(4)(4)$  & $351016$ \\
			Borasoy \textit{et al.}~\cite{Borasoy:2005du}             & $-0.116(11)$ & $-0.042(34)$ & $0$ & $0.010(19)$  & --- \\
			Escribano \textit{et al.}~\cite{Escribano:2010wt}         & $-0.098(48)$ & $-0.050(1)$  & $0$ & $-0.092(8)$  & --- \\
			Escribano \textit{et al.}~\cite{Escribano:2010wt}         & $-0.098(48)$ & $-0.033(1)$  & $0$ & $-0.072(1)$  & --- \\
			Kubis \textit{et al.}~\cite{Isken:2017dkw}                & $-0.041(9)$  & $-0.088(7)$  & $0$ & $-0.068(11)$ & --- \\
			Kubis \textit{et al.}~\cite{Isken:2017dkw}                & $-0.148(18)$ & $-0.082(14)$ & $0$ & $-0.086(22)$ & --- \\
			Gonzalez-Solis \textit{et al.}~\cite{Gonzalez-Solis:2018xnw} & $-0.065(7)(8)$ & $-0.048(1)(2)$ & $0$ & $-0.045(7)(5)$ & --- \\
			\midrule
			$\eta^{\prime}\to\eta\pi^{0}\pi^{0}$ & $a$ & $b$ & $c$ & $d$ & \# events \\
			\midrule
			GAMS-$4\pi$~\cite{Blik:2009zz}      & $-0.067(16)(4)$ & $-0.064(29)(5)$ & $0$            & $-0.067(20)(3)$ & $15000$ \\ 
			GAMS-$4\pi$~\cite{Blik:2009zz}      & $-0.066(16)(4)$ & $-0.063(28)(4)$ & $-0.107(96)(3)$ & $0.018(78)(6)$  & $15000$ \\ 
			A2~\cite{Adlarson:2017wlz}          & $-0.074(8)(6)$  & $-0.063(14)(5)$ & ---             & $-0.050(9)(5)$  & $\sim 1.23\times 10^{5}$ \\
			BESIII~\cite{BESIII:2017djm}        & $-0.087(9)(6)$  & $-0.073(14)(5)$ & $0$             & $-0.074(9)(4)$  & $56249$ \\
			BESIII~\cite{BESIII:2022tas}        & $-0.077(3)(1)$  & $-0.066(6)(1)$  & $0$             & $-0.068(4)(1)$  & $\sim 4.3\times 10^5$ \\
			Borasoy \textit{et al.}~\cite{Borasoy:2005du}             & $-0.127(9)$    & $-0.049(36)$   & --- & $0.011(21)$    & --- \\
			Gonzalez-Solis \textit{et al.}~\cite{Gonzalez-Solis:2018xnw} & $-0.072(7)(8)$ & $-0.052(1)(2)$ & $0$ & $-0.051(8)(6)$ & --- \\
			\bottomrule
		\end{tabular}%
	}
\caption{Experimental and theoretical determinations of the Dalitz-plot parameters for the $\eta^{\prime}\to\eta\pi^{+}\pi^{-}$ (upper panel) and $\eta^{\prime}\to\eta\pi^{0}\pi^{0}$ (lower panel) decays. \label{tab:DalitzExTheory}}
\end{table}

\subsection{Analytical structure of the amplitudes}

The decay amplitude of $\eta^\prime \to \eta \pi \pi$ can be expanded according to the parameter $\delta$ within the framework of large-$N_C$ ChPT as
\begin{align}
	\mathcal{M} = \mathcal{M}^{(\delta=0)} + \mathcal{M}^{(\delta=1)} + \mathcal{M}^{(\delta=2)}\,,
\end{align}
where $\mathcal{M}^{(\delta=0)}$, $\mathcal{M}^{(\delta=1)}$, and $\mathcal{M}^{(\delta=2)}$ represent the decay amplitudes at orders $\delta=0$, $1$ and $2$, respectively. The LO decay amplitude is given by~\cite{Cronin:1967jq,Schwinger:1968zz,Schechter:1971efy,Weisz:1972gm,DiVecchia:1980vpx,Fajfer:1987ij,Akhoury:1988as,Herrera-Siklody:1999sbe}
\begin{align}
	\mathcal{M}^{(\delta=0)} = \frac{m_\pi^2}{6 F_0^2}\left[2 \sqrt{2} \cos(2 \theta^{(0)} ) - \sin(2 \theta^{(0)} ) \right]\,. \label{eq:AmpLO}
\end{align}
Here we take $F_0 = F_\pi = 92.2~\text{MeV}$. Using the LO mixing angle $\theta^{(0)} = -19.5^\circ$, we find that the resulting decay branching ratio reads approximately $0.6\%$, which accounts only about $2.7\%$ of the experimental value of $22.4\%$~\cite{ParticleDataGroup:2024cfk} and is therefore far from dominant. As can be seen from eq.~\eqref{eq:AmpLO}, the LO contribution is proportional to the pion mass squared and hence suppressed in the chiral limit. Consequently, higher-order corrections in $\delta$ are expected to play a dominant role in these processes.

The NLO decay amplitude can be calculated using the Lagrangian given by eq.~\eqref{eq:NLO}. This includes tree-level contributions at NLO, as well as corrections arising from wave-function renormalization, pion masses, decay constants, and field redefinitions. For convenience, we collect in appendix~\ref{app:decayamp} the NNLO corrections to the pion mass, decay constant, and wave-function renormalization. The resulting final expression can be written as
\begin{align} \label{eq:AmpNLO}
    \mathcal{M}^{(\delta=1)} =& \frac{1}{3F_\pi^4} \bigg\{ \left(2 \sqrt{2} \cos (2 \theta^{(1)}) - \sin (2 \theta^{(1)}) \right) \Big[ \tfrac{1}{2} F_{\pi}^2 m_{\pi}^2 + 2\left(3 L_2+L_3\right) \nonumber \\[0.15cm]
    & \quad \times \left(s^2 + t^2 + u^2 - m_{\eta'}^4 - m_{\eta}^4 - 2 m_{\pi}^4\right) - 2 L_5 m_{\pi}^2 (m_{\eta'}^2 + m_{\eta}^2 + 2 m_{\pi}^2) + 24L_8 m_{\pi}^4 \Big] \nonumber \\[0.15cm]
    & \quad - \tfrac{1}{2} F_{\pi}^2 \left(\Lambda_1 - 2 \Lambda_2\right) m_{\pi}^2 \left(\sqrt{2} \cos (2 \theta^{(1)}) - 2 \sin (2 \theta^{(1)})\right) \bigg\}\,.
\end{align}
Our NLO result derived above is identical to the one presented in ref.~\cite{Escribano:2010wt}, where the amplitude is conveniently expressed in terms of the so-called Okubo–Zweig–Iizuka (OZI)-allowed and OZI-suppressed contributions, respectively. Keeping in mind that the $\eta-\eta^\prime$ system can be equivalently described in terms of the quark-flavor $(\eta_q, \eta_s)$~\cite{Feldmann:1998vh} and the octet-singlet $(\eta_8, \eta_0)$ basis, which are connected with each other through $\eta_8=(\eta_q-\sqrt{2}\eta_s)/\sqrt{3}$ and $\eta_0=(\sqrt{2}\eta_q+\eta_s)/\sqrt{3}$, we can easily reproduce the individual OZI-allowed and OZI-suppressed contributions, with the latter characterized by the terms proportional to $\Lambda_1$ and $\Lambda_2$ in eq.~\eqref{eq:AmpNLO}. It is also observed from eq.~\eqref{eq:AmpNLO} that even in the chiral limit, the amplitude is not suppressed, with the dominant contribution arising from the terms characterized by $L_2$ and $L_3$.

There are three sources that contribute to the NNLO decay amplitude. According to the power-counting rules of the large-$N_C$ ChPT, the one-loop diagrams generated from the $\delta=0$ Lagrangian contribute to the amplitude at order $\delta=2$. It also receives tree-level contributions from the NNLO Lagrangian, as well as corrections from wave-function renormalization, pion masses, decay constants, and field redefinitions. Its complete expression, being quite lengthy, is relegated to appendix~\ref{app:decayamp}. 

\subsection{\texorpdfstring{$\pi\pi$}{pipi} final-state interactions}

For a three-body decay, the two-body FSI can occur either by means of a rescattering where two out of the three final-state particles rescatter an arbitrary number of times in each of the two-particle channels considering the third particle as a spectator or by interactions between one of the two rescattering particles and the third spectatoring particle. Compared to these dominant two-body FSI, the three-body FSI are suppressed by power counting and phase space. In the context of $\eta^\prime \to \eta \pi \pi$, previous studies~\cite{Gonzalez-Solis:2018xnw,Escribano:2010wt} have demonstrated that the two-body FSI, especially the $\pi\pi$ rescattering, play a crucial role. In this work, we will incorporate the $\pi\pi$ FSI using the $N/D$ unitarization method~\cite{Oller:1998zr,Pallante:1999qf,Oset:1999cq,Nieves:1999bx,Oller:2000fj,Jamin:2000wn,Oller:2000ma,Guo:2011pa,Chew:1960iv},\footnote{Another unitarization procedure relying on the Khuri-Treinman framework can be found in ref.~\cite{Isken:2017dkw}.} while neglecting the $\eta\pi$ FSI as their effects are expected to be subdominant. 

The invariant amplitude of $\eta^\prime\to\eta\pi\pi$ can be decomposed into partial waves with definite isospin $I$ and angular momentum $J$ via
\begin{align}
	\mathcal{M}^{I}(s,t,u) = \sum_{J} 32\pi(2J+1) P_J(\cos\theta_\pi) \mathcal{M}^{IJ}(s)\,,
\end{align}
where $P_J$ is the Legendre polynomial of order $J$, and $\theta_\pi$ is the angle between the three-momentum of the pion, $\bm{p}_\pi$, and that of the $\eta$ meson, $\bm{p}_\eta$, in the $\pi\pi$ rest frame. The partial-wave amplitude $\mathcal{M}^{IJ}(s)$ is obtained by performing the following projection:
\begin{align}
	\mathcal{M}^{IJ}(s) = \frac{1}{32\pi} \frac{s}{\lambda^{1/2}(s,m_{\eta^\prime}^2,m_\eta^2)\lambda^{1/2}(s,m_\pi^2,m_\pi^2)} \int_{t_{\text{min}}}^{t_{\text{max}}} dt\, P_J(\cos\theta_\pi) \mathcal{M}^{I}(s,t,u)\,,
\end{align}
where the integration limits $t_{\text{max}/\text{min}}$ have already been defined in eq.~\eqref{eq:intlimt1}, and the angle $\theta_\pi$ is expressed in terms of the Mandelstam variables as
\begin{align}
	\cos \theta_\pi = -\frac{s\left(m_{\eta^{\prime}}^2+m_\eta^2+2 m_\pi^2-s-2 t\right)}{\lambda^{1 / 2}(s, m_\eta^2, m_{\eta^{\prime}}^2) \lambda^{1 / 2}\left(s, m_\pi^2, m_\pi^2\right)}\,.
\end{align}
Similarly, the $\pi\pi$ rescattering amplitude with isospin $I$ is expanded into partial waves as
\begin{align}
	T^{I}(s,t,u) = \sum_{J} 32\pi (2J+1) P_J\!\left(1+\frac{2t}{s-4m_\pi^2}\right) \mathcal{T}^{IJ}(s)\,,
\end{align}
with the partial-wave amplitude $\mathcal{T}^{IJ}$ given by
\begin{align}
	\mathcal{T}^{IJ}(s) = \frac{1}{32 \pi} \frac{1}{s-4 m_\pi^2} 
	\int_{4 m_\pi^2-s}^0 dt \, P_J\!\left(1+\frac{2 t}{s-4 m_\pi^2}\right) T^I(s, t, u)\,. \label{eq:pwa}
\end{align}
Here, the isospin amplitude $T^I$ is expressed in terms of the invariant $\pi\pi$ rescattering amplitude $A(s,t,u)$. Due to isospin conservation, the total isospin $I$ of the $\pi\pi$ final state must be zero, and one has $T^0(s,t,u) = 3A(s,t,u) + A(t,u,s) + A(u,s,t)$~\cite{Gasser:1983yg}. The explicit expression of the rescattering amplitude $A(s,t,u)$ can be found in appendix~\ref{app:pionscattering}. As the higher-order partial waves are suppressed by phase space, in this work we calculate only the $S$- and $D$-wave amplitudes $\mathcal{T}^{00}(s)$ and $\mathcal{T}^{02}(s)$ using eq.~\eqref{eq:pwa}.

For the decays $\eta^\prime\to\eta\pi\pi$ considered here, the partial-wave amplitudes $\mathcal{M}^{IJ}(s)$ satisfy the unitarity relation~\cite{Gonzalez-Solis:2018xnw}
\begin{align}
	\text{Im}\mathcal{M}^{IJ}(s) = \sigma_\pi(s) \left(\mathcal{T}^{IJ}(s)\right)^* \mathcal{M}^{IJ}(s)\,, \label{eq:unirelation}
\end{align}
where $\sigma_\pi(s) = \sqrt{1-4m_\pi^2/s}$ is the two-pion phase-space factor. This relation ensures that the decay amplitudes properly include the effects of elastic $\pi\pi$ rescattering. We employ the $N/D$ unitarization method~\cite{Oller:1998zr,Pallante:1999qf,Oset:1999cq,Nieves:1999bx,Oller:2000fj,Jamin:2000wn,Oller:2000ma,Guo:2011pa,Chew:1960iv} to make the amplitudes satisfy eq.~\eqref{eq:unirelation}. The core idea of the $N/D$ method is to factor the amplitude into a numerator function containing only the left-hand cut contributions, and a denominator function containing only the right-hand cut contributions. Within this framework, the partial-wave amplitudes for $\eta^\prime\to\eta\pi\pi$ can be written as~\cite{Guo:2011pa,Gonzalez-Solis:2018xnw}
\begin{align}
	\mathcal{M}^{IJ}_{\eta^\prime\to\eta\pi\pi}(s) = \left[1 + 16\pi g_{\pi\pi}(s) N_{\pi\pi}^{IJ}(s) \right]^{-1} R^{IJ}_{\eta^\prime\to\eta\pi\pi}(s)\,, \label{eq:NDamp}
\end{align}
with
\begin{align}
	N_{\pi\pi}^{IJ}(s) &= \left[\mathcal{T}^{IJ}_{\pi\pi}(s)\right]^{(0)+(1)+(2,\,\text{Tree})+(2,\, \text{Loop})}\,, \\[0.2cm]
	R^{IJ}_{\eta^\prime\to\eta\pi\pi}(s) &= \left[\mathcal{M}^{IJ}_{\eta^\prime\to\eta\pi\pi}(s)\right]^{(0)+(1)+(2,\text{Tree})+(2,\text{Loop})}\,,
\end{align}
where $\left[\mathcal{T}^{IJ}_{\pi\pi}(s)\right]^{(0)+\cdots}$ and $\left[\mathcal{M}^{IJ}_{\eta^\prime\to\eta\pi\pi}(s)\right]^{(0)+\cdots}$ denote the partial-wave amplitudes of the $\pi\pi$ rescattering and $\eta^\prime\to\eta\pi\pi$ decay, respectively. They are both calculated perturbatively within the framework of large-$N_C$ ChPT. The superscripts $(0)$, $(1)$, and $(2, \text{Tree})$ denote the tree-level contributions at orders $\delta = 0, 1, 2$ respectively, while $(2, \text{Loop})$ represents the loop contributions from the $t$- and $u$-channels as well as the inelastic $s$-channel loops. In eq.~\eqref{eq:NDamp}, the function $g_{\pi\pi}(s)$ encodes the discontinuity of the amplitude along the right-hand cut arising from two-pion intermediate states, and is given by~\cite{Guo:2011pa,Gonzalez-Solis:2018xnw}
\begin{align}
	g_{\pi\pi}(s) = \frac{1}{16\pi^2} \left(a_{\pi\pi}(\mu) + \ln\frac{m_\pi^2}{\mu^2} - \sigma_\pi(s)\ln\frac{\sigma_\pi(s)-1}{\sigma_\pi(s)+1}\right)\,, \label{eq:gfun}
\end{align}
where $a_{\pi\pi}(\mu)$ is a subtraction constant that is not directly determined by the unitarization procedure itself, but should be fixed elsewhere, \textit{e.g.}, by fitting to the experimental data. The function $g_{\pi\pi}(s)$ satisfies $\text{Im}\, g_{\pi\pi}(s) = -\sigma_\pi(s)/16\pi$ for $s > 4m_\pi^2$~\cite{Guo:2011pa}. With these prescriptions, we can ensure that eq.~\eqref{eq:NDamp} satisfies the unitarity condition in eq.~\eqref{eq:unirelation} approximately. The reason behind such an approximation is that we restrict the analysis only to elastic $\pi\pi$ FSI, while neglecting contributions such as $\pi\eta$ rescattering or inelastic channels above higher mass thresholds. Although this would induce slight deviations from exact unitarity, our numerical checks confirm that these effects are indeed negligible~\cite{Gonzalez-Solis:2018xnw}.

Finally, we express the unitarized decay amplitude of $\eta^\prime \to \eta \pi\pi$ in terms of its $S$- and $D$-wave components as
\begin{align}
	\mathcal{M}^{I=0}(s,t,u) &= \sum_{J=0,2} 32\pi(2J+1)\, P_J(\cos\theta_\pi)\, \mathcal{M}^{0J}(s) \nonumber\\[0.15cm]
	&= 32\pi P_0(\cos\theta_\pi)\mathcal{M}_{\eta^\prime\to\eta\pi\pi}^{00}(s) + 160\pi P_2(\cos\theta_\pi)\mathcal{M}_{\eta^\prime\to\eta\pi\pi}^{02}(s)\,, \label{eq:ampFSI}
\end{align}
where $\mathcal{M}_{\eta^\prime \to \eta \pi \pi}^{00}(s)$ and $\mathcal{M}_{\eta^\prime \to \eta \pi \pi}^{02}(s)$ denote the corresponding partial-wave amplitudes of the $S$ and $D$ waves, respectively. Once the decay amplitude is determined, the corresponding Dalitz-plot parameters can be extracted by comparison with the experimental data. However, due to the presence of unknown LECs, unambiguous predictions are still not possible. In the next section, we first determine these LECs and then the Dalitz-plot parameters by fitting our theoretical predictions to the available experimental data, with the $\pi\pi$ FSI effects properly taken into account; the quality of this fit will serve as a test of the validity of our theoretical framework.

\section{Fit to experimental data}
\label{sec:fitdata}

\subsection{Dalitz-plot data and fitting procedure}

The binned Dalitz-plot distributions of the $\eta^\prime \to \eta \pi^0 \pi^0$ decay in $(X, Y)$ space are provided by the A2 collaboration~\cite{Adlarson:2017wlz}, with the number of events in each bin related to the theoretically calculated differential decay width $d^2\Gamma(\eta^\prime \to \eta \pi^0 \pi^0)/dX dY$ by~\cite{Gonzalez-Solis:2018xnw}
\begin{align}
	\frac{d^2 N_{\text{events}}}{dX dY} = 2 \frac{N_{\text{events}}}{\Gamma_{\eta^\prime} \text{Br}(\eta^\prime \to \eta \pi^0 \pi^0)} \frac{d^2 \Gamma(\eta^\prime \to \eta \pi^0 \pi^0)}{dX dY} \Delta X \Delta Y\,,
\end{align}
where the factor of 2 accounts for the fact that the experimental data corresponds to only half of the symmetric Dalitz-plot distribution, and $N_{\text{events}}$ denotes the total number of events. $\Gamma_{\eta^\prime}$ is the total decay width of $\eta^\prime$, and $\text{Br}(\eta^\prime \to \eta \pi^0 \pi^0)$ is the branching ratio of the $\eta^\prime \to \eta \pi^0 \pi^0$ decay, for which we adopt the world-averaged value of $22.4\%$~\cite{ParticleDataGroup:2024cfk}. The doubly differential decay width $d^2\Gamma(\eta^\prime \to \eta \pi^0 \pi^0)/dX dY$ can be inferred from eq.~\eqref{eq:decaywidth}. The bin widths of the two variables $X$ and $Y$ are fixed at $\Delta X = \Delta Y= 0.1$~\cite{Adlarson:2017wlz}. 

Depending on the analysis frameworks and the selections of the data samples, the A2 collaboration~\cite{Adlarson:2017wlz} presented two sets of analyses of the same data, with the corresponding efficiency-corrected total event numbers of $463066$ (Analysis~I) and $473044$ (Analysis~II), respectively. In our analysis, both the Analysis~I and Analysis~II datasets are utilized to verify the stability of the parameters fitted in different reconstruction strategies, when the $\pi\pi$ FSI are taken into account. In contrast, for fits performed without the $\pi\pi$ FSI, we primarily use the Analysis~I dataset, as it provides a nearly background-free Dalitz plot and hence minimizes potential systematic uncertainties. The $\chi^2$ function is defined as
\begin{align}
	\chi^2 = \sum_{X,Y} \left(\frac{ N_{X Y}^{\text{th}} - N_{X Y}^{\text{exp}} }{\sigma_{X Y}^{\text{exp}}}\right)^2\,,
\end{align}
where $N_{X Y}^{\text{th}}$ and $ N_{X Y}^{\text{exp}}$ represent the theoretically predicted and experimentally measured number of events in each $(X,Y)$ bin respectively, and $\sigma_{XY}^{\text{exp}}$ is the associated experimental uncertainty in the same bin. The fit is performed over a total of 200 data points~\cite{Adlarson:2017wlz}.

Using the A2 data, our objective is to determine the unknown LECs within the framework of large-$N_C$ ChPT. Although the number of data points is sufficient, the quality of the fit tends to deteriorate if all $U(3)$ LECs are treated as free parameters. To stabilize the fit, we adopt appropriate input values from the literature. Although these LECs are originally determined in the $SU(3)$ framework, the numerical differences between $U(3)$ and $SU(3)$ are expected to be small; thus, using these values introduces only negligible uncertainties. In our analysis, the renormalization scale is set to $\mu = 770~\mathrm{MeV}$. The $\mathcal{O}(p^4)$ LECs are taken as $L_2 = 1.6 \times 10^{-3}$, $L_3 = -3.8 \times 10^{-3}$, $L_4 = 0$, $L_6 = 0$, and $L_7 = -0.3 \times 10^{-3}$~\cite{Bijnens:2014lea}. For the $\mathcal{O}(p^6)$ LECs, we adopt the following values from ref.~\cite{Pan:2023qja} (in units of $\mathrm{GeV}^{-2}$): 
\begin{align}
C_1 &= 1.75 \times 10^{-3}\,,&  C_5 &= -4.25 \times 10^{-4}\,, & C_8 &= 1.24 \times 10^{-3}\,,\notag\\[0.15cm]
C_{10} &= -4.13 \times 10^{-4}\,, & C_{12} &= -7.08 \times 10^{-4}\,, & C_{14} &= -3.91 \times 10^{-4}\,,\notag\\[0.15cm]
C_{17} &= 0.32 \times 10^{-4}\,, & C_{19} &= -6.38 \times 10^{-4}\,, &
C_{22} &= -0.69 \times 10^{-3}\,,\notag\\[0.15cm]
C_{26} &= -0.51 \times 10^{-3}\,, & C_{29} &= -2.73 \times 10^{-3}\,, & C_{31} &= 2.53 \times 10^{-4}\,.
\end{align}
We then perform two types of fits, one based on the perturbative amplitude and the other including the $\pi\pi$ FSI effects.

\subsection{Fit with \texorpdfstring{large-$N_C$}{large-NC} perturbative amplitude}
\label{sec:fit1}

We first consider the fit to the decay amplitude calculated perturbatively within the large-$N_C$ ChPT, which yields
\begin{align}
	L_{13} = 0.97(9) \times 10^{-3}\,, \qquad L_{21} = 8.1(6) \times 10^{-3}\,,\qquad  v_1^{(2)} = -0.24(2)\,,  \label{eq:fitpar1}
\end{align}
with a $\chi^2/\text{d.o.f} = 224.5/197 \simeq 1.14$. The error bars of the fitted parameters in eq.~\eqref{eq:fitpar1} represent only the statistical uncertainties during the fit. Since the resulting uncertainties are generally asymmetric, they are already symmetrized and rounded when presenting these numerical values. For a $\chi^2$ distribution with a large number of degrees of freedom, we determine the allowed parameter intervals using the following criterion~\cite{Etkin:1981sg}:
\begin{align}
	\Delta\chi^2 \le n_\sigma \sqrt{2\chi_0^2}\,, \label{eq:uncert}
\end{align}
where $n_\sigma$ denotes the number of standard deviations, $\chi_0^2$ is the minimum value of the $\chi^2$ function obtained at the best-fit point, and $\Delta\chi^2 = \chi^2 - \chi_0^2$. Using eq.~\eqref{eq:uncert}, we determine the allowed intervals of the LECs and evaluate the corresponding observables within this parameter space. At the $68\%$ confidence level, the uncertainty of each observable is estimated as the maximal deviation from its value at the best-fit point. Using this procedure, we obtain the Dalitz-plot parameters as
\begin{align}
	a = -0.076(15)\,, \qquad b = -0.083(1)\,, \qquad d = -0.059(5)\,. \label{eq:Daplot1}
\end{align}
These results can be compared with those extracted by the A2 collaboration~\cite{Adlarson:2017wlz}
\begin{equation}
a = -0.074(8){\text{stat}}(6){\text{syst}}\,, ~~
b = -0.063(14){\text{stat}}(5){\text{syst}}\,, ~~
d = -0.050(9){\text{stat}}(5){\text{syst}}\,.
\end{equation} 
It can be seen that the differences between the theoretical and experimental central values are small for $a$ and $d$, while the deviation is somewhat larger for $b$. Nevertheless, once both theoretical and experimental uncertainties are taken into account, all three parameters remain statistically compatible with the corresponding experimental data. 

\begin{figure}[t]
	\centering
	\includegraphics[width=0.49\textwidth]{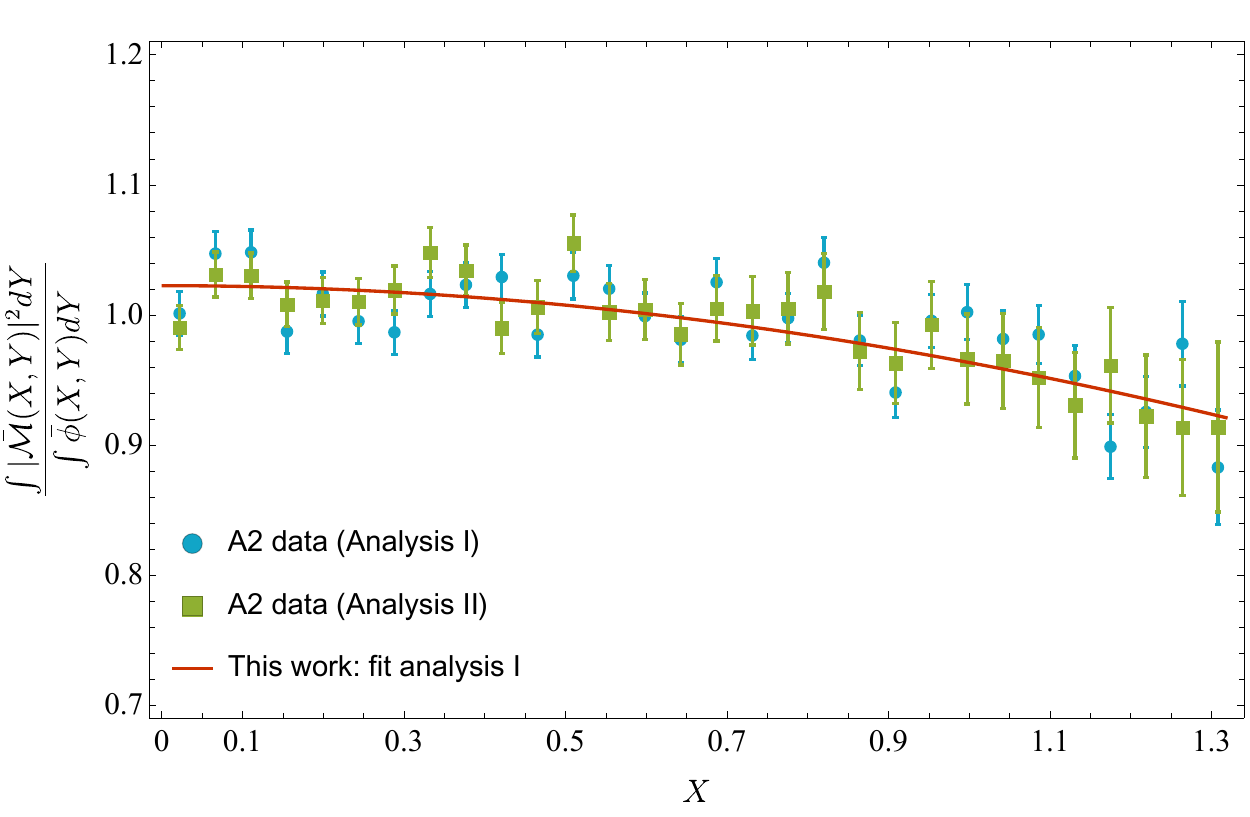}\hfill
	\includegraphics[width=0.49\textwidth]{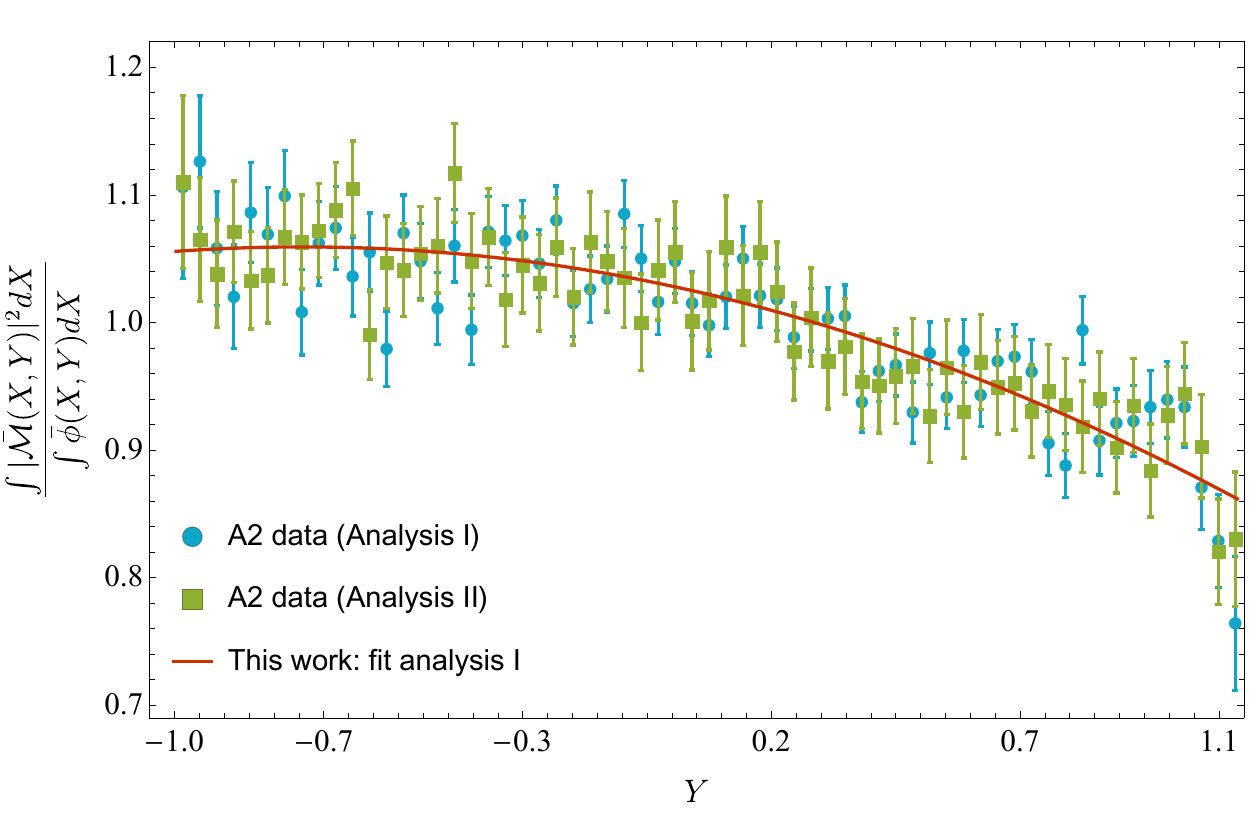} \\[5pt]
	\includegraphics[width=0.49\textwidth]{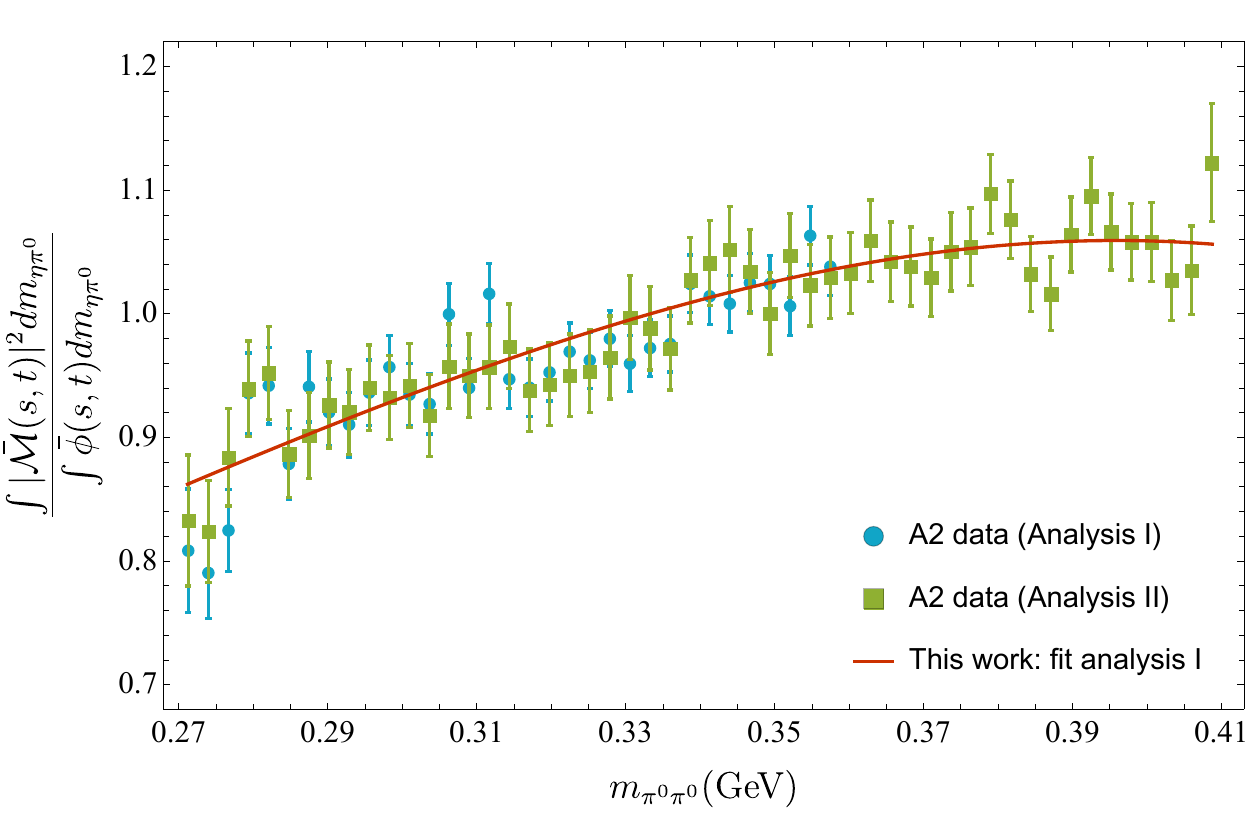}\hfill
	\includegraphics[width=0.49\textwidth]{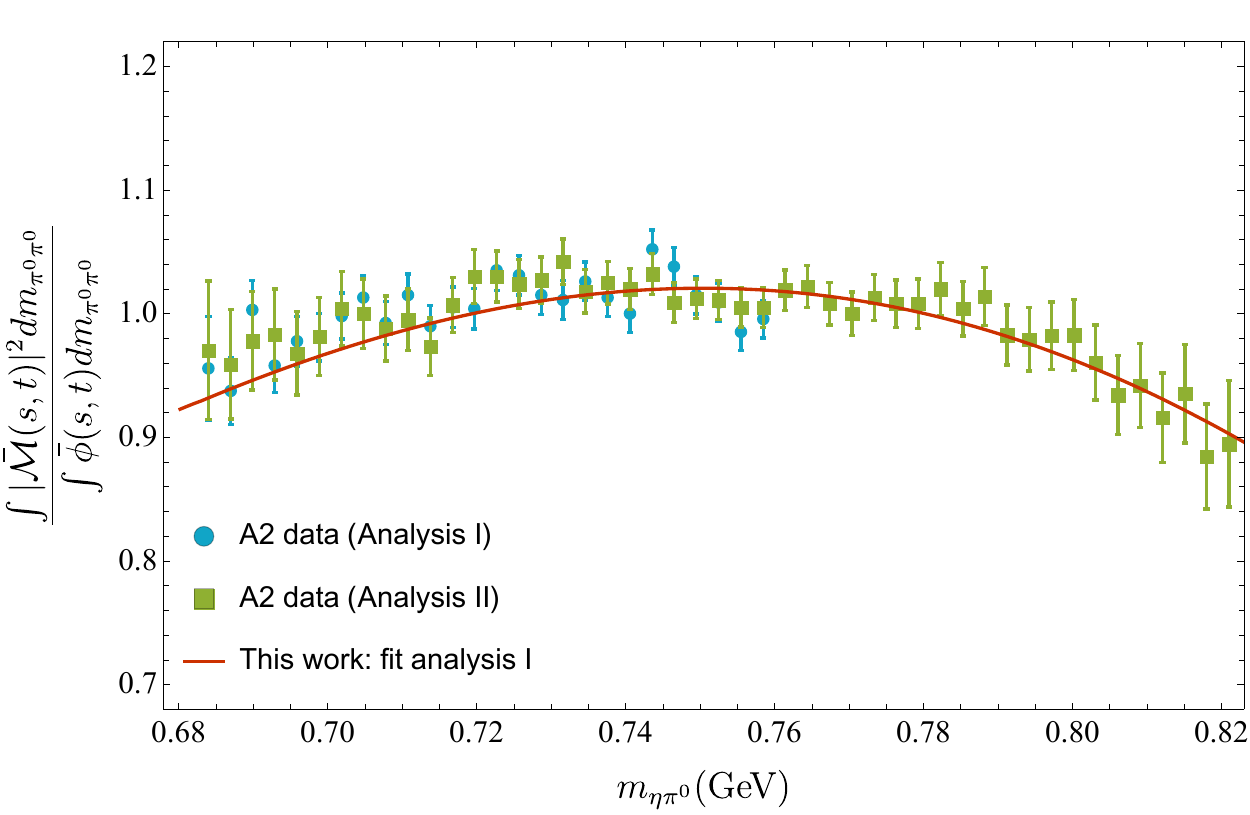}
	\caption{Ratio of the normalized differential decay-width to the normalized phase-space distribution for the $\eta^\prime \to \eta \pi^0 \pi^0$ decay, as a function of the Dalitz variables $X$ (top left) and $Y$ (top right), as well as the invariant masses $m_{\pi^0\pi^0}$ (bottom left) and $m_{\eta\pi^0}$ (bottom right). The experimental data from the A2 collaboration~\cite{Adlarson:2017wlz} is included for comparison. \label{fig:Nofsi}}
\end{figure}

To make a direct comparison with the A2 measurement~\cite{Adlarson:2017wlz}, we show in figure~\ref{fig:Nofsi} the ratio of the differential decay-width to the phase-space distribution for the $\eta^\prime \to \eta \pi^0 \pi^0$ decay. It is presented as a function of the Dalitz variables $X$ (top left) and $Y$ (top right), as well as the invariant masses $m_{\pi^0\pi^0}$ (bottom left) and $m_{\eta\pi^0}$ (bottom right), respectively. Here, for convenience, we have already normalized both the decay-width and phase-space distributions properly, with the normalized amplitude and phase space denoted by $\bar{\mathcal{M}}$ and $\bar{\phi}$, respectively. Evidence for a cusp effect around the charged pion mass threshold has been reported by both the A2~\cite{Adlarson:2017wlz} and BESIII~\cite{BESIII:2022tas} collaborations. However, as can be seen from the top-right and bottom-left plots of figure~\ref{fig:Nofsi}, while our fit is in good agreement with the experimental data, no significant cusp structure is observed at the $\pi^+\pi^-$ mass threshold in such a perturbative description. This indicates that while the large-$N_C$ ChPT provides a reasonable theoretical description of the $\eta^\prime \to \eta \pi^0 \pi^0$ decay, a purely perturbative treatment is insufficient to account for such a cusp effect. In the next subsection, we will improve our theoretical description by including the $\pi\pi$ FSI, to obtain a better agreement with the experimental data and, especially, the cusp structure.

\begin{figure}[t]
	\centering
	\includegraphics[width=0.48\textwidth]{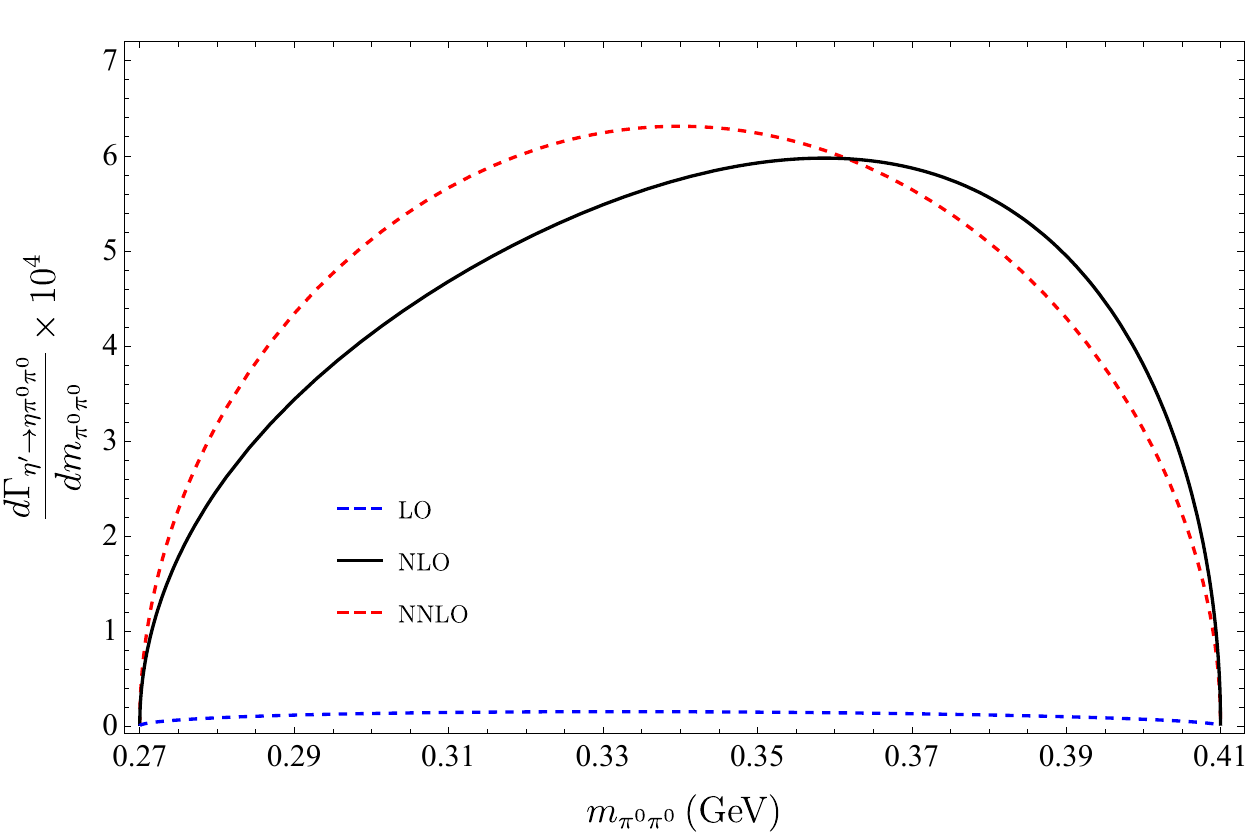}\hfill
	\includegraphics[width=0.49\textwidth]{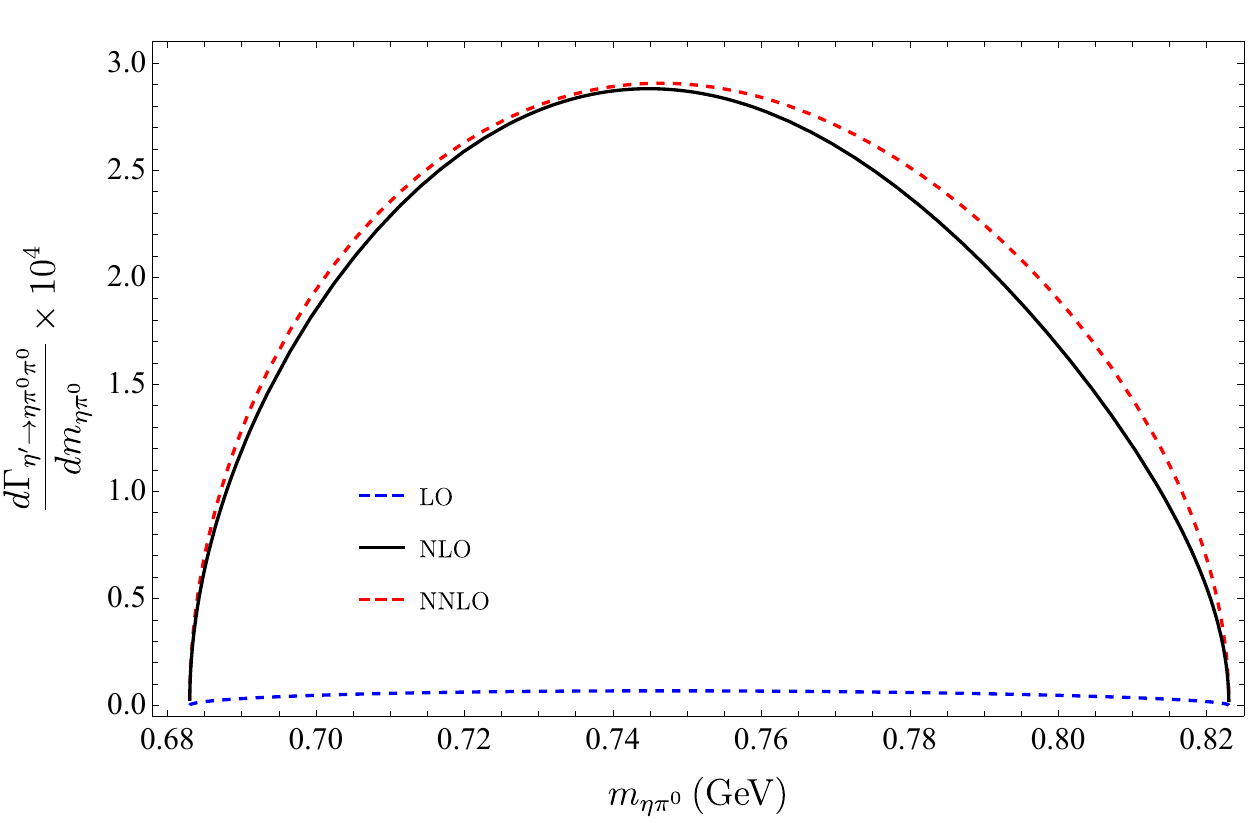} 
	\caption{Differential decay rate of the $\eta^\prime \to \eta \pi^0 \pi^0$ decay as a function of the $m_{\pi^0\pi^0}$ (left) and $m_{\eta\pi^0}$ (right) invariant masses. The blue dashed, black solid, and red dotted curves represent the theoretical predictions at LO, NLO, and NNLO, respectively. \label{fig:Diffdecay}}
\end{figure}

To provide our theoretical predictions for the experimentally accessible kinematic distributions, we use the LECs determined in eq.~\eqref{eq:fitpar1} to evaluate the differential decay rate. The resulting distributions are shown in figure~\ref{fig:Diffdecay}, where the left panel displays the differential decay rate as a function of the invariant mass $m_{\pi^0\pi^0}$, while the right panel shows the distribution in $m_{\eta\pi^0}$. The blue dashed, black solid, and red dotted curves represent the LO, NLO, and NNLO results, respectively. Furthermore, the branching ratio calculated at LO is $0.6\%$, while the results that include the contributions of LO~+~NLO and LO~+~NLO~+~NNLO are given by $23.7\%$ and $24.8(3)\%$, respectively. This indicates that the LO contribution is tiny, and the decay is largely dominated by the loop contributions, with the NNLO correction visible especially in the invariant mass distribution of $\pi^0\pi^0$, as shown by the left plot in figure~\ref{fig:Diffdecay}.

\begin{figure}[t]
	\centering
	\includegraphics[width=0.488\textwidth]{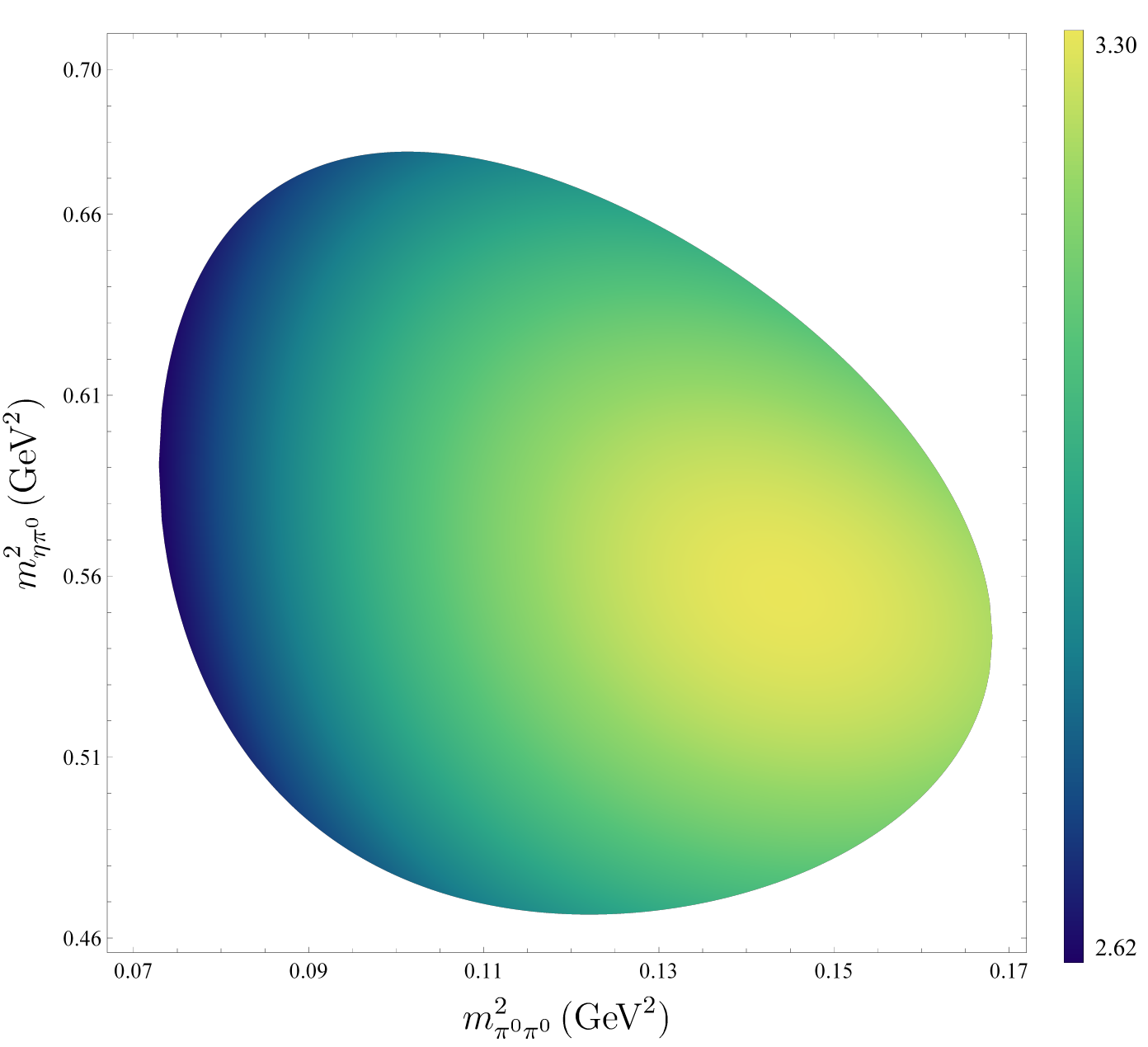}\hfill
	\includegraphics[width=0.49\textwidth]{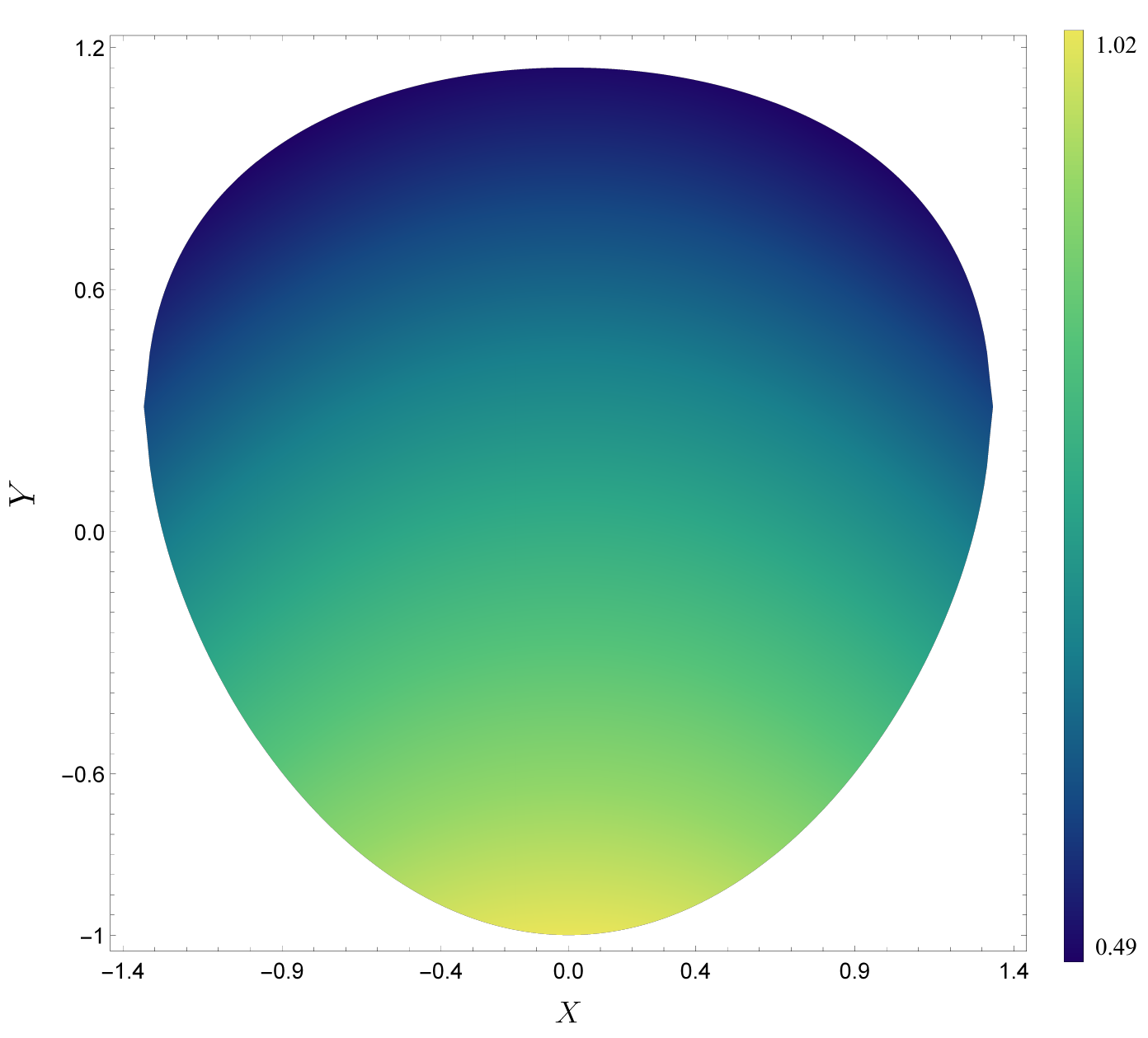} 
	\caption{Dalitz-plot distributions of the $\eta^\prime \to \eta \pi^0 \pi^0$ decay as a function of the invariant masses squared $m_{\pi\pi}^2$ and $m_{\eta\pi}^2$ (left) as well as the Dalitz variables $X$ and $Y$ (right). The distributions correspond to the differential decay widths $10^3 \times d^2\Gamma/dsdt$ and $10^4 \times d^2\Gamma/dXdY$, respectively. \label{fig:Dalitz}}
\end{figure}

For future experimental reference, we show in figure~\ref{fig:Dalitz} the Dalitz-plot distributions of the $\eta^\prime \to \eta \pi^0 \pi^0$ decay, both in the $m_{\pi\pi}^2$ and $m_{\eta\pi}^2$ (left) as well as in the $X$ and $Y$ (right) planes. As is generally expected, the distributions are symmetric under the exchange of two identical pions (corresponding to the exchange $X \leftrightarrow -X$). The event density is predominantly concentrated around the central region of the Dalitz variable $X$, where the $\eta$ meson is at rest and the two pions move back-to-back. Furthermore, clear deviations from a uniform phase-space distribution reflect the nontrivial dynamical structure of the decay amplitude.

\subsection{Fit including \texorpdfstring{$\pi\pi$}{pipi} final-state interactions}
\label{sec:FSI}

In this subsection, we investigate the effects of $\pi\pi$ FSI on $\eta^\prime \to \eta \pi \pi$ decays. We first consider a simplified scenario where only the $S$-wave contribution~---~corresponding to the first term in eq.~\eqref{eq:ampFSI}~---~is retained. Within this approximation, the fit to the A2 Analysis~I dataset yields
\begin{align}
L_{13} &= 1.28(1) \times 10^{-3}\,,& L_{21} &= -3.08(10) \times 10^{-3}\,,\notag\\[0.15cm]
	v_1^{(2)} &= -0.23(6)\,, & a_{\pi\pi} &= -0.16_{-0.51}^{+1.09}\,, \label{eq:fitSwave}
\end{align}
with a $\chi^2/\text{d.o.f} = 249.5/196 \simeq 1.27$. The corresponding Dalitz-plot parameters are given, respectively, by
\begin{align}
	a = -1.02(12)\,, \qquad b = -0.059(5)\,. \label{eq:Daplot2}
\end{align}
In such an $S$-wave-only approximation, the parameter $d$ (associated with the term $X^2$ in eq.~\eqref{eq:amplitude_expanded}) cannot be determined because it is governed by higher-order partial waves. The larger $\chi^2/\text{d.o.f}$ compared to that in the perturbative case ($\chi^2/\text{d.o.f} = 224.5/197 \simeq 1.14$) suggests that the $S$ wave alone is insufficient to describe the experimental data.

Let us now turn to the case that includes both the $S$- and $D$-wave contributions. Our fit results in this case read
\begin{align}
	L_{13} &= 1.89(9) \times 10^{-3}\,, & L_{21} &= 1.40(70) \times 10^{-3}\,, \notag\\[0.15cm]
	v_1^{(2)} &= -0.11(2)\,, &
	  a_{\pi\pi} &= -0.62_{-0.26}^{+0.37}\,, \label{eq:fitFSI1}
\end{align}
with a $\chi^2/\text{d.o.f} = 220.4/196 \simeq 1.12$. Thus, we can see that the inclusion of the $\pi\pi$ FSI results in a better description of the experimental data. The corresponding Dalitz-plot parameters are given, respectively, by
\begin{align}
	a = -0.087(18)\,, \qquad b = -0.078(10)\,, \qquad d = -0.041(6)\,. \label{eq:dalitz1}
\end{align}
To further investigate the influence of $\pi\pi$ FSI and higher-order partial waves on the decay dynamics, we compare the Dalitz-plot parameters derived from the full amplitude (incorporating both the $S$- and $D$-wave contributions) with those from the perturbative calculation but without the $\pi\pi$ FSI (c.f. eq.~\eqref{eq:Daplot1}). Our analysis demonstrates clearly that the inclusion of the $\pi\pi$ FSI significantly modifies the Dalitz-plot parameters. In particular, the parameters $a$ and $d$~---~associated with the terms $Y$ and $X^2$ in eq.~\eqref{eq:amplitude_expanded}, respectively~---~exhibit the most pronounced changes. Furthermore, comparing these results with those obtained by retaining only the $S$-wave contribution (eq.~\eqref{eq:Daplot2}), we can see that the $D$-wave contribution not only dominates the determination of the parameter $d$, but also affects the other two parameters $a$ and $b$. In particular, since $d$ reflects the leading $(t-u)$-dependent structure allowed by the $t \leftrightarrow u$ symmetry of the amplitude, it provides the most direct sensitivity to the $D$-wave dynamics. 

\begin{figure}[t]
	\centering
	\includegraphics[width=0.486\textwidth]{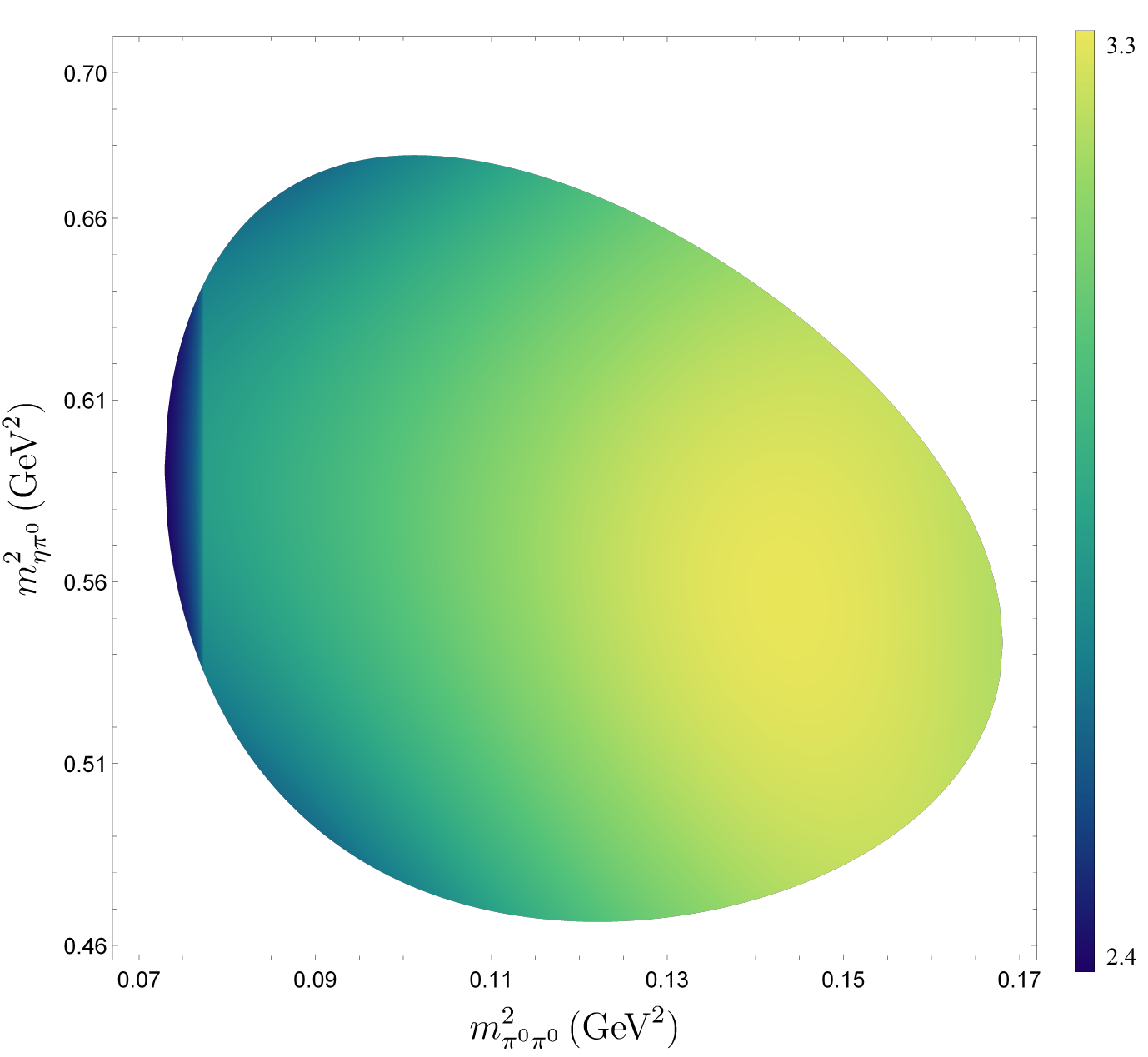}\hfill
	\includegraphics[width=0.492\textwidth]{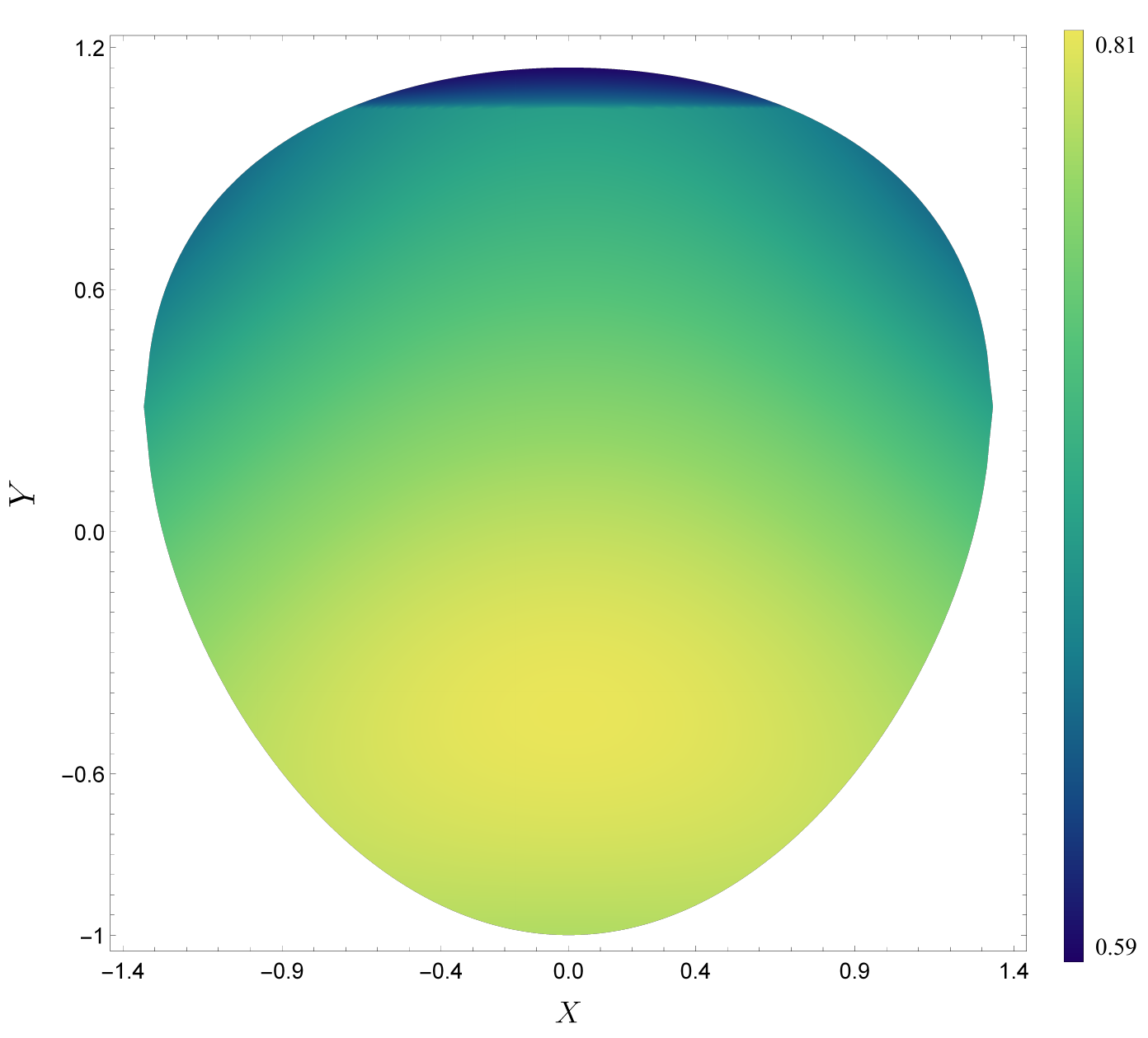} 
	\caption{Same as in figure~\ref{fig:Dalitz}, but with the influence of $\pi\pi$ FSI taken into account. \label{fig:Dalitz1}}
\end{figure}

We show in figure~\ref{fig:Dalitz1} the Dalitz-plot distributions of the $\eta^\prime \to \eta \pi^0 \pi^0$ decay based on the fit results given by eq.~\eqref{eq:fitFSI1}. Compared to the perturbative case shown in figure~\ref{fig:Dalitz}, the inclusion of $\pi\pi$ FSI leads to a distinctly smoother amplitude distribution, with pronounced changes observed near the $\pi^+\pi^-$ mass threshold. It is also observed that the $\pi\pi$ rescattering effects are clearly seen by the enhancement of the distribution in the center of the Dalitz plot. Furthermore, due to Bose symmetry of the two identical neutral pions in the final state, the $X$ distribution is symmetric under the exchange of the Dalitz variable $X$, $X \leftrightarrow -X$, as dictated by $C$ and $CP$ invariances.

\begin{figure}[t]
	\centering
	\includegraphics[width=0.50\textwidth]{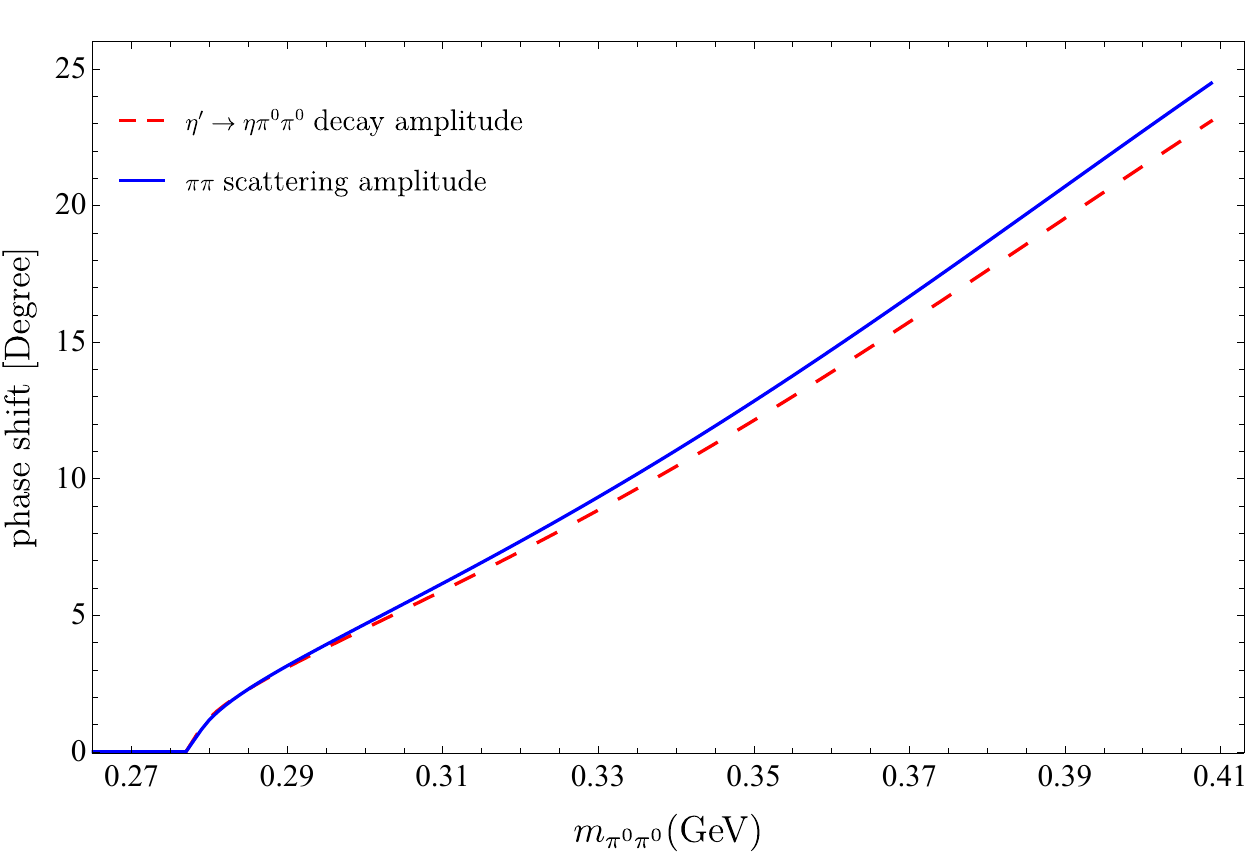}
	\caption{Comparison between the phase of the $\eta^\prime \to \eta \pi^0 \pi^0$ decay amplitude (red dashed curve) and the elastic $\pi\pi$ rescattering phase shift (solid blue curve). The minimal deviation between the two curves provides a quantitative measure of the unitarity violation in the decay amplitude. \label{fig:univi}}
\end{figure}

At this stage, it is necessary to discuss the domain of validity and the limitation of the unitarity relation defined by eq.~\eqref{eq:unirelation}. This relation is strictly applicable only within the elastic region, where the dynamics is governed by the $\pi\pi$ rescattering. In this regime, Watson's theorem~\cite{Watson:1954uc} dictates that the phase of the decay amplitude must coincide with the corresponding $\pi\pi$ rescattering phase shift. However, minor departures from the exact unitarity relation can occur even below the first inelastic threshold (\textit{i.e.} the $K\bar{K}$ mass threshold). A primary source of these deviations is the left-hand cut contributions arising from crossed-channel $\pi\eta$ loops~\cite{Gonzalez-Solis:2018xnw}; their imaginary parts may partially overlap with the physical region, generating spurious absorptive components. There is another source of unitarity violation generated from the on-shell approximation within the $N/D$ unitarization method by the $t$- and $u$-channel $\pi\pi$ loops entering the $\pi\pi$ rescattering amplitude~\cite{Gonzalez-Solis:2018xnw}. We show in figure~\ref{fig:univi} the comparison between the phase of the $\eta^\prime \to \eta \pi^0 \pi^0$ decay amplitude (blue solid line) and the elastic $\pi\pi$ rescattering phase shift (red dashed line). The minimal discrepancy between the two phases within the relevant kinematic range indicates that the unitarity-violating effects remain suppressed, ensuring a level of precision consistent with the objectives of this study. Furthermore, our predicted phase shift of the $\pi\pi$ rescattering shows excellent agreement with the low-energy constraints~\cite{Garcia-Martin:2011iqs} derived from the Roy equations~\cite{Roy:1971tc} for the $\pi\pi$ scattering.

\subsection{Final fit results and further predictions}
\label{sec:final_fit}

To assess the sensitivity of our results to the different data samples, we fit both the Analysis~I and Analysis~II datasets provided by the A2 collaboration~\cite{Adlarson:2017wlz}. Along the same line as in eq.~\eqref{eq:fitFSI1}, our fit results derived from the Analysis~II dataset are given by
\begin{align}\label{eq:fitFSI2}
	L_{13} &= 1.98(9) \times 10^{-3}\,, &  L_{21} &= 2.07_{-0.64}^{+0.65} \times 10^{-3}\,,\notag\\[0.2cm]
	v_1^{(2)} &= -0.09(2)\,, & a_{\pi\pi} &= -0.71^{+0.31}_{-0.23}\,, 
\end{align}
with $\chi^2/\mathrm{d.o.f} = 239.2/196 \simeq 1.22$. The slightly larger $\chi^2/\mathrm{d.o.f.}$ relative to that obtained with the Analysis~I dataset is attributed to the fact that the background subtraction in the Analysis~II case is less robust than in the Analysis~I case~\cite{Adlarson:2017wlz}. Using these fit results, the corresponding Dalitz-plot parameters are extracted as
\begin{align}\label{eq:dalitz2}
	a = -0.083(19)\,, \qquad b = -0.085(12)\,, \qquad d = -0.049(5)\,.
\end{align}
These values exhibit a consistent pattern with those obtained from the Analysis~I dataset (c.f. eq.~\eqref{eq:dalitz1}), sharing identical signs and comparable magnitudes. 

\begin{figure}[t]
	\centering
	\includegraphics[width=0.49\textwidth]{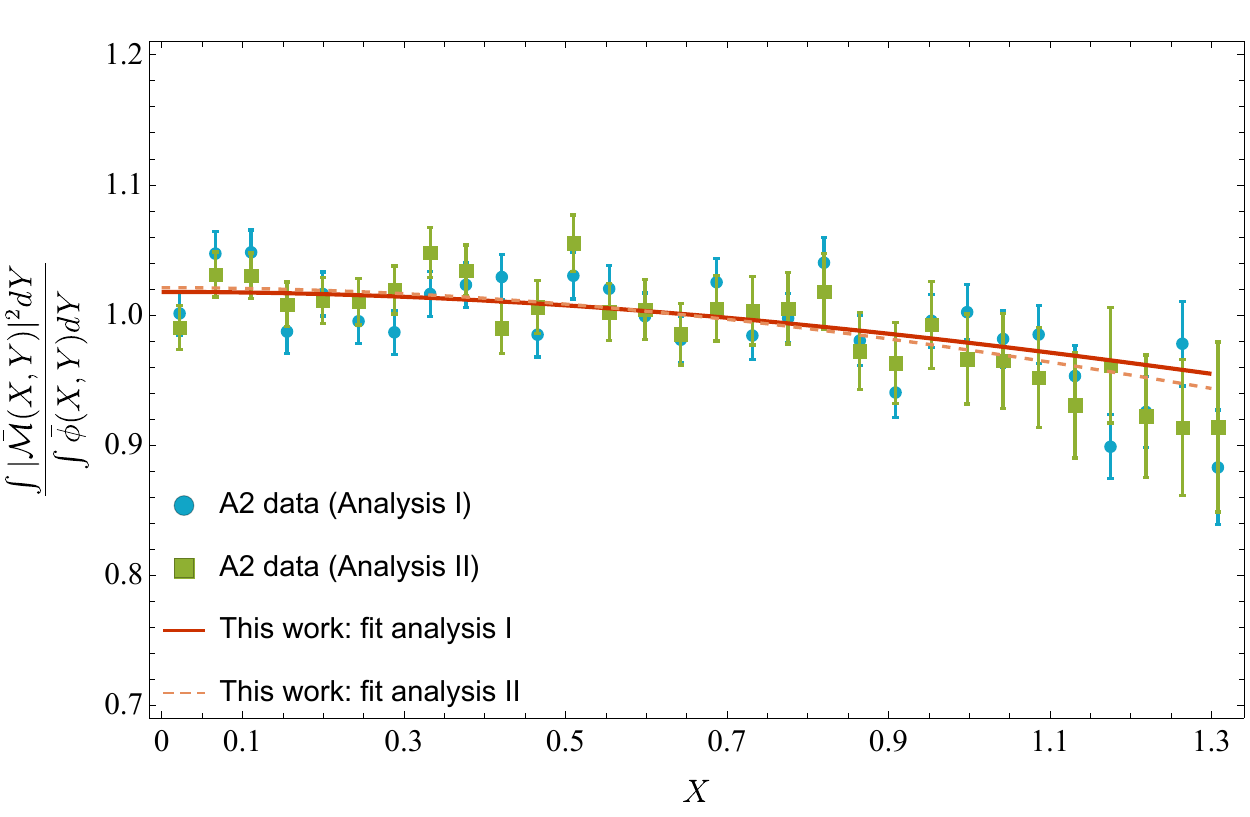}\hfill
	\includegraphics[width=0.49\textwidth]{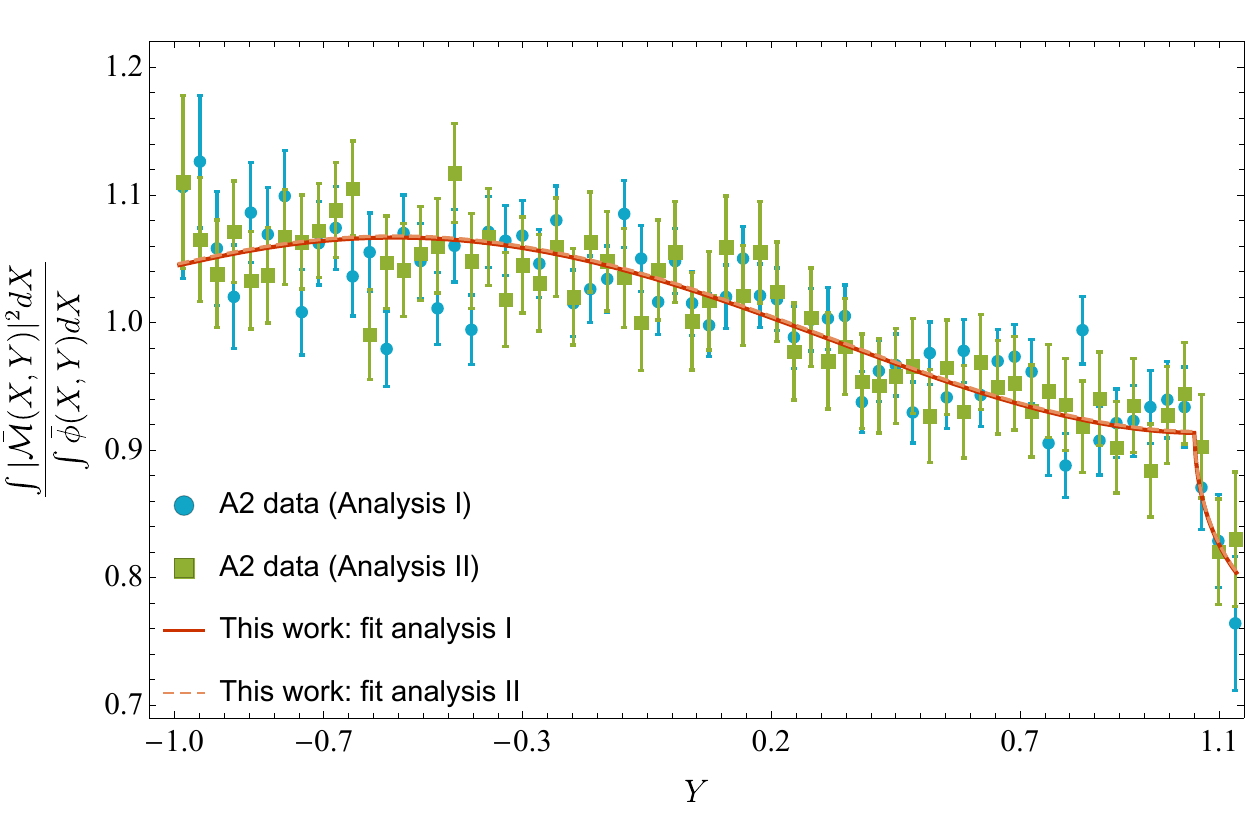} \\[5pt]
	\includegraphics[width=0.49\textwidth]{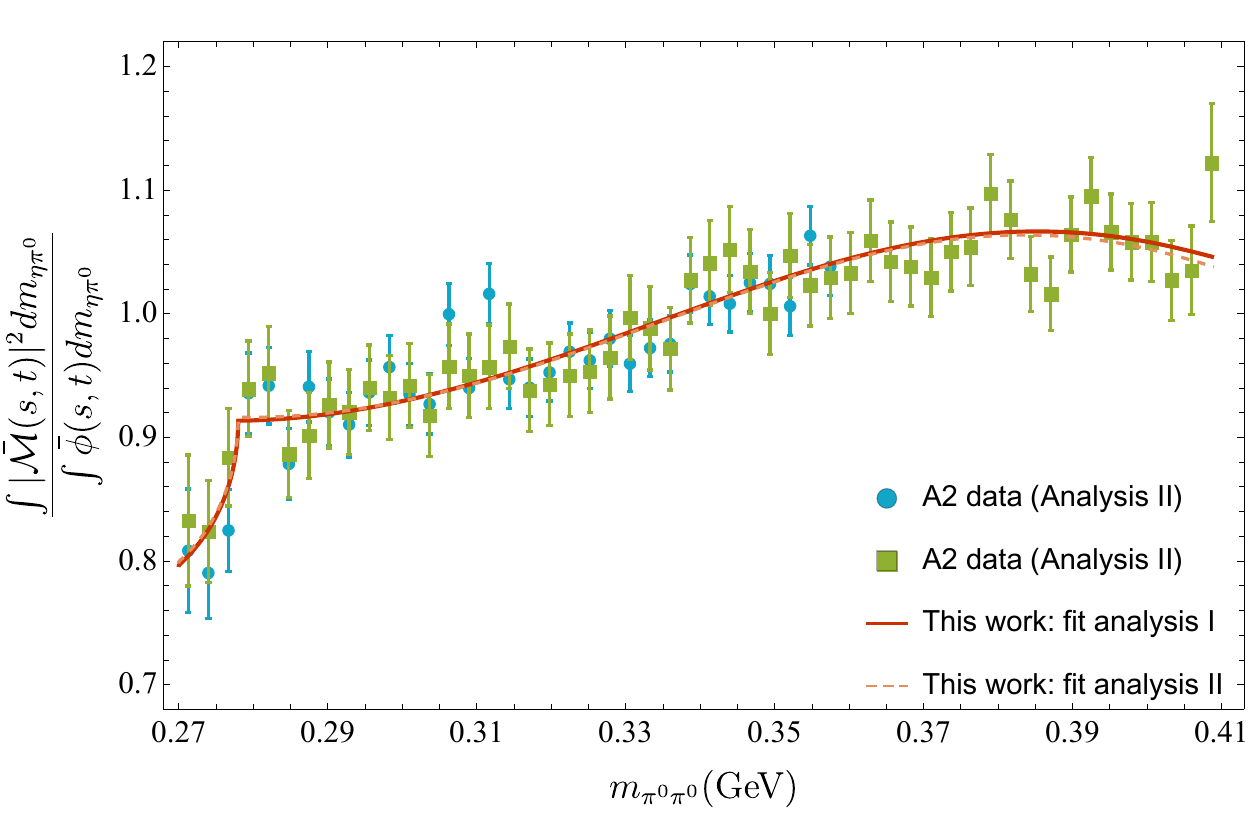}\hfill
	\includegraphics[width=0.49\textwidth]{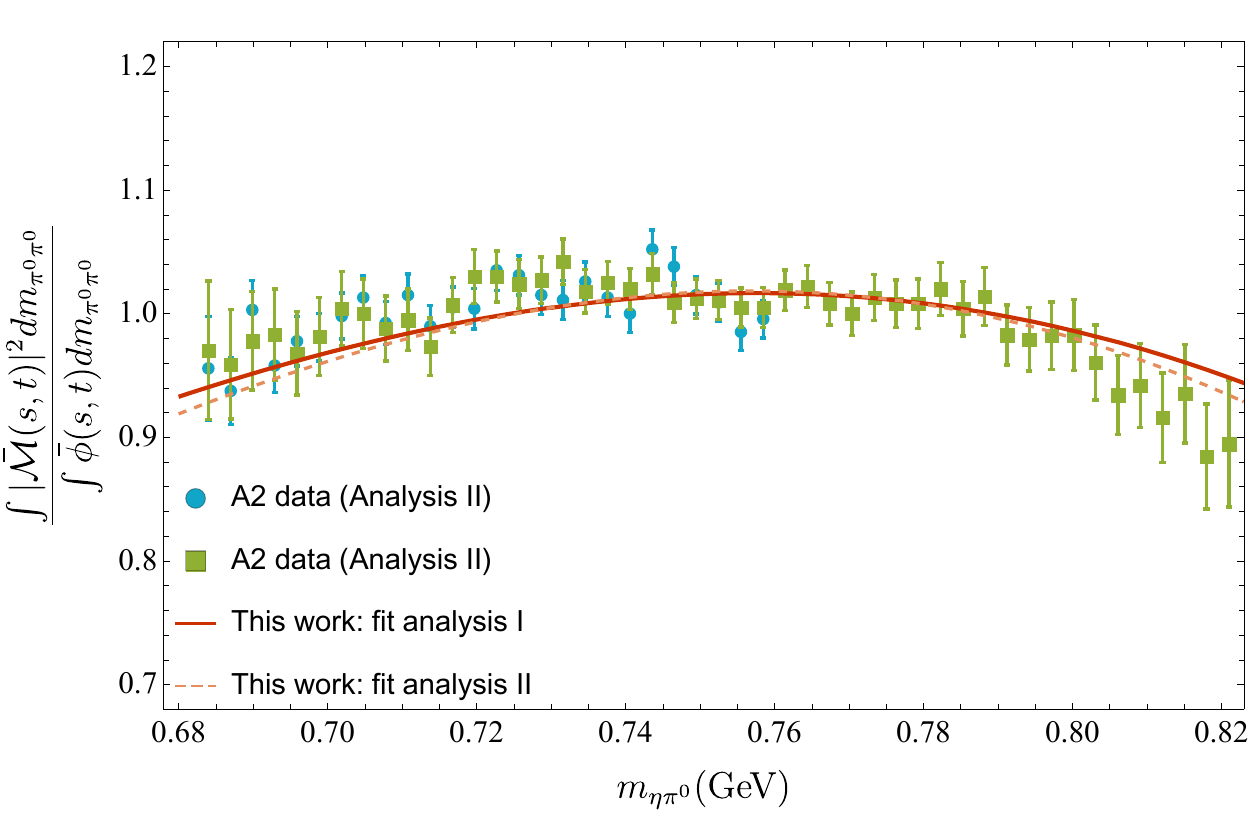}
	\caption{Ratio of the normalized differential decay-rate to the normalized phase-space distribution for the $\eta^\prime \to \eta \pi^0 \pi^0$ decay, as a functions of $X$ (top left), $Y$ (top right), $m_{\pi^0\pi^0}$ (bottom left), and $m_{\eta\pi^0}$ (bottom right). The solid and dashed curves correspond to our predictions obtained with the fit results of the Analysis~I and Analysis~II datasets from the A2 collaboration~\cite{Adlarson:2017wlz}, respectively. \label{fig:fsi}}
\end{figure}

In figure~\ref{fig:fsi}, we make a comparison between the A2 data and our predictions for the differential distribution of the $\eta^\prime \to \eta \pi^0 \pi^0$ decay rate divided by the phase space distribution. The solid and dashed curves represent our predictions obtained with the fit results of the Analysis~I and Analysis~II datasets from the A2 collaboration~\cite{Adlarson:2017wlz}, respectively. In contrast to the perturbative results shown in figure~\ref{fig:Nofsi}, a pronounced cusp structure emerges at the $\pi^+\pi^-$ mass threshold, as can be seen from the top-right and bottom-left panels.\footnote{Here, the charged-pion mass is used in eq.~\eqref{eq:gfun} and for the $t$- and $u$-channel contributions to the $\pi\pi$ scattering, while the neutral-pion mass is used elsewhere.} This confirms that the unitarized amplitude, with explicit $\pi\pi$ FSI effects taken into account, provides a satisfactory description of the $\eta' \to \eta \pi^0 \pi^0$ decay distribution in the physical region. In addition, the Dalitz-plot distributions are well reproduced in both cases, with the two curves practically overlapping with each other. Thus, the large-$N_C$ ChPT, after including the higher-order corrections, together with the aid of the $N/D$ method to restore unitarity and account for the $\pi\pi$ rescattering effects, provides a robust description of both the experimental distribution~\cite{Adlarson:2017wlz} and the cusp effect at the $\pi^+\pi^-$ mass threshold~\cite{Adlarson:2017wlz,BESIII:2022tas}.

\begin{figure}[t]
    \centering
    \includegraphics[width=0.65\textwidth]{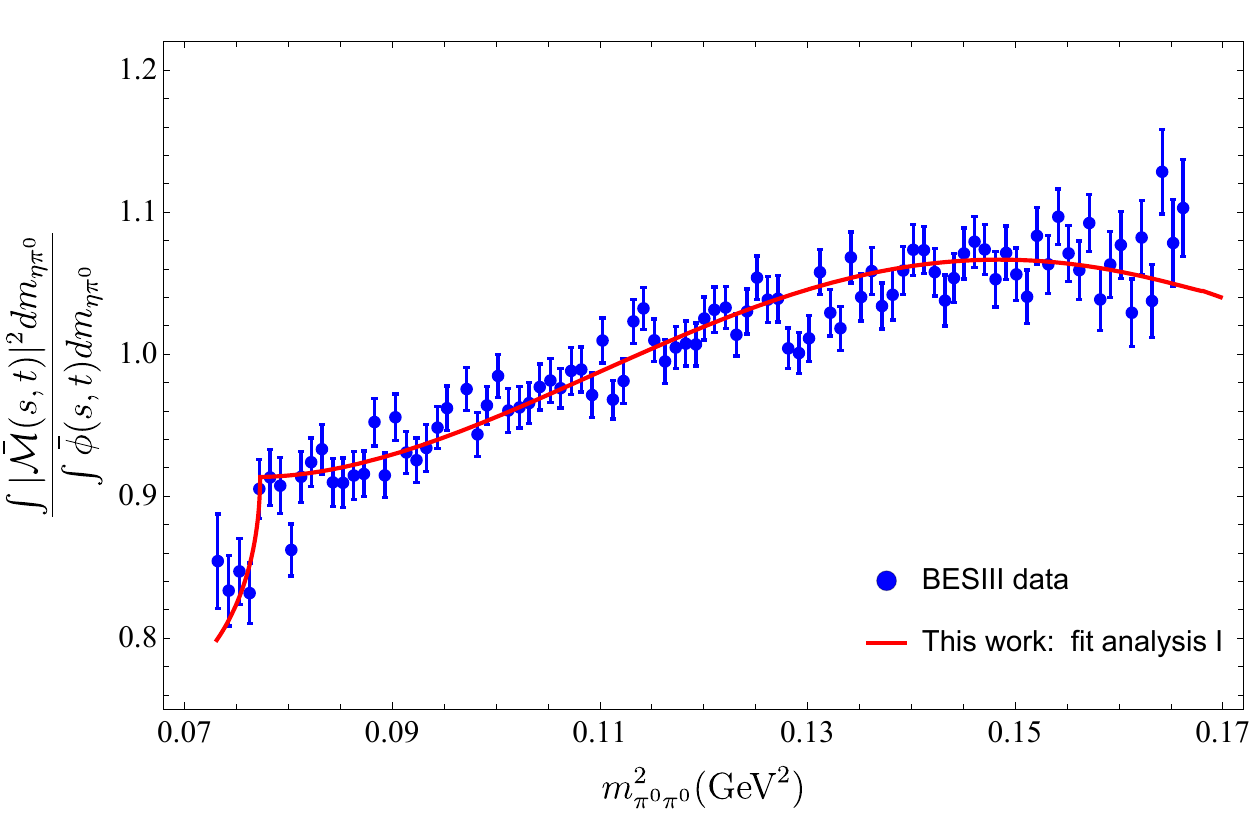}
    \caption{Comparison between the BESIII data~\cite{BESIII:2022tas} (blue points) and our theoretical result based on the unitarized amplitude given by eq.~\eqref{eq:fitFSI1} (red solid line) for the normalized ratio of the differential decay rate to the phase-space distribution. \label{fig:BESIIIFSI}}
\end{figure}

The BESIII collaboration also reported evidence for a cusp structure at the $\pi^+\pi^-$ mass threshold in the invariant mass spectrum of $\pi^0\pi^0$ for the $\eta^\prime \to \eta \pi^0 \pi^0$ decay~\cite{BESIII:2022tas}. As another check of the robustness of our theoretical framework, let us now compare our prediction with the BESIII data~\cite{BESIII:2022tas} (c.f. the right plot of FIG.~5 in this reference). As can be seen from figure~\ref{fig:BESIIIFSI}, our theoretical prediction successfully reproduces the characteristic behavior of the cusp structure observed by the BESIII collaboration. To quantify this consistency, we perform a standard $\chi^2$ test using the experimental statistical uncertainties, and obtain $\chi^2/\mathrm{d.o.f.} = 110.1/93 \simeq 1.18$, indicating that our theoretical framework provides a statistically satisfactory description of the BESIII data including the cusp structure.

As our final fit result for each Dalitz-plot parameter, we take the average of the values given by eqs.~\eqref{eq:dalitz1} and \eqref{eq:dalitz2} as the central value. The spread between the two analyses is assigned as the systematic uncertainty, while the uncertainty obtained in the Analysis~I case is adopted as the statistical one. In this way, we obtain
\begin{equation} \label{eq:dalitz3}
    a=-0.085(18)_{\mathrm{stat}}(4)_{\mathrm{syst}}\,, \quad b=-0.081(10)_{\mathrm{stat}}(6)_{\mathrm{syst}}\,, \quad d=-0.045(6)_{\mathrm{stat}}(8)_{\mathrm{syst}}\,.
\end{equation}
Comparing eq.~\eqref{eq:dalitz3} with the A2 measurement~\cite{Adlarson:2017wlz} listed in table~\ref{tab:DalitzExTheory}, we can see that, when the statistical and systematic uncertainties are combined, our results remain consistent with the A2 data within one standard deviation for all three Dalitz-plot parameters.

Although both the $\eta^\prime \to \eta \pi^0 \pi^0$ and $\eta^\prime \to \eta \pi^+ \pi^-$ decay amplitudes are the same in the exact isospin limit, they become different due to strong FSI. Thus, when using the fitted parameters from the neutral $\pi^0\pi^0$ mode to predict the decay-rate distributions for the charged $\pi^+\pi^-$ mode and determine the corresponding Dalitz-plot parameters, we must take into account the isospin-breaking effects. Within our framework, the dominant isospin-breaking contribution originates from the pion mass difference in the definition of the Dalitz-plot variables. Consequently, the Dalitz-plot parameters for the charged and neutral channels are no longer identical. Here we adopt the following relations derived in ref.~\cite{Isken:2017dkw} to connect these parameters in the charged (with the superscript ``$c$'') and neutral (without the superscript ``$c$'') channels:
\begin{align}
	a = a^c + \epsilon_{\text{iso}}(a^c + 2b^c)\,, \quad
	b = b^c(1 + 2\epsilon_{\text{iso}})\,, \quad
	d = d^c\left(\frac{Q_\eta^0}{Q_\eta}\right)\,, \label{eq:dalitzR}
\end{align}
where the isospin-breaking parameter $\epsilon_{\mathrm{iso}}$ accounts for the phase-space difference and is defined by~\cite{Isken:2017dkw}
\begin{align}
    \epsilon_{\mathrm{iso}} = \frac{(m_\eta + 2m_{\pi^+}) m_{\pi^0} Q_{\eta^\prime}^0}{(m_\eta + 2m_{\pi^0}) m_{\pi^+} Q_{\eta^\prime}^+} - 1 \simeq 4.7\%\,,
\end{align}
with $Q_{\eta^\prime}^{0,+} = m_{\eta^\prime} - m_\eta - 2m_{\pi^{0,+}}$. Using eq.~\eqref{eq:dalitzR} and our fit results given by eq.~\eqref{eq:dalitz3}, the inferred parameters for the charged $\pi^+\pi^-$ channel read
\begin{equation}
\hspace{-0.1cm}
a^c = -0.075(17)_{\text{stat}}(4)_{\text{syst}}\,, \quad b^c = -0.074(9)_{\text{stat}}(5)_{\text{syst}}\,, \quad d^c = -0.040(6)_{\text{stat}}(7)_{\text{syst}}\,. \label{eq:dalitz4}
\end{equation}
These results can be directly compared with the BESIII measurement~\cite{BESIII:2017djm}
\begin{align}
a^{c} = -0.056(4)_{\text{stat}}(3)_{\text{syst}}\,, \quad b^{c} = -0.049(6)_{\text{stat}}(6)_{\text{syst}}\,, \quad d^{c} = -0.063(4)_{\text{stat}}(4)_{\text{syst}}\,.
\end{align}
A slight tension is observed between the two determinations, with $a^{c}$ by approximately $1\sigma$, while $b^{c}$ and $d^{c}$ by approximately $2\sigma$, where the statistical and systematic uncertainties have been combined in quadrature.

\begin{figure}[t]
    \centering
    \includegraphics[width=0.50\textwidth]{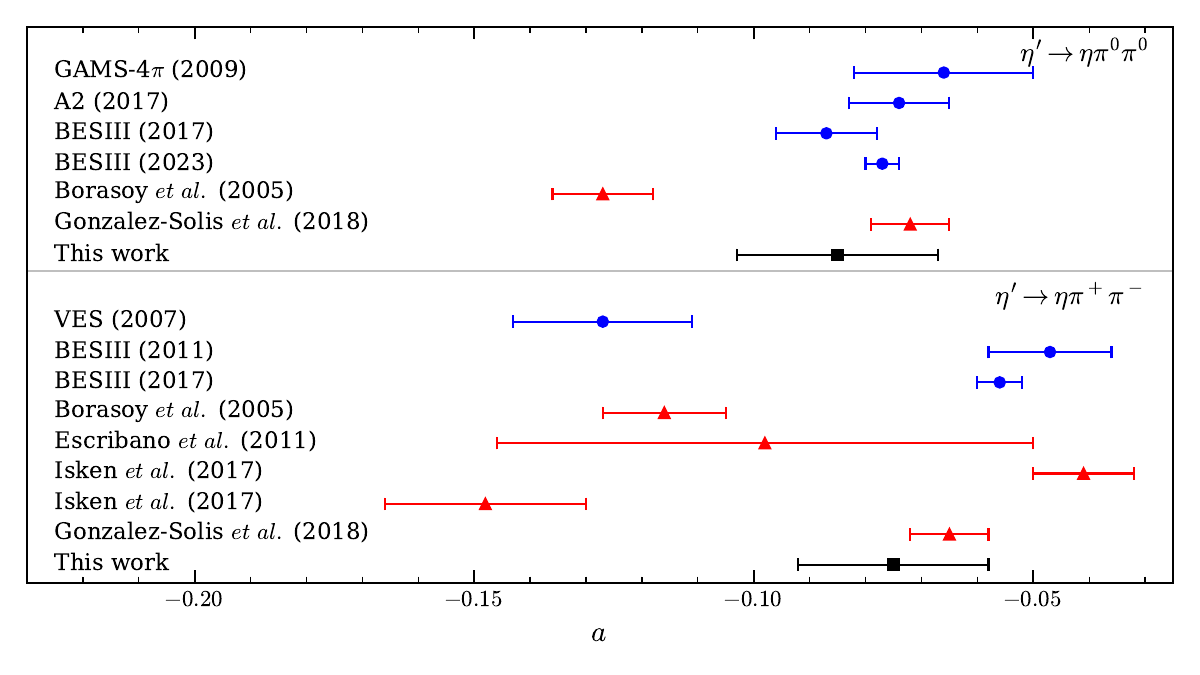}\hfill
    \includegraphics[width=0.50\textwidth]{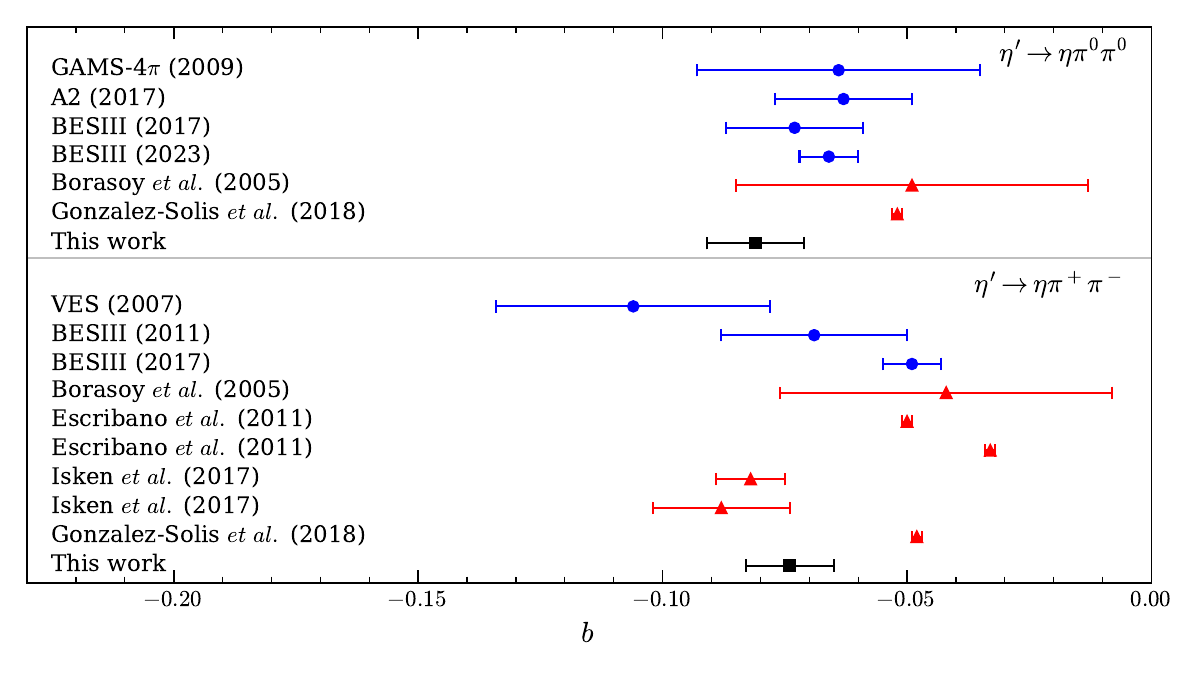}\\[5pt]
    \includegraphics[width=0.50\textwidth]{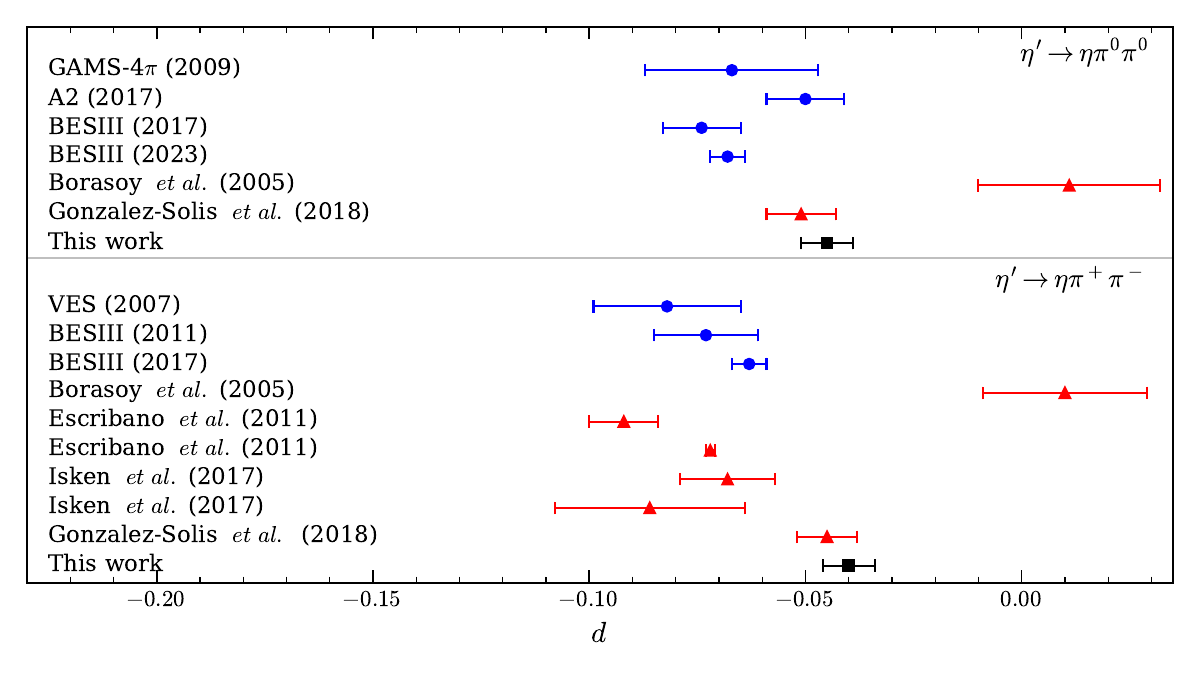}
    \caption{Comparison of our results (\textcolor{black}{\tiny $\blacksquare$}) with available experimental data (\textcolor{blue}{\Large $\bullet$}) and other theoretical determinations (\textcolor{red}{$\blacktriangle$}) for the Dalitz-plot parameters $a, b,$ and $d$. The error bars only reflect the statistical uncertainties, and correspond to the first uncertainties given in eqs.~\eqref{eq:dalitz3} and \eqref{eq:dalitz4}. \label{fig:dalitzTotal}}
\end{figure}

Finally, for convenience, we present in figure~\ref{fig:dalitzTotal} a comprehensive comparison of our final fit results for the Dalitz-plot parameters (c.f. eqs.~\eqref{eq:dalitz3} and \eqref{eq:dalitz4}) with the available experimental data and other theoretical determinations summarized in table~\ref{tab:DalitzExTheory}.

\section{Conclusion}
\label{sec:conclusion}

In this paper, we have performed a detailed study of $\eta^\prime \to \eta \pi \pi$ decays within the framework of large-$N_C$ ChPT up to NNLO in a simultaneous expansion in powers of external momenta $p^2$, quark masses $m_q$, and $1/N_C$. We first determined the $\eta-\eta^\prime$ mixing consistently in the large-$N_C$ expansion and constructed the decay amplitudes order by order. Using the obtained NNLO perturbative amplitude, we then performed a fit to the A2 data. It is found that, while the perturbative approach provides an overall reasonable description of the Dalitz-plot distributions, it fails to reproduce the experimentally observed cusp structure at the $\pi^+\pi^-$ mass threshold, indicating therefore the necessity of including non-perturbative effects. To this end, we adopted the $N/D$ method to construct an amplitude that incorporates the $\pi\pi$ FSI effects. This unitarized amplitude was subsequently fitted to the A2 datasets. We found that the inclusion of the $\pi\pi$ FSI significantly improves the quality of the fit. Despite the increase in the number of fit parameters, the reduced $\chi^2/\text{d.o.f}$ demonstrates that the unitarized amplitude provides a statistically superior description of the experimental data compared to the case with the purely perturbative result. Based on the fitted non-perturbative amplitude, we have also extracted the Dalitz-plot parameters for the neutral decay channel $\eta^\prime \to \eta \pi^0 \pi^0$, with our final results given by
\begin{align*}
    a = -0.085(18)_{\text{stat}}(4)_{\text{syst}}\,, \quad
    b = -0.081(10)_{\text{stat}}(6)_{\text{syst}}\,, \quad
    d = -0.045(6)_{\text{stat}}(8)_{\text{syst}}\,.
\end{align*}
These values agree well with the experimental measurement reported by the A2 collaboration within the uncertainties quoted, confirming the reliability of our approach and highlighting the essential role played by the $\pi\pi$ FSI in a precise theoretical description of the $\eta^\prime \to \eta \pi \pi$ decays. Furthermore, taking into account the isospin-breaking effects, we provided predictions for the Dalitz-plot parameters of the charged decay channel $\eta^\prime \to \eta \pi^+ \pi^-$, which read
\begin{align*}
    a^{c} = -0.075(17)_{\text{stat}}(4)_{\text{syst}}\,, \quad
    b^{c} = -0.074(9)_{\text{stat}}(5)_{\text{syst}}\,, \quad
    d^{c} = -0.040(6)_{\text{stat}}(7)_{\text{syst}}\,.
\end{align*}
These predictions may serve as useful benchmarks for future high-precision measurements at the JLab, HIAF and STCF facilities.

\acknowledgments

This work is supported in part by the National Natural Science Foundation of China under Grant Nos.~12475094, 12475095 and 12135006, as well as the Science and Technology Innovation Leading Talent Support Program of Henan Province under Grant No.~254000510039.

\appendix

\section{\texorpdfstring{$\boldsymbol{\eta^\prime\to\eta\pi\pi}$}{eta-prime to eta pi pi} decay amplitudes up to NNLO}
\label{app:decayamp}

In this appendix, we give the explicit expressions for the $\eta^\prime\to\eta\pi\pi$ decay amplitudes up to NNLO within the framework of large-$N_C$ ChPT, which are utilized in the numerical analysis presented in the main text. The total amplitude is decomposed as
\begin{equation}
    \mathcal{M}^{\mathrm{NNLO}}=\mathcal{M}^{\mathrm{Tree}}+\mathcal{M}^{\mathrm{Loop}}+\mathcal{M}^{\mathrm{Tadpole}}\,,
\end{equation}
where $\mathcal{M}^{\mathrm{Tree}}$ denotes the tree-level contribution, $\mathcal{M}^{\mathrm{Loop}}$ collects the one-loop correction generated by the LO vertices, while $\mathcal{M}^{\mathrm{Tadpole}}$ encompasses the NNLO tadpole contribution along with the mass and wave-function renormalizations of the $\pi$ and $K$ mesons. 

The explicit expression for the tree-level contribution $\mathcal{M}^{\mathrm{Tree}}$ is given by
\begin{align}
    \mathcal{M}^{\mathrm{Tree}} &= \frac{2 \sqrt{2} \cos(2\theta^{(2)}) - \sin(2\theta^{(2)})}{3F_{\pi }^4} \Bigg\{2 (3 L_2+L_3) \big( s^2-m_{\eta }^4-m_{\eta '}^4-2 m_{\pi }^4+t^2+u^2 \big) \nonumber\\[0.15cm]
    &\quad + 4 L_4 \bigl[ m_K^2 (6 m_{\pi }^2-2s)+2s m_{\pi }^2-3 m_{\pi }^4 \bigr] - 2 L_5 m_{\pi }^2 (m_{\eta }^2+m_{\eta '}^2+2 m_{\pi }^2) \nonumber\\[0.15cm]
    &\quad + 32 L_6 m_{\pi }^2 ( m_{\pi }^2- m_K^2) + 24 L_8 m_{\pi }^4 + 2 C_1 \bigg[ m_{\eta }^2 \Big( 5 s^2 + 2 m_{\eta '}^2 (-3 s+8 m_{\pi }^2-4 t) \nonumber\\[0.15cm]
    &\quad  - 2 m_{\pi }^2 (9 s+8 t) + 4 s t + 3 m_{\eta '}^4 + 16 m_{\pi }^4 + 5 t^2 - u^2 \Big) + m_{\eta '}^2 \Big( 5 s^2 - 2 m_{\pi }^2 (9 s+8 t) \nonumber\\[0.15cm]
    &\qquad + 4 s t + 16 m_{\pi }^4 + 5 t^2 - u^2 \Big) + 2 \Bigl(-s^3 + m_{\pi }^2 (5 s^2+4 s t+5 t^2-u^2) \nonumber\\[0.15cm]
    &\qquad - m_{\pi }^4 (7 s+8 t) + s u^2 + 4 m_{\pi }^6 - t^3 + t u^2 \Bigr) + m_{\eta }^4 \bigl( -4 (s+t) + 3 m_{\eta '}^2 \nonumber\\[0.15cm]
    &\qquad + 8 m_{\pi }^2 \bigr) + m_{\eta '}^4 (8 m_{\pi }^2 - 4 (s+t)) + m_{\eta }^6 + m_{\eta '}^6 \bigg] \nonumber\\[0.15cm]
    &\quad - 4 (C_5+C_8+C_{10}) m_{\pi }^2 (-s^2+m_{\eta }^4+m_{\eta '}^4+2 m_{\pi }^4-t^2-u^2) \nonumber\\[0.15cm]
    &\quad - 4 C_{12} m_{\pi }^2 \bigg[ 4 \big( s^2-2 m_{\pi }^2 (s+t)+s t+2 m_{\pi }^4+t^2 \big) + 2 m_{\eta }^2 \big( -2 s+m_{\eta '}^2 \nonumber\\[0.05cm]
    &\qquad + 2 m_{\pi }^2-2 t \big) + 4 m_{\eta '}^2 (-s+m_{\pi }^2-t) + 3 m_{\eta }^4 + 3 m_{\eta '}^4 \bigg] \nonumber\\[0.15cm]
    &\quad - 8 (C_{14}+C_{17}) m_{\pi }^4 (m_{\eta }^2+m_{\eta '}^2+m_{\pi }^2) + 144 (C_{19}+C_{31}) m_{\pi }^6 \nonumber\\[0.15cm]
    &\quad + 2 (2 C_{22}+C_{25}) m_{\pi }^2 \bigg[ 4 \big( s^2-2 m_{\pi }^2 (s+t)+s t+m_{\pi }^4+t^2 \big) \nonumber\\[0.05cm]
    &\qquad + 2 m_{\eta }^2 (-2 s+m_{\eta '}^2+2 m_{\pi }^2-2 t) + 4 m_{\eta '}^2 (-s+m_{\pi }^2-t) + m_{\eta }^4 + m_{\eta '}^4 \bigg] \nonumber\\[0.15cm]
    &\quad + \frac{F_{\pi }^2 m_{\pi }^2}{2} - 8 (C_{26}+C_{29}) m_{\pi }^4 (m_{\eta }^2+m_{\eta '}^2+2 m_{\pi }^2) - \frac{16 (3 L_2+L_3)}{F_{\pi }^2} \bigg[ L_5 m_{\pi }^2 \big( -s^2 \nonumber\\[0.05cm]
    &\qquad + m_{\eta }^4 - 2 m_{\eta }^2 m_{\pi }^2 + m_{\eta '}^4 - 2 m_{\eta '}^2 m_{\pi }^2 + 2 m_{\pi }^4 - t^2 - u^2 \big) + 4 L_8 m_{\pi }^4 (m_{\eta }^2+m_{\eta '}^2) \bigg] \nonumber\\[0.15cm]
    &\quad - \frac{16 m_{\pi }^4}{F_{\pi }^2} \Big( L_5^2 (2 m_{\eta }^2+2 m_{\eta '}^2+13 m_{\pi }^2) - 2 L_8 L_5 (m_{\eta }^2+m_{\eta '}^2+26 m_{\pi }^2) + 48 L_8^2 m_{\pi }^2 \Big) \Bigg\} \nonumber\\[0.05cm]
    &+ \frac{16 L_7 m_{\pi }^2 \Bigl[ 2 \sin(2\theta^{(2)}) (m_{\pi }^2-4 m_K^2) - \sqrt{2} \cos(2\theta^{(2)}) (2 m_K^2-5 m_{\pi }^2) \Bigr]}{3 F_{\pi }^4} \nonumber\\[0.05cm]
    &+ \frac{L_{13} \bigl( \sqrt{2} \cos(2\theta^{(2)}) - 2 \sin(2\theta^{(2)}) \bigr) \bigl(-s^2+m_{\eta }^4+m_{\eta '}^4+2 m_{\pi }^4-t^2-u^2 \bigr)}{F_{\pi }^4} \nonumber\\[0.05cm]
    &+ \frac{4 L_{18} m_{\pi }^2 \Bigl[ \sqrt{2} m_{\eta }^2 \sin^2 \theta^{(2)} + \sin(2\theta^{(2)}) (m_{\eta }^2+m_{\eta '}^2) - \sqrt{2} m_{\eta '}^2 \cos^2 \theta^{(2)} \Bigr]}{F_{\pi }^4} \nonumber\\[0.05cm]
    &+ \frac{4 L_{21} m_{\pi }^2 \Bigl[ \sqrt{2} (m_{\eta }^2-m_{\eta '}^2) + 2 m_{\pi }^2 (\sqrt{2} \cos(2\theta^{(2)}) - 2 \sin(2\theta^{(2)})) \Bigr]}{F_{\pi }^4} \nonumber\\[0.05cm]
    &- \frac{32 L_{25} m_{\pi }^4 \bigl(\sqrt{2} \cos(2\theta^{(2)}) - 2 \sin(2\theta^{(2)})\bigr)}{F_{\pi }^4} + \frac{3 \sin(2\theta^{(2)}) \Bigl[ v_1^{(2)} (s-2 m_{\pi }^2) - 2 m_{\pi }^2 v_2^{(2)} \Bigr]}{F_{\pi }^2} \nonumber\\[0.05cm]
    &- \frac{8 m_{\pi }^4 \bigl(\sqrt{2} \cos(2\theta^{(2)}) - 2 \sin(2\theta^{(2)})\bigr) \Bigl[ L_5 (\Lambda _1-\Lambda _2) - L_8 (\Lambda _1-2 \Lambda _2) \Bigr]}{3 F_{\pi }^4} \nonumber\\[0.05cm]
    &- \frac{m_{\pi }^2 (\Lambda _1-2 \Lambda _2) \bigl(\sqrt{2} \cos(2\theta^{(2)}) - 2 \sin(2\theta^{(2)})\bigr)}{6 F_{\pi }^2}\,. \label{eq:NNLOTramp}
\end{align}
The one-loop contribution $\mathcal{M}^{\mathrm{Loop}}$ reads
\begin{align}
    \mathcal{M}^{\mathrm{Loop}} &= \frac{m_{\pi }^2 }{54 F_\pi^4} \big( \cos\theta^{(2)} - \sqrt{2} \sin\theta^{(2)} \big)^2 \bigg[ 4 \sqrt{2} \sin^4\theta^{(2)} \big( m_K^2 - m_{\pi }^2 \big) \nonumber\\[0.15cm]
    &\quad + \cos^4\theta^{(2)} \big( 5 \sqrt{2} m_{\pi }^2 - 8 \sqrt{2} m_K^2 \big) + \sin\theta^{(2)} \cos^3\theta^{(2)} \big( -8 m_K^2 - m_{\pi }^2 \big) \nonumber\\[0.15cm]
    &\quad + \sin^2\theta^{(2)} \cos^2\theta^{(2)} \big( 12 \sqrt{2} m_K^2 - 3 \sqrt{2} m_{\pi }^2 \big) \nonumber\\[0.15cm]
    &\quad + \sin^3\theta^{(2)} \cos\theta^{(2)} \big( 20 m_K^2 - 8 m_{\pi }^2 \big) \bigg] B_0(s, m_{\eta }^2, m_{\eta }^2) \nonumber\\[0.15cm]
    &+ \frac{m_{\pi }^2 }{54 F_\pi^4} \big( \sin\theta^{(2)} + \sqrt{2} \cos\theta^{(2)} \big)^2 \bigg[ \sqrt{2} \sin^4\theta^{(2)} \big( 8 m_K^2 - 5 m_{\pi }^2 \big) \nonumber\\[0.15cm]
    &\quad + 4 \sqrt{2} \cos^4\theta^{(2)} \big( m_{\pi }^2 - m_K^2 \big) + 4 \sin\theta^{(2)} \cos^3\theta^{(2)} \big( 5 m_K^2 - 2 m_{\pi }^2 \big) \nonumber\\[0.15cm]
    &\quad + 3 \sqrt{2} \sin^2\theta^{(2)} \cos^2\theta^{(2)} \big( m_{\pi }^2 - 4 m_K^2 \big) \nonumber\\[0.15cm]
    &\quad - \sin^3\theta^{(2)} \cos\theta^{(2)} \big( 8 m_K^2 + m_{\pi }^2 \big) \bigg] B_0(s, m_{\eta '}^2, m_{\eta '}^2) \nonumber\\[0.15cm]
    &- \frac{m_{\pi }^2 (m_{\pi }^2-2 s)}{12 F_\pi^4} \Bigl[ 2 \sqrt{2} \cos(2\theta^{(2)}) - \sin(2\theta^{(2)}) \Bigr] B_0(s, m_{\pi }^2, m_{\pi }^2) \nonumber\\[0.15cm]
    &- \frac{s}{24 F_\pi^4} \bigg[ 2 \sqrt{2} \sin^2\theta^{(2)} \big( 2 m_K^2 - m_{\pi }^2 \big) + 2 \sqrt{2} \cos^2\theta^{(2)} \big( m_{\pi }^2 - 2 m_K^2 \big) \nonumber\\[0.15cm]
    &\quad - \sin\theta^{(2)} \cos\theta^{(2)} \big( 8 m_K^2 + 3 m_{\eta '}^2 + 3 m_{\eta }^2 + 2 m_{\pi }^2 - 9 s \big) \bigg] B_0(s, m_K^2, m_K^2) \nonumber\\[0.15cm]
    &+ \frac{m_{\pi }^2(4 m_K^2-m_{\pi }^2)}{27 F_\pi^4} \big( \cos\theta^{(2)} - \sqrt{2} \sin\theta^{(2)} \big)^2 \Big[ -\sqrt{2} \sin^4\theta^{(2)} + 2 \sqrt{2} \cos^4\theta^{(2)} \nonumber\\[0.15cm]
    &\quad + 2 \sin\theta^{(2)} \cos^3\theta^{(2)} - 3 \sqrt{2} \sin^2\theta^{(2)} \cos^2\theta^{(2)} - 5 \sin^3\theta^{(2)} \cos\theta^{(2)} \Big] B_0(s, m_{\eta }^2, m_{\eta '}^2) \nonumber\\[0.05cm]
    &+ \Bigg\{ \frac{1}{216 F_\pi^4} \bigg[ -2 \sqrt{2} \cos^2\theta^{(2)} (2 m_K^2+m_{\pi }^2) \big( 8 m_K^2 + 3 m_{\eta }^2 + m_{\pi }^2 - 9 t \big) \nonumber\\[0.10cm]
    &\quad + 2 \sqrt{2} \sin^2\theta^{(2)} (2 m_K^2+m_{\pi }^2) \big( 8 m_K^2 + 3 m_{\eta '}^2 + m_{\pi }^2 - 9 t \big) \nonumber\\[0.20cm]
    &\quad + \sin\theta^{(2)} \cos\theta^{(2)} \Big( 3 m_{\eta }^2 (8 m_K^2 + 3 m_{\eta '}^2 + m_{\pi }^2 - 9 t) + 3 m_{\eta '}^2 (8 m_K^2 + m_{\pi }^2 - 9 t) \nonumber\\[0.10cm]
    &\quad - 144m_K^2t + 32 m_K^4 - 16 m_{\pi }^2 m_K^2 - 18 m_{\pi }^2 t - 7 m_{\pi }^4 + 81 t^2 \Big) \bigg] B_0(t, m_K^2, m_K^2) \nonumber\\[0.10cm]
    &\quad + \frac{m_{\pi }^4 }{9 F_\pi^4} \big( \cos\theta^{(2)} - \sqrt{2} \sin\theta^{(2)} \big)^2 \Big[ -\sqrt{2} \sin^2\theta^{(2)} + \sqrt{2} \cos^2\theta^{(2)} \nonumber\\[0.10cm]
    &\quad - \sin\theta^{(2)} \cos\theta^{(2)} \Big] B_0(t, m_{\eta }^2, m_{\pi }^2) + \frac{m_{\pi }^4 }{9 F_\pi^4} \big( \sin\theta^{(2)} + \sqrt{2} \cos\theta^{(2)} \big)^2 \Big[ -\sqrt{2} \sin^2\theta^{(2)} \nonumber\\[0.05cm]
    &\quad + \sqrt{2} \cos^2\theta^{(2)} - \sin\theta^{(2)} \cos\theta^{(2)} \Big] B_0(t, m_{\eta'}^2, m_{\pi}^2) \Bigg\} + (t \leftrightarrow u)\,.
\end{align}
The resulting expression for the amplitude $\mathcal{M}^{\mathrm{Tadpole}}$ is given by
\begin{align}
    \mathcal{M}^{\mathrm{Tadpole}} &= \frac{2 m_{\pi}^2 A_0(m_{\pi}^2)}{3 F_{\pi}^4} \Big( -\sqrt{2} \sin^2\theta^{(2)} + \sqrt{2} \cos^2\theta^{(2)} - \sin\theta^{(2)} \cos\theta^{(2)} \Big) \nonumber\\[0.05cm]
    &\quad - \frac{1}{15 F_{\pi}^4} \bigg[ 3 \sin\theta^{(2)} \cos\theta^{(2)} \big( m_{\eta}^2 + m_{\eta '}^2 \big) \nonumber\\[0.15cm]
    &\quad + 2 m_K^2 \big( -\sqrt{2} \sin^2\theta^{(2)} + \sqrt{2} \cos^2\theta^{(2)} - 4 \sin\theta^{(2)} \cos\theta^{(2)} \big) 
    \nonumber\\[0.15cm]
    &\quad + m_{\pi}^2 \big( 2 \sqrt{2} \sin^2\theta^{(2)} - 2 \sqrt{2} \cos^2\theta^{(2)} + 17 \sin\theta^{(2)} \cos\theta^{(2)} \big) \bigg] A_0(m_K^2)\,.
\end{align}
Here the loop functions are evaluated with dimensional regularization and renormalized in the $\overline{\mathrm{MS}}-1$ scheme~\cite{Gasser:1983yg,Gasser:1984gg}, with
\begin{align}
	A_0(m^2) &= -\frac{m^2}{16\pi^2}\ln\frac{m^2}{\mu^2}, \nonumber\\[0.15cm]
	B_0(s, m_a^2, m_b^2) &= \frac{1}{16 \pi^2}\left(1-\ln \frac{m_b^2}{\mu^2}+x_{+} \ln \frac{x_{+}-1}{x_{+}}+x_{-} \ln \frac{x_{-}-1}{x_{-}}\right)\,,
\end{align}
where $\mu$ is the renormalization scale, and
\begin{equation}
	x_{ \pm}=\frac{s+m_a^2-m_b^2}{2 s} \pm \frac{1}{2 s} \sqrt{\left(s+m_a^2-m_b^2\right)^2 - 4 s\left(m_a^2-i 0^{+}\right)}\,.
\end{equation}
In the case of equal masses, the one-loop two-point function reduces to 
\begin{align}
    B_0(s, m^2) = \frac{1}{16 \pi^2}\left[1-\ln \frac{m^2}{\mu^2}+\sigma(s) \ln \frac{\sigma(s)-1}{\sigma(s)+1}\right]\,,
\end{align}
with $\sigma(s) = \sqrt{1-4m^2/s}$.

\section{\texorpdfstring{$\boldsymbol{\pi\pi}$}{pipi} rescattering amplitudes within \texorpdfstring{large-$\boldsymbol{N_C}$}{large-NC} ChPT}
\label{app:pionscattering} 
 
In this appendix, we present the complete $\pi\pi$ rescattering amplitude $A(s,t,u)$ up to NNLO within the framework of large-$N_C$ ChPT. To this end, let us decompose the invariant amplitude $A(s,t,u)$ as
\begin{align}
    A^{\mathrm{NNLO}}(s,t,u) = A^{\mathrm{Tree}}(s,t,u) + A^{\mathrm{Loop}}(s,t,u) + A^{\mathrm{Tadpole}}(s,t,u)\,,
\end{align}
where $A^{\mathrm{Tree}}(s,t,u)$ denotes the purely tree-level contribution evaluated up to NNLO, while $A^{\mathrm{Loop}}(s,t,u)$ and $A^{\mathrm{Tadpole}}(s,t,u)$ represent the one-loop and NNLO tadpole contributions, respectively. Their explicit expressions are given, respectively, by
\begin{align}
    A^{\mathrm{Tree}}(s,t,u) &=  \frac{1}{3F_{\pi}^4} \Biggl[ 12 (2 L_1+L_3) (s-2 m_{\pi}^2)^2 + 24 (2L_4+L_5) m_{\pi}^2 (s-2m_\pi^2) \nonumber \\[-0.15cm]
	&\quad - 12 L_2 \bigl( -s^2 - 4 s m_{\pi}^2 + 12 m_{\pi}^4 - 2 (t^2+u^2) \bigr) + 48( 2 L_6+L_8) m_{\pi}^4 \nonumber \\[0.15cm]
	&\quad - 24 C_1 (s - 2 m_{\pi}^2)^3 + 24 (C_5+C_8) m_{\pi}^2 (s - 2 m_{\pi}^2)^2 \nonumber \\[0.15cm]
	&\quad + 96 (C_{14}+C_{17}+C_{26}) m_{\pi}^4 (s - 2 m_{\pi}^2) - 192 C_{29} m_{\pi}^4 (s - m_{\pi}^2) \nonumber \\[0.15cm]
	&\quad - 24 C_{10} m_{\pi}^2 (s^2 - 8 s m_{\pi}^2 + 12 m_{\pi}^4 - t^2 - u^2) + 288( C_{19}+C_{31}) m_{\pi}^6 \nonumber \\[0.15cm]
	&\quad - 96 C_{12} \bigl( -2 m_{\pi}^4 (s+2 t) + m_{\pi}^2 t (s+t) + 5 m_{\pi}^6 \bigr) + 3 F_{\pi}^2 (s - m_{\pi}^2) \nonumber \\[0.15cm]
	&\quad + 96 C_{22} m_{\pi}^2 (2 m_{\pi}^2 - t) (-s + 2 m_{\pi}^2 - t) \nonumber \\[0.15cm]
	&\quad - 48 C_{25} m_{\pi}^2 (-s^2 + s t + 4 m_{\pi}^4 - 4 m_{\pi}^2 t + t^2) \nonumber \\[0.15cm]
	&\quad + \frac{384 (L_5-2 L_8) m_{\pi}^4 \bigl( L_3 (2 m_{\pi}^2-s) + 2 L_2 m_{\pi}^2 \bigr)}{F_{\pi}^2} \nonumber \\[-0.10cm]
	&\quad - \frac{96 m_{\pi}^4 \bigl( L_5^2 (9 m_{\pi}^2-3 s) + 4 L_8 L_5 (3 s-8 m_{\pi}^2) + 16 L_8^2 m_{\pi}^2 \bigr)}{F_{\pi}^2} \Biggr]\,,\\[0.2cm]
    A^{\mathrm{Loop}}(s,t,u) &= \frac{m_\pi^4 (\cos\theta^{(2)} - \sqrt{2}\sin\theta^{(2)})^4}{18F_\pi^4} B_0(s,m_\eta^2,m_\eta^2) \nonumber \\[0.15cm]
	&\quad + \frac{m_\pi^4 (\cos\theta^{(2)} - \sqrt{2}\sin\theta^{(2)})^2 (\sin\theta^{(2)} + \sqrt{2}\cos\theta^{(2)})^2}{9 F_\pi^4} B_0(s,m_\eta^2,m_{\eta'}^2) \nonumber \\[0.15cm]
	&\quad + \frac{m_\pi^4 (\sin\theta^{(2)} + \sqrt{2}\cos\theta^{(2)})^4}{18 F_\pi^4} B_0(s,m_{\eta'}^2,m_{\eta'}^2) \nonumber \\[0.15cm]
	&\quad - \frac{(m_\pi^4-s^2) B_0(s,m_\pi^2,m_\pi^2)}{2 F_\pi^4} + \frac{s^2 B_0(s,m_K^2,m_K^2)}{8 F_\pi^4} \nonumber \\[0.15cm]
	&\quad + \frac{\bigl( -2 m_\pi^2 (2 s+5 t) + 14 m_\pi^4 + t (s+2 t) \bigr) B_0(t,m_\pi^2,m_\pi^2)}{3 F_\pi^4} \nonumber \\[0.15cm]
	&\quad + \frac{\bigl( 2 m_\pi^2 (s+t) - 10 m_\pi^4 + s t + t^2 + u^2 \bigr) B_0(u,m_\pi^2,m_\pi^2)}{3 F_\pi^4} \nonumber \\[0.15cm]
	&\quad - \frac{(4 m_K^2-t) (-4 m_\pi^2 + 2 s + t) B_0(t,m_K^2,m_K^2)}{12 F_\pi^4} \nonumber \\[0.15cm]
	&\quad - \frac{(s-t) (4 m_K^2-u) B_0(u,m_K^2,m_K^2)}{12 F_\pi^4}\,,\\[0.2cm]
    A^{\mathrm{Tadpole}}(s,t,u) &= \frac{(9s - 12m_\pi^2) A_0(m_K^2) + 4(7s - 4m_\pi^2) A_0(m_\pi^2)}{36 F_\pi^4}\,.
\end{align}

\bibliographystyle{JHEP}
\bibliography{reference}

\providecommand{\href}[2]{#2}\begingroup\raggedright\begin{thebibliography}{10}

\bibitem{Weinberg:1978kz}
S.~Weinberg, \emph{{Phenomenological Lagrangians}}, \href{https://doi.org/10.1016/0378-4371(79)90223-1}{\emph{Physica A} {\bfseries 96} (1979) 327}.

\bibitem{Gasser:1982ap}
J.~Gasser and H.~Leutwyler, \emph{{Quark Masses}}, \href{https://doi.org/10.1016/0370-1573(82)90035-7}{\emph{Phys. Rept.} {\bfseries 87} (1982) 77}.

\bibitem{Gasser:1984gg}
J.~Gasser and H.~Leutwyler, \emph{{Chiral Perturbation Theory: Expansions in the Mass of the Strange Quark}}, \href{https://doi.org/10.1016/0550-3213(85)90492-4}{\emph{Nucl. Phys. B} {\bfseries 250} (1985) 465}.

\bibitem{Gasser:1983yg}
J.~Gasser and H.~Leutwyler, \emph{{Chiral Perturbation Theory to One Loop}}, \href{https://doi.org/10.1016/0003-4916(84)90242-2}{\emph{Annals Phys.} {\bfseries 158} (1984) 142}.

\bibitem{Pich:1995bw}
A.~Pich, \emph{{Chiral perturbation theory}}, \href{https://doi.org/10.1088/0034-4885/58/6/001}{\emph{Rept. Prog. Phys.} {\bfseries 58} (1995) 563} [\href{https://arxiv.org/abs/hep-ph/9502366}{{\ttfamily hep-ph/9502366}}].

\bibitem{Ecker:1994gg}
G.~Ecker, \emph{{Chiral perturbation theory}}, \href{https://doi.org/10.1016/0146-6410(95)00041-G}{\emph{Prog. Part. Nucl. Phys.} {\bfseries 35} (1995) 1} [\href{https://arxiv.org/abs/hep-ph/9501357}{{\ttfamily hep-ph/9501357}}].

\bibitem{Scherer:2002tk}
S.~Scherer, \emph{{Introduction to chiral perturbation theory}}, {\emph{Adv. Nucl. Phys.} {\bfseries 27} (2003) 277} [\href{https://arxiv.org/abs/hep-ph/0210398}{{\ttfamily hep-ph/0210398}}].

\bibitem{Bernard:2006gx}
V.~Bernard and U.-G.~Meissner, \emph{{Chiral perturbation theory}}, \href{https://doi.org/10.1146/annurev.nucl.56.080805.140449}{\emph{Ann. Rev. Nucl. Part. Sci.} {\bfseries 57} (2007) 33} [\href{https://arxiv.org/abs/hep-ph/0611231}{{\ttfamily hep-ph/0611231}}].

\bibitem{Kaiser:2000gs}
R.~Kaiser and H.~Leutwyler, \emph{{Large $N_c$ in chiral perturbation theory}}, \href{https://doi.org/10.1007/s100520000499}{\emph{Eur. Phys. J. C} {\bfseries 17} (2000) 623} [\href{https://arxiv.org/abs/hep-ph/0007101}{{\ttfamily hep-ph/0007101}}].

\bibitem{Escribano:2010wt}
R.~Escribano, P.~Masjuan and J.J.~Sanz-Cillero, \emph{{Chiral dynamics predictions for $\eta'\to\eta \pi \pi$}}, \href{https://doi.org/10.1007/JHEP05(2011)094}{\emph{JHEP} {\bfseries 05} (2011) 094} [\href{https://arxiv.org/abs/1011.5884}{{\ttfamily 1011.5884}}].

\bibitem{Gonzalez-Solis:2018xnw}
S.~Gonz{\`a}lez-Sol{\'\i}s and E.~Passemar, \emph{{$\eta ^{\prime }\rightarrow \eta \pi \pi $ decays in unitarized resonance chiral theory}}, \href{https://doi.org/10.1140/epjc/s10052-018-6238-6}{\emph{Eur. Phys. J. C} {\bfseries 78} (2018) 758} [\href{https://arxiv.org/abs/1807.04313}{{\ttfamily 1807.04313}}].

\bibitem{Beisert:2002ad}
N.~Beisert and B.~Borasoy, \emph{{The $\eta^\prime \to \eta \pi \pi$ decay in $U(3)$ chiral perturbation theory}}, \href{https://doi.org/10.1016/S0375-9474(02)00652-8}{\emph{Nucl. Phys. A} {\bfseries 705} (2002) 433} [\href{https://arxiv.org/abs/hep-ph/0201289}{{\ttfamily hep-ph/0201289}}].

\bibitem{Borasoy:2005du}
B.~Borasoy and R.~Nissler, \emph{{Hadronic $\eta$ and $\eta^\prime$ decays}}, \href{https://doi.org/10.1140/epja/i2005-10188-9}{\emph{Eur. Phys. J. A} {\bfseries 26} (2005) 383} [\href{https://arxiv.org/abs/hep-ph/0510384}{{\ttfamily hep-ph/0510384}}].

\bibitem{Borasoy:2006uv}
B.~Borasoy, U.-G.~Meissner and R.~Nissler, \emph{{On the extraction of the quark mass ratio $(m_d - m_u)/m_s$ from $\Gamma(\eta^{\prime} \to \pi^0 \pi^+ \pi^-)/\Gamma(\eta^\prime \to \eta \pi^+ \pi^-)$}}, \href{https://doi.org/10.1016/j.physletb.2006.10.020}{\emph{Phys. Lett. B} {\bfseries 643} (2006) 41} [\href{https://arxiv.org/abs/hep-ph/0609010}{{\ttfamily hep-ph/0609010}}].

\bibitem{Kubis:2009sb}
B.~Kubis and S.P.~Schneider, \emph{{The Cusp effect in $\eta^\prime \to \eta \pi \pi$ decays}}, \href{https://doi.org/10.1140/epjc/s10052-009-1054-7}{\emph{Eur. Phys. J. C} {\bfseries 62} (2009) 511} [\href{https://arxiv.org/abs/0904.1320}{{\ttfamily 0904.1320}}].

\bibitem{Isken:2017dkw}
T.~Isken, B.~Kubis, S.P.~Schneider and P.~Stoffer, \emph{{Dispersion relations for $\eta'\rightarrow \eta \pi \pi $}}, \href{https://doi.org/10.1140/epjc/s10052-017-5024-1}{\emph{Eur. Phys. J. C} {\bfseries 77} (2017) 489} [\href{https://arxiv.org/abs/1705.04339}{{\ttfamily 1705.04339}}].

\bibitem{Akdag:2021efj}
H.~Akdag, T.~Isken and B.~Kubis, \emph{{Patterns of C- and CP-violation in hadronic $\eta$ and $\eta^\prime$ three-body decays}}, \href{https://doi.org/10.1007/JHEP02(2022)137}{\emph{JHEP} {\bfseries 02} (2022) 137} [\href{https://arxiv.org/abs/2111.02417}{{\ttfamily 2111.02417}}].

\bibitem{Fariborz:1999gr}
A.H.~Fariborz and J.~Schechter, \emph{{$\eta^\prime\to\eta\pi\pi$ decay as a probe of a possible lowest lying scalar nonet}}, \href{https://doi.org/10.1103/PhysRevD.60.034002}{\emph{Phys. Rev. D} {\bfseries 60} (1999) 034002} [\href{https://arxiv.org/abs/hep-ph/9902238}{{\ttfamily hep-ph/9902238}}].

\bibitem{Singh:1975aq}
C.A.~Singh and J.~Pasupathy, \emph{{On the Decay Modes of the Meson $\eta^{\prime}(958)$ and Chiral Symmetry Breaking}}, \href{https://doi.org/10.1103/PhysRevLett.35.1193}{\emph{Phys. Rev. Lett.} {\bfseries 35} (1975) 1193}.

\bibitem{Herrera-Siklody:1999sbe}
P.~Herrera-Siklody, \emph{{$\eta$ and $\eta^\prime$ hadronic decays in $U_L(3)\otimes U_R(3)$ chiral perturbation theory}},  \href{https://arxiv.org/abs/hep-ph/9902446}{{\ttfamily hep-ph/9902446}}.

\bibitem{Guo:2015xva}
X.-K.~Guo, Z.-H.~Guo, J.A.~Oller and J.J.~Sanz-Cillero, \emph{{Scrutinizing the $\eta$-$\eta'$ mixing, masses and pseudoscalar decay constants in the framework of U(3) chiral effective field theory}}, \href{https://doi.org/10.1007/JHEP06(2015)175}{\emph{JHEP} {\bfseries 06} (2015) 175} [\href{https://arxiv.org/abs/1503.02248}{{\ttfamily 1503.02248}}].

\bibitem{Bickert:2016fgy}
P.~Bickert, P.~Masjuan and S.~Scherer, \emph{{$\eta$-$\eta'$ Mixing in Large-$N_c$ Chiral Perturbation Theory}}, \href{https://doi.org/10.1103/PhysRevD.95.054023}{\emph{Phys. Rev. D} {\bfseries 95} (2017) 054023} [\href{https://arxiv.org/abs/1612.05473}{{\ttfamily 1612.05473}}].

\bibitem{Feldmann:1998vh}
T.~Feldmann, P.~Kroll and B.~Stech, \emph{{Mixing and decay constants of pseudoscalar mesons}}, \href{https://doi.org/10.1103/PhysRevD.58.114006}{\emph{Phys. Rev. D} {\bfseries 58} (1998) 114006} [\href{https://arxiv.org/abs/hep-ph/9802409}{{\ttfamily hep-ph/9802409}}].

\bibitem{Guevara:2025weo}
A.~Guevara, F.-K.~Guo and H.-J.~Jing, \emph{{Novel method for determining the light quark mass ratio using $\eta'\to\eta \pi\pi$ decays}},  \href{https://arxiv.org/abs/2502.02837}{{\ttfamily 2502.02837}}.

\bibitem{Dorofeev:2006fb}
V.~Dorofeev et~al., \emph{{Study of $\eta^\prime\to\eta\pi^+\pi^-$ Dalitz plot}}, \href{https://doi.org/10.1016/j.physletb.2007.05.060}{\emph{Phys. Lett. B} {\bfseries 651} (2007) 22} [\href{https://arxiv.org/abs/hep-ph/0607044}{{\ttfamily hep-ph/0607044}}].

\bibitem{BESIII:2010niv}
{\scshape BESIII} collaboration, \emph{{Measurement of the Matrix Element for the Decay $\eta^{\prime} \to \eta \pi^+\pi^-$}}, \href{https://doi.org/10.1103/PhysRevD.83.012003}{\emph{Phys. Rev. D} {\bfseries 83} (2011) 012003} [\href{https://arxiv.org/abs/1012.1117}{{\ttfamily 1012.1117}}].

\bibitem{BESIII:2017djm}
{\scshape BESIII} collaboration, \emph{{Measurement of the matrix elements for the decays $\eta^{\prime}\rightarrow\eta\pi^+\pi^-$ and $\eta^{\prime}\rightarrow\eta\pi^0\pi^0$}}, \href{https://doi.org/10.1103/PhysRevD.97.012003}{\emph{Phys. Rev. D} {\bfseries 97} (2018) 012003} [\href{https://arxiv.org/abs/1709.04627}{{\ttfamily 1709.04627}}].

\bibitem{Blik:2009zz}
A.M.~Blik et~al., \emph{{Measurement of the matrix element for the decay $\eta^\prime\to\eta\pi^0\pi^0$ with the GAMS-$4\pi$ spectrometer}}, \href{https://doi.org/10.1134/S1063778809020045}{\emph{Phys. Atom. Nucl.} {\bfseries 72} (2009) 231}.

\bibitem{Adlarson:2017wlz}
P.~Adlarson et~al., \emph{{Measurement of the decay $\eta^{\prime}\to\pi^{0}\pi^{0}\eta$ at MAMI}}, \href{https://doi.org/10.1103/PhysRevD.98.012001}{\emph{Phys. Rev. D} {\bfseries 98} (2018) 012001} [\href{https://arxiv.org/abs/1709.04230}{{\ttfamily 1709.04230}}].

\bibitem{BESIII:2022tas}
{\scshape BESIII} collaboration, \emph{{Evidence for the Cusp Effect in $\eta'$ Decays into $\eta\pi^0\pi^0$}}, \href{https://doi.org/10.1103/PhysRevLett.130.081901}{\emph{Phys. Rev. Lett.} {\bfseries 130} (2023) 081901} [\href{https://arxiv.org/abs/2207.01004}{{\ttfamily 2207.01004}}].

\bibitem{CLAS:2003umf}
{\scshape CLAS} collaboration, \emph{{The CEBAF Large Acceptance Spectrometer (CLAS)}}, \href{https://doi.org/10.1016/S0168-9002(03)01001-5}{\emph{Nucl. Instrum. Meth. A} {\bfseries 503} (2003) 513}.

\bibitem{GlueX:2020idb}
{\scshape GlueX} collaboration, \emph{{The GLUEX beamline and detector}}, \href{https://doi.org/10.1016/j.nima.2020.164807}{\emph{Nucl. Instrum. Meth. A} {\bfseries 987} (2021) 164807} [\href{https://arxiv.org/abs/2005.14272}{{\ttfamily 2005.14272}}].

\bibitem{Dudek:2012vr}
J.~Dudek et~al., \emph{{Physics Opportunities with the 12 GeV Upgrade at Jefferson Lab}}, \href{https://doi.org/10.1140/epja/i2012-12187-1}{\emph{Eur. Phys. J. A} {\bfseries 48} (2012) 187} [\href{https://arxiv.org/abs/1208.1244}{{\ttfamily 1208.1244}}].

\bibitem{Gan:2020aco}
L.~Gan, B.~Kubis, E.~Passemar and S.~Tulin, \emph{{Precision tests of fundamental physics with $\eta$ and $\eta'$ mesons}}, \href{https://doi.org/10.1016/j.physrep.2021.11.001}{\emph{Phys. Rept.} {\bfseries 945} (2022) 1} [\href{https://arxiv.org/abs/2007.00664}{{\ttfamily 2007.00664}}].

\bibitem{Chen:2024wad}
X.~Chen et~al., \emph{{Plan for a super $\eta$ factory at Huizhou accelerator complex}}, \href{https://doi.org/10.1007/s41365-025-01708-1}{\emph{Nucl. Sci. Tech.} {\bfseries 36} (2025) 137} [\href{https://arxiv.org/abs/2407.00874}{{\ttfamily 2407.00874}}].

\bibitem{Achasov:2023gey}
M.~Achasov et~al., \emph{{STCF conceptual design report (Volume 1): Physics $\&$ detector}}, \href{https://doi.org/10.1007/s11467-023-1333-z}{\emph{Front. Phys. (Beijing)} {\bfseries 19} (2024) 14701} [\href{https://arxiv.org/abs/2303.15790}{{\ttfamily 2303.15790}}].

\bibitem{Charm-TauFactory:2013cnj}
{\scshape Charm-Tau Factory} collaboration, \emph{{Project of a Super Charm-Tau factory at the Budker Institute of Nuclear Physics in Novosibirsk}}, \href{https://doi.org/10.1134/S1063778813090032}{\emph{Phys. Atom. Nucl.} {\bfseries 76} (2013) 1072}.

\bibitem{Fang:2021wes}
S.-s.~Fang, B.~Kubis and A.~Kupsc, \emph{{What can we learn about light-meson interactions at electron{\textendash}positron colliders?}}, \href{https://doi.org/10.1016/j.ppnp.2021.103884}{\emph{Prog. Part. Nucl. Phys.} {\bfseries 120} (2021) 103884} [\href{https://arxiv.org/abs/2102.05922}{{\ttfamily 2102.05922}}].

\bibitem{Oller:1998zr}
J.A.~Oller and E.~Oset, \emph{{N/D description of two meson amplitudes and chiral symmetry}}, \href{https://doi.org/10.1103/PhysRevD.60.074023}{\emph{Phys. Rev. D} {\bfseries 60} (1999) 074023} [\href{https://arxiv.org/abs/hep-ph/9809337}{{\ttfamily hep-ph/9809337}}].

\bibitem{Pallante:1999qf}
E.~Pallante and A.~Pich, \emph{{Strong enhancement of $\epsilon^\prime / \epsilon$ through final state interactions}}, \href{https://doi.org/10.1103/PhysRevLett.84.2568}{\emph{Phys. Rev. Lett.} {\bfseries 84} (2000) 2568} [\href{https://arxiv.org/abs/hep-ph/9911233}{{\ttfamily hep-ph/9911233}}].

\bibitem{Oset:1999cq}
E.~Oset, S.~Hirenzaki, E.~Marco, J.A.~Oller, J.R.~Pelaez and H.~Toki, \emph{{The $\phi\to\pi^+\pi^-$ and $\phi$ radiative decays within a chiral unitary approach}}, {\emph{Frascati Phys. Ser.} {\bfseries 16} (1999) 581} [\href{https://arxiv.org/abs/hep-ph/0002189}{{\ttfamily hep-ph/0002189}}].

\bibitem{Nieves:1999bx}
J.~Nieves and E.~Ruiz~Arriola, \emph{{Bethe-Salpeter approach for unitarized chiral perturbation theory}}, \href{https://doi.org/10.1016/S0375-9474(00)00321-3}{\emph{Nucl. Phys. A} {\bfseries 679} (2000) 57} [\href{https://arxiv.org/abs/hep-ph/9907469}{{\ttfamily hep-ph/9907469}}].

\bibitem{Oller:2000fj}
J.A.~Oller and U.G.~Meissner, \emph{{Chiral dynamics in the presence of bound states: Kaon nucleon interactions revisited}}, \href{https://doi.org/10.1016/S0370-2693(01)00078-8}{\emph{Phys. Lett. B} {\bfseries 500} (2001) 263} [\href{https://arxiv.org/abs/hep-ph/0011146}{{\ttfamily hep-ph/0011146}}].

\bibitem{Jamin:2000wn}
M.~Jamin, J.A.~Oller and A.~Pich, \emph{{S-wave $K\pi$ scattering in chiral perturbation theory with resonances}}, \href{https://doi.org/10.1016/S0550-3213(00)00479-X}{\emph{Nucl. Phys. B} {\bfseries 587} (2000) 331} [\href{https://arxiv.org/abs/hep-ph/0006045}{{\ttfamily hep-ph/0006045}}].

\bibitem{Oller:2000ma}
J.A.~Oller, E.~Oset and A.~Ramos, \emph{{Chiral unitary approach to meson meson and meson - baryon interactions and nuclear applications}}, \href{https://doi.org/10.1016/S0146-6410(00)00104-6}{\emph{Prog. Part. Nucl. Phys.} {\bfseries 45} (2000) 157} [\href{https://arxiv.org/abs/hep-ph/0002193}{{\ttfamily hep-ph/0002193}}].

\bibitem{Guo:2011pa}
Z.-H.~Guo and J.A.~Oller, \emph{{Resonances from meson-meson scattering in $U(3)$ $\chi$PT}}, \href{https://doi.org/10.1103/PhysRevD.84.034005}{\emph{Phys. Rev. D} {\bfseries 84} (2011) 034005} [\href{https://arxiv.org/abs/1104.2849}{{\ttfamily 1104.2849}}].

\bibitem{Chew:1960iv}
G.F.~Chew and S.~Mandelstam, \emph{{Theory of low-energy pion pion interactions}}, \href{https://doi.org/10.1103/PhysRev.119.467}{\emph{Phys. Rev.} {\bfseries 119} (1960) 467}.

\bibitem{Witten:1979vv}
E.~Witten, \emph{{Current Algebra Theorems for the U(1) Goldstone Boson}}, \href{https://doi.org/10.1016/0550-3213(79)90031-2}{\emph{Nucl. Phys. B} {\bfseries 156} (1979) 269}.

\bibitem{Veneziano:1979ec}
G.~Veneziano, \emph{{U(1) Without Instantons}}, \href{https://doi.org/10.1016/0550-3213(79)90332-8}{\emph{Nucl. Phys. B} {\bfseries 159} (1979) 213}.

\bibitem{Leutwyler:1996sa}
H.~Leutwyler, \emph{{Bounds on the light quark masses}}, \href{https://doi.org/10.1016/0370-2693(96)85876-X}{\emph{Phys. Lett. B} {\bfseries 374} (1996) 163} [\href{https://arxiv.org/abs/hep-ph/9601234}{{\ttfamily hep-ph/9601234}}].

\bibitem{Bijnens:2001bb}
J.~Bijnens, L.~Girlanda and P.~Talavera, \emph{{The Anomalous chiral Lagrangian of order $p^6$}}, \href{https://doi.org/10.1007/s100520100887}{\emph{Eur. Phys. J. C} {\bfseries 23} (2002) 539} [\href{https://arxiv.org/abs/hep-ph/0110400}{{\ttfamily hep-ph/0110400}}].

\bibitem{Ebertshauser:2001nj}
T.~Ebertshauser, H.W.~Fearing and S.~Scherer, \emph{{The Anomalous chiral perturbation theory meson Lagrangian to order $p^6$ revisited}}, \href{https://doi.org/10.1103/PhysRevD.65.054033}{\emph{Phys. Rev. D} {\bfseries 65} (2002) 054033} [\href{https://arxiv.org/abs/hep-ph/0110261}{{\ttfamily hep-ph/0110261}}].

\bibitem{Bijnens:1999sh}
J.~Bijnens, G.~Colangelo and G.~Ecker, \emph{{The Mesonic chiral Lagrangian of order $p^6$}}, \href{https://doi.org/10.1088/1126-6708/1999/02/020}{\emph{JHEP} {\bfseries 02} (1999) 020} [\href{https://arxiv.org/abs/hep-ph/9902437}{{\ttfamily hep-ph/9902437}}].

\bibitem{Herrera-Siklody:1996tqr}
P.~Herrera-Siklody, J.I.~Latorre, P.~Pascual and J.~Taron, \emph{{Chiral effective Lagrangian in the large-$N_c$ limit: The Nonet case}}, \href{https://doi.org/10.1016/S0550-3213(97)00260-5}{\emph{Nucl. Phys. B} {\bfseries 497} (1997) 345} [\href{https://arxiv.org/abs/hep-ph/9610549}{{\ttfamily hep-ph/9610549}}].

\bibitem{ParticleDataGroup:2024cfk}
{\scshape Particle Data Group} collaboration, \emph{{Review of particle physics}}, \href{https://doi.org/10.1103/PhysRevD.110.030001}{\emph{Phys. Rev. D} {\bfseries 110} (2024) 030001}.

\bibitem{KLOE:2006guu}
{\scshape KLOE} collaboration, \emph{{Measurement of the pseudoscalar mixing angle and $\eta^\prime$ gluonium content with KLOE detector}}, \href{https://doi.org/10.1016/j.physletb.2007.03.032}{\emph{Phys. Lett. B} {\bfseries 648} (2007) 267} [\href{https://arxiv.org/abs/hep-ex/0612029}{{\ttfamily hep-ex/0612029}}].

\bibitem{Cronin:1967jq}
J.A.~Cronin, \emph{{Phenomenological model of strong and weak interactions in chiral $U(3) \times U(3)$}}, \href{https://doi.org/10.1103/PhysRev.161.1483}{\emph{Phys. Rev.} {\bfseries 161} (1967) 1483}.

\bibitem{Schwinger:1968zz}
J.~Schwinger, \emph{{Chiral Transformations}}, \href{https://doi.org/10.1103/PhysRev.167.1432}{\emph{Phys. Rev.} {\bfseries 167} (1968) 1432}.

\bibitem{Schechter:1971efy}
J.~Schechter and Y.~Ueda, \emph{{General treatment of the breaking of chiral symmetry and scale invariance in the $SU(3)$ $\sigma$ model}}, \href{https://doi.org/10.1103/PhysRevD.3.2874}{\emph{Phys. Rev. D} {\bfseries 3} (1971) 2874}.

\bibitem{Weisz:1972gm}
P.~Weisz, Riazuddin and S.~Oneda, \emph{{Chiral and dilational symmetry breaking and the $\eta^\prime \to \eta \pi \pi$ decay rate}}, \href{https://doi.org/10.1103/PhysRevD.5.2264}{\emph{Phys. Rev. D} {\bfseries 5} (1972) 2264}.

\bibitem{DiVecchia:1980vpx}
P.~Di~Vecchia, F.~Nicodemi, R.~Pettorino and G.~Veneziano, \emph{{Large $N$, Chiral Approach to Pseudoscalar Masses, Mixings and Decays}}, \href{https://doi.org/10.1016/0550-3213(81)90356-4}{\emph{Nucl. Phys. B} {\bfseries 181} (1981) 318}.

\bibitem{Fajfer:1987ij}
S.~Fajfer and J.M.~Gerard, \emph{{Hadronic Decays of $\eta$ and $\eta^\prime$ in the Large $N$ Limit}}, \href{https://doi.org/10.1007/BF01548448}{\emph{Z. Phys. C} {\bfseries 42} (1989) 431}.

\bibitem{Akhoury:1988as}
R.~Akhoury and M.~Leurer, \emph{{Low-energy Effective Lagrangian Description of $\eta$ and $\eta^\prime$ Decays}}, \href{https://doi.org/10.1007/BF02430620}{\emph{Z. Phys. C} {\bfseries 43} (1989) 145}.

\bibitem{Bijnens:2014lea}
J.~Bijnens and G.~Ecker, \emph{{Mesonic low-energy constants}}, \href{https://doi.org/10.1146/annurev-nucl-102313-025528}{\emph{Ann. Rev. Nucl. Part. Sci.} {\bfseries 64} (2014) 149} [\href{https://arxiv.org/abs/1405.6488}{{\ttfamily 1405.6488}}].

\bibitem{Pan:2023qja}
H.-X.~Pan, D.-K.~Kong, Q.-Y.~Wen and S.-Z.~Jiang, \emph{{Bayesian method for fitting the low-energy constants in chiral perturbation theory}}, \href{https://doi.org/10.1007/s11467-024-1430-7}{\emph{Front. Phys. (Beijing)} {\bfseries 19} (2024) 64203} [\href{https://arxiv.org/abs/2311.10423}{{\ttfamily 2311.10423}}].

\bibitem{Etkin:1981sg}
A.~Etkin et~al., \emph{{Amplitude Analysis of the $K^0_S K^0_S$ System Produced in the Reaction $\pi^- p \to K^0_S K^0_S n$ at $23$ GeV/c}}, \href{https://doi.org/10.1103/PhysRevD.25.1786}{\emph{Phys. Rev. D} {\bfseries 25} (1982) 1786}.

\bibitem{Watson:1954uc}
K.M.~Watson, \emph{{Some general relations between the photoproduction and scattering of $\pi$ mesons}}, \href{https://doi.org/10.1103/PhysRev.95.228}{\emph{Phys. Rev.} {\bfseries 95} (1954) 228}.

\bibitem{Garcia-Martin:2011iqs}
R.~Garcia-Martin, R.~Kaminski, J.R.~Pelaez, J.~Ruiz~de Elvira and F.J.~Yndurain, \emph{{The Pion-pion scattering amplitude. IV: Improved analysis with once subtracted Roy-like equations up to 1100 MeV}}, \href{https://doi.org/10.1103/PhysRevD.83.074004}{\emph{Phys. Rev. D} {\bfseries 83} (2011) 074004} [\href{https://arxiv.org/abs/1102.2183}{{\ttfamily 1102.2183}}].

\bibitem{Roy:1971tc}
S.M.~Roy, \emph{{Exact integral equation for pion pion scattering involving only physical region partial waves}}, \href{https://doi.org/10.1016/0370-2693(71)90724-6}{\emph{Phys. Lett. B} {\bfseries 36} (1971) 353}.

\end{thebibliography}\endgroup

\end{document}